\documentclass[manuscript,screen]{acmart}

\usepackage{tikz}
\usepackage{pgfplots}
\pgfplotsset{compat=newest}
\usepackage{pgfplotstable}

\usepackage{multirow}
\usepackage{booktabs}
\usepackage{tabularx}
\usepackage{longtable}
\usepackage{array}
\newcolumntype{M}[1]{>{\centering\arraybackslash}m{#1}}
\newcolumntype{R}[1]{>{\raggedleft\arraybackslash}m{#1}}
\newcolumntype{L}[1]{>{\raggedright\arraybackslash}m{#1}}
\newcolumntype{Y}{>{\raggedright\arraybackslash}X}

\usepackage{enumitem}

\acmJournal{CSUR}

\setcopyright{acmcopyright}
\copyrightyear{2025}
\acmYear{2025}
\acmDOI{XXXXXXX.XXXXXXX}

\begin{document}

\title{Cyber Security Educational Games for Children: A Systematic Literature Review}

\author{Temesgen Kitaw Damenu}
\authornote{Co-corresponding authors}
\email{td329@kent.ac.uk}
\orcid{0009-0002-2599-4286}
\affiliation{
  \department[0]{School of Computing}
  \department[1]{Institute of Cyber Security for Society (iCSS)}
  \institution{University of Kent}
  \city{Canterbury}
  \country{UK}
  \postcode{CT2 7FS}
}
\author{İnci Zaim Gökbay}
\authornote{Part of her work was done when she was visiting the University of Kent, UK.}
\email{inci.gokbay@istanbul.edu.tr}
\orcid{0000-0002-4488-1642}
\affiliation{
  \department{Department of Artificial Intelligence and Data Engineering}
  \department{Faculty of Computer and Information Technologies}
  \institution{Istanbul University}
  \city{Istanbul}
  \country{Türkiye}
  \postcode{34134}
}

\author{Alexandra Covaci}
\email{a.covaci@kent.ac.uk}
\orcid{0000-0002-3205-2273}
\affiliation{
  \department[0]{School of Engineering, Mathematics and Physics}
  \department[1]{Institute of Cyber Security for Society (iCSS)}
  \institution{University of Kent}
  \city{Canterbury}
  \country{UK}
  \postcode{CT2 7FS}
}
\author{Shujun Li}
\authornotemark[1]
\email{s.j.li@kent.ac.uk}
\orcid{0000-0001-5628-7328}
\affiliation{
  \department[0]{School of Computing}
  \department[1]{Institute of Cyber Security for Society (iCSS)}
  \institution{University of Kent}
  \city{Canterbury}
  \country{UK}
  \postcode{CT2 7FS}
}

\renewcommand{\shortauthors}{T.~K. Damenu et al.}

\begin{abstract}
Educational games have been widely used to teach children about cyber security. This systematic literature review reveals evidence of positive learning outcomes, after analysing 91 such games reported in 68 papers published between 2010 and 2024. However, critical gaps have also been identified regarding the design processes and the methodological rigour, including lack of systematic design, misalignment between proposed and achieved learning outcomes, rare use of control groups, limited discussions on ethical considerations, and underutilisation of emerging technologies. We recommend multiple future research directions, e.g., a hybrid approach to game design and evaluation that combines bottom-up and top-down approaches.
\end{abstract}

% The applicable cyber security concepts accoding to ACM classification



%\received{XX XX XXXX}
%\received[revised]{XX XX XXXX}
%\received[accepted]{XX XX XXXX}

\maketitle

\section{Introduction}
\label{sec:introduction} 

The Internet significantly improves children's daily activities. It facilitates education related tasks, communication with others, exploration of personal interests and other activities. While online engagement supports children's development, it also presents risks. These risks include phishing, cyber bullying, access to age-inappropriate content, misinformation, and privacy breaches~\cite{maqsood_design_2021}. Cyber threats targeting children are increasing significantly, largely due to their growing online presence, limited cyber security awareness, and the increasing sophistication of online threats across various platforms, including social media~\cite{quayyum2021cyber, Ofcom2024}. A report from the UK's online safety regulator (Ofcom)~\cite{Ofcom2024} revealed that nearly all (99\%) children in the UK spend time online and 68\% of 8--17 years old children use social media applications. The majority (51\%) of children under 13 years of age have their own social media accounts and a higher proportion of 5--7 years old children have their own profiles despite the minimum age limit set for account creation on these platforms (which is 13 for most of them).

This context underscore the importance of enhancing cyber security awareness among children and cultivating positive attitudes towards cyber security. Educating them about cyber security is not only vital for their personal well-being but also has a broader impact on the cyber security of society. Furthermore, children can play an important role in shaping the cyber security culture of the future generations.

Cyber security education for children can be provided through various techniques including teacher-led classroom training, e-learning, text-based training, video-based training and game-based learning~\cite{tschakert2019effectiveness, saglam_systematic_2023}. Game-based learning is often mentioned as an ideal approach to provide cyber security education for children~\cite{maqsood_design_2021}. It offers a safe environment for practising favourable behaviours and developing critical thinking and reasoning skills related with secure use of technology~\cite{maqsood_design_2021}. However, the current state of cyber security educational games for children is not systematically analysed.

This systematic literature review (SLR) aims to critically analyses and synthesises existing research, offering a comprehensive review of the current landscape of cyber security educational games for children. To the best of our knowledge, no prior SLR provides a thorough and up-to-date analysis of the literature in this area. Our work addresses this gap by incorporating the latest research published between 2010 and 2024. We identified the following research questions (RQs), each of which specifically focuses on cyber security educational games designed for children, referred to as cyber games for children in the questions.

\newlist{URRQ}{enumerate}{1}
\setlist[URRQ,1]{label=\textbf{RQ\arabic*:}, left=\parindent}
\begin{URRQ}
\item What are the design elements of cyber games for children?

\item How were cyber games for children evaluated and were they effective?

\item What are the theories and frameworks used in the existing cyber games for children?

\item What are the current research gaps and future directions related to cyber games for children?
\end{URRQ}

Through this review, we identified key findings, highlighted notable gaps in existing literature, and proposed recommendations to guide future research and development in this field. The main findings include:
\begin{itemize}
\item Key educational design elements of the games, including proposed and achieved learning outcomes, cyber security concepts, content development and delivery methods, were identified.

\item Non-educational design elements affecting the design and effectiveness of the games were analysed.

\item The reviewed papers have significant gaps in the systematic application of relevant social science theories and frameworks in the design, implementation, and evaluation of the educational games.

\item The methods applied to evaluate the effectiveness of the games lack methodological rigour, raising concerns about the credibility of the reported learning outcomes. 
\end{itemize} 

To address the identified gaps, we recommend theory-informed and hybrid (combining bottom-up and top-down) game design approach which enhance educational impact. In addition, we provide further recommendations to guide future research and development in this area.

The rest of the paper is organised as follows. The next section presents the background. Section~\ref{sec:related_work} presents the related works and the research questions of this SLR. The methodology used to conduct this SLR is presented in Section~\ref{sec:methodology}. Section~\ref{sec:SLR_results} presents the analysis of the findings addressing the first three RQs. Section~\ref{sec:further_discussions} presents further discussion of these results, identifies the gaps in the existing literature, and suggests directions for future research, covering the last RQ. The final section concludes the review.

\section{Background}
\label{sec:background}

\subsection{Key Concepts and Scope}

This SLR is shaped by four key concepts: cyber security, children, education, and games. These concepts, along with their associated terms, are defined in this section.

Cyber security has become an increasingly highlighted topic in our everyday life because of the wide spread of cyber attacks in all sectors. There exist various definitions of the term cyber security. According to \citet[p.~101]{vonsolms2013information}, cyber security is broadly defined as ``the protection of cyberspace itself, the electronic information, the ICTs that support cyberspace, and the users of cyberspace in their personal, societal and national capacity, including any of their interests, either tangible or intangible, that are vulnerable to attacks originating in cyberspace''. The UK's National Cyber Security Centre (NCSC)~\cite{NCSC2023what} defines cyber security as ``how individuals and organisations reduce the risk of cyber attacks''. Online safety is a term often related with cyber security although both are applied in different contexts. Online safety refers to the protection of individuals while they are online or engaging with computing systems~\cite{waldock2022pre-univeristy}. Based on the provided definitions, this paper considers online safety to be a component of the broader concept of cyber security due to their significant overlap, aligning with the terminology used by \citet{waldock2022pre-univeristy}. Further explanation on the relationship between cyber security and online safety can be found in \cite{waldock2022pre-univeristy}.

According to the United Nations, children are defined as individuals under the age of 18~\cite{unicef_uncrc_2016}. While the existing literature does not consistently define subcategories of children, this general group is typically divided into `young children', referring to those younger than 13 years, and `teenagers', who are between the ages of 13 and 18. This classification aligns with the categorisation by the National Health Service (NHS)~\cite{age_nhs} and terminology found in technology-related research~\cite{brink2020robot, Guha2004mixing}.

Cyber security awareness, training, and education are closely related but distinct concepts that explain initiatives aimed at enabling children to protect themselves while using digital media. Awareness programmes focus on explaining what cyber security is and aim to draw children's attention to the importance of cyber security. In contrast, training programmes focus on teaching how children can protect themselves, equipping them with the knowledge and skills necessary to defend themselves against online harms and cyber attacks. Education programmes go a step further by enhancing children's understanding of why the skills learnt in training are crucial, and aim to create a deep understanding of the required practices and procedures~\cite{nist2003special800-50}. On the other hand, learning is defined as a more holistic and self-directed process that encompasses training and education, through which children acquire new knowledge and skills~\cite{masadeh2012training}.

According to \citet[p.~231]{vogel2006computer}, ``a computer game is defined as such by the author, or inferred by the reader because the activity has goals, is interactive, and is rewarding (gives feedback).'' This definition acknowledges the diverse and evolving nature of computer games. The phrase `inferred by the reader' implies that users or researchers may interpret an activity as a game if it exhibits key characteristics such as defined goals, interactivity, and feedback mechanisms. Building on this inclusive perspective, our definition of games encompasses emerging immersive technologies, including virtual reality (VR), augmented reality (AR), mixed reality (MR), extended reality (XR), and the metaverse. While not all applications within these technologies are formally categorised as games, we include them in our broader definition due to their capacity to deliver experiences that incorporate defined goals, active user interaction, and responsive feedback. These elements reflect the core attributes of games as described by \citet{vogel2006computer}, and are relevant in the context of cyber security education.

The scope of this SLR is defined by the concepts outlined above. It includes cyber security educational games designed for children. These games encompass both serious games and gamified educational tools that aim to educate cyber security concepts for children.

\subsection{Cyber Security Education for Children}

Research has shown that increased use of ICT technologies can negatively influence children's cyber security behaviours. As children increasingly use these technologies and experience benefits, they tend to perceive them as less risky~\cite{vanSchaik2018security}. This growing confidence in technology can sometimes lead to less cautious behaviour toward cyber security practices~\cite{gkioulos2017security}. Research on Facebook users has shown that those with more experience (years of Facebook use) have less risk perception and are generally less cautious as their familiarity with the platform grows~\cite{vanSchaik2018security}. Children can share information that may include personal details as they might lack awareness and skills in protecting themselves online~\cite{shen_cyber_2021}. In addition, their lack of experience can make them more vulnerable to cyber threats, and they might not always recognise the severity of these threats~\cite{shen_cyber_2021}.

Cyber security awareness, training and education are usually provided through various methods, including presentations, workshops, multimedia packages, email reminders, screen savers, and other techniques~\cite{alahmari2022moving}. Different platforms are used to conduct these sessions and the commonly used techniques and their associated platforms include instructor-led classroom training, e-learning, text-based training, and video-based training~\cite{tschakert2019effectiveness}. While traditional methods have benefits, they struggle to create lasting impact and immersion in training~\cite{saunders2019validating, shen_cyber_2021, ulsamer2021immersive}. This can lead to boredom and poor retention, making cyber security awareness and behaviour difficult to sustain~\cite{shen_cyber_2021}. Practical and physical training also requires significant resources, faces geographic limitations, and incurs ongoing costs~\cite{saunders2019validating}. These problems and limitations of traditional methods have led to a growing demand for alternative techniques and platforms that can supplement the traditional methods to effectively create long-lasting cyber security practices for children. Technology-enhanced learning (TEL), which involves using technology to improve the learning process, is a well-established component of modern education systems and can be effectively applied in this context~\cite{gallud2023technology}. Among TEL strategies, gamification and game-based learning offer promising alternative to meet this demand~\cite{qusa_cyber-hero_2021}.

Studies have shown that game-based learning is highly effective in educating children about cyber security~\cite{coenraad_experiencing_2020, giannakas_comprehensive_2019, quayyum_collaboration_2023}. Game-based learning has been increasingly adopted in schools to effectively deliver cyber security education for children~\cite{giannakas_comprehensive_2019, maqsood_design_2021}. Studies have applied game-based learning to achieve various cyber security learning outcomes, as shown in Section~\ref{sec:SLR_results}. This SLR aims to thoroughly review these studies, as detailed in the sections that follow.

\section{Related Work}
\label{sec:related_work}

Literature reviews on cyber security educational games for children are limited. In particular, we found no SLR that focus specifically on this area.  While some studies have explored related areas, they focus on different aspects. \citet{coenraad_experiencing_2020} systematically reviewed digital cyber security game products available on app stores and the web, concentrating on commercial games but not covering academic literature. \citet{roepke_teaching_2020} examined both academic and commercial cyber security games but primarily targeted non-professional end users across all age groups, with little emphasis on games specifically designed for children.

\citet{saglam_systematic_2023} conducted a systematic review on cyber security education for children, focusing on content development approaches and educational methods. While their discussion on content development is valuable, game-based learning was only briefly mentioned as an alternative method, leaving its role in cyber security education under-explored. Additionally, \citet{clark_digital_2016} performed a meta-analysis and systematic review of educational games across various disciplines. Although cyber security games were not their primary focus, their coding framework for analysing educational games was comprehensive.

Building on these studies, our SLR specifically addresses the gap by focusing on cyber security educational games for children as discussed in academic literature. We adopt the content development approaches highlighted by \citet{saglam_systematic_2023} and the coding framework from \citet{clark_digital_2016} to strengthen our analysis. Our approach provides a more in-depth understanding of how educational games have been designed and evaluated in the context of cyber security learning for children. Accordingly, we identified the four RQs presented in Section~\ref{sec:introduction}.

\section{Methodology}
\label{sec:methodology}

This literature review was conducted using a SLR method. The screening of papers eligible for the review was conducted following the SLR process defined in the PRISMA 2020 guideline~\cite{prisma2020}. The process we followed is shown in Figure~\ref{fig:prisma}.

\subsection{Selected Databases}

Potential databases for conducting the systematic search were identified based on their relevance to the research domain. The following databases were selected: Scopus\footnote{\url{https://www.scopus.com/}}, Web of Science (WoS)\footnote{\url{https://www.webofscience.com/}}, Association for Computing Machinery Digital Library (ACM DL)\footnote{\url{https://dl.acm.org/}}, and Institute of Electrical and Electronics Engineers (IEEE) Xplore\footnote{\url{https://ieeexplore.ieee.org/}}. 

\subsection{Search Keywords}

A search query was developed by considering multiple aspects of the SLR. It consists of key terms related to the core topic, the target group (children), educational aspects, and the delivery platform.

\begin{enumerate}
\item \textbf{Topic-related terms}: \texttt{cybersecurity OR "cyber security" OR "information security" OR "data security" OR "computer security" OR "system security" OR "e-security" OR privacy OR "online safety" OR "internet safety" OR "digital safety" OR "cyber safety" OR "e-safety"};

\item \textbf{Children-related terms}: \texttt{toddler OR "pre-schooler" OR child OR kid OR teenager OR adolescent OR student OR pupil OR minor OR youth OR young OR human OR user OR participant OR people};

\item \textbf{Education-related terms}: \texttt{awareness OR education* OR learning OR training OR teaching};

\item \textbf{Game-related terms}: \texttt{game OR gami* OR "virtual reality" OR "augmented reality" OR "mixed reality" OR "extended reality" OR metaverse};
\end{enumerate}

Some of the terms listed above are deliberately broad to ensure we capture relevant papers that may include such terms in their title, abstract, or keywords. The search query was derived using AND-combinations of the above four groups of search terms.  

\subsection{Exclusion Criteria}

The following exclusion criteria were applied to exclude ineligible papers based on a review of their titles and abstracts. If there was any ambiguity about potential eligibility based on the abstract, we erred on the side of including such papers at this stage. We also avoided excluding any paper that provides insufficient information to make a decision based solely on the title and abstract. Papers that were excluded from the systematic review are those which meet at least one of the following exclusion criteria:
\begin{enumerate}
\item do not address cyber security, privacy, and digital safety,

\item do not address awareness, training, or education,

\item do not utilise games, virtual reality, augmented reality, mixed reality, extended reality, or metaverse,

\item clearly mention that the targets are not children,

\item have only the abstract but not the full text,

\item do not report original research,

\item were not peer reviewed,

\item are not written in English,

\item were published before 2010 or after 2024.
\end{enumerate}

\subsection{Inclusion Criteria}

This step involved reviewing the full text of each paper remaining from the previous stage, to assess its relevance based on predefined inclusion criteria. Eligible studies were identified accordingly, and reasons for the exclusion of ineligible papers were documented. Papers that met inclusion criteria were selected for the review.
Included papers are those that:
\begin{enumerate}
\item have at least one section dedicated to discussion on educating or raising the awareness of children on cyber security; and
  
\item report at least one of the following:
    \begin{enumerate}
    \item a design of a game in a way that is playable based on the description,
    
    \item a developed game that can be played, and
    
    \item a design of a game evaluated using some method.
    \end{enumerate}
\end{enumerate}

In addition, in the case of multiple papers with overlapping content, the more detailed version was included.

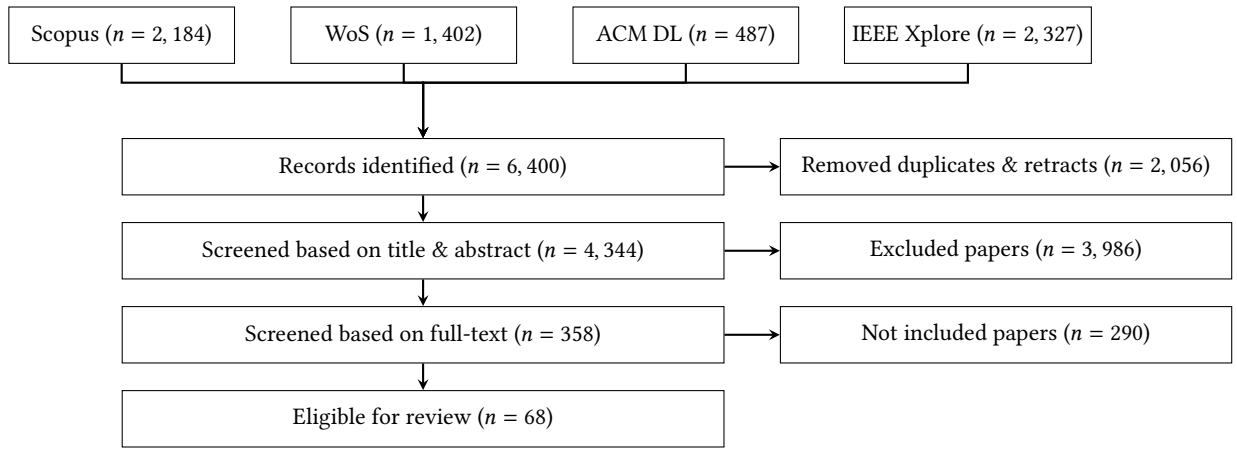
\begin{figure}[!htb]
\centering
\begin{tikzpicture}[node distance=1.25cm,>=latex]

\tikzstyle{startstop} = [rectangle, minimum width=8cm, minimum height=0.75cm, text centered, draw=black, fill=white]
\tikzstyle{process} = [rectangle, minimum width=8cm, minimum height=0.75cm, text centered, draw=black, fill=white]
\tikzstyle{decision} = [rectangle, minimum width=8cm, minimum height=0.75cm, text centered, draw=black, fill=white]
\tikzstyle{smallbox} = [rectangle, minimum width=3cm, minimum height=0.75cm, text centered, draw=black, fill=white]
\tikzstyle{mediumbox} = [rectangle, minimum width=6cm, minimum height=0.75cm, text centered, draw=black, fill=white]
\tikzstyle{arrow} = [thick,->,>=stealth]

% Four source boxes at the top
\node (scopus) [smallbox] {Scopus ($n = 2,184$)};
\node (wos) [smallbox, right of=scopus, xshift=2.5cm] {WoS ($n = 1,402$)};
\node (acmdl) [smallbox, right of=wos, xshift=2.5cm] {ACM DL ($n = 487$)};
\node (ieee) [smallbox, right of=acmdl, xshift=2.5cm] {IEEE Xplore ($n = 2,327$)};

% Combined box 
\node (combined) [startstop, below of=scopus, yshift=-0.5cm, anchor=west, xshift=0cm] {Records identified ($n = 6,400$)};

% Remaining process boxes 

\node (removedduplicates) [mediumbox, right of=combined, xshift=6.5cm] {Removed duplicates \& retracts ($n = 2,056$)};
\node (screenexclusion) [process, below of=combined, yshift=-0.25cm, anchor=south] {Screened based on title \& abstract ($n = 4,344$)};
\node (excludedpapers) [mediumbox, right of=screenexclusion, xshift=6.5cm] {Excluded papers ($n = 3,986$)};
\node (screeninclusion) [process, below of=screenexclusion, yshift=-0.25cm, anchor=south] {Screened based on full-text ($n = 358$)};
\node (notincludedpapers) [mediumbox, right of=screeninclusion, xshift=6.5cm] {Not included papers ($n = 290$)};
\node (eligible) [process, below of=screeninclusion, yshift=-0.25cm, anchor=south] {Eligible for review ($n = 68$)};

% Arrows from sources to combined box
\draw [arrow] (scopus.south) -- ++(0,-0.25cm) -| (combined.north);
\draw [arrow] (wos.south) -- ++(0,-0.25cm) -| (combined.north);
\draw [arrow] (acmdl.south) -- ++(0,-0.25cm) -| (combined.north);
\draw [arrow] (ieee.south) -- ++(0,-0.25cm) -| (combined.north);

% Arrows connecting the main flow 
\draw [arrow] (combined.south) -- (screenexclusion.north);
\draw [arrow] (screenexclusion.east) -- (excludedpapers.west);
\draw [arrow] (screenexclusion.south) -- (screeninclusion.north);
\draw [arrow] (screeninclusion.east) -- (notincludedpapers.west);
\draw [arrow] (screeninclusion.south) -- (eligible.north);
\draw [arrow] (combined.east) -- (removedduplicates.west);

\end{tikzpicture}
\caption{The PRISMA procedure used for the SLR}
\label{fig:prisma}
\end{figure}

\subsection{Data Encoding}

Data encoding was informed by a coding framework derived from previous studies~\cite{clark_digital_2016, coenraad_experiencing_2020, saglam_systematic_2023, huang2022systematic}. The primary foundation for this framework is the systematic literature review and meta-analysis conducted by \citet{clark_digital_2016}, which examined educational games across a broad range of disciplines. While many of the codes and code values identified in their work are applicable to our context, they lack categories applicable to cyber security educational games. Therefore, we derived additional codes to cover such missing concepts from the works of \citet{coenraad_experiencing_2020} and \citet{saglam_systematic_2023}. Theory related codes are derived from the work of \citet{huang2022systematic}. In addition, we included new codes that have not been addressed in previous studies. These codes are referred to as `New codes' in the list of codes shown in Table~\ref{table:SLR_codes_categories}. The new codes are necessary to address concepts not covered in existing papers but deemed essential for our SLR.

The codes are grouped into eight categories, as shown in Table~\ref{table:SLR_codes_categories}. The possible values of each code were determined in two ways: some are preset based on previous research and our prior knowledge, while the others are derived from the papers reviewed. The first author was responsible for extracting the second sub-set of code values from all papers reviewed. All codes and code values were discussed with the other authors to achieve a consensus among them. Some minor adjustments of the codes and values were also refined incrementally during the encoding process, and such adjustments were also discussed with all authors to get a consensus.

The encoding was done by the first two co-authors, one being the primary encoder and the other serving as an independent validator. After the initial encoding was finalised, the second author independently reviewed the encoding results of all papers to validate the work of the first encoder. These variations were discussed with the first encoder, and most of the initial encoding results were confirmed, with only a few changes agreed upon.

\begin{table}[!htb]
\caption{Codes Under Each Category and Their Sources}
\label{table:SLR_codes_categories}
\centering
\begin{tabular}{p{\textwidth}}
\toprule

\textbf{Study and Participant Characteristics}\\
\cite{clark_digital_2016}: publication year, timing of post-test measurement, age group(s) of participants\\
New codes: number of male participants, number of female participants, number of participants with undeclared gender, educational level of participants\\
\midrule

\textbf{Game Basics}\\
\cite{coenraad_experiencing_2020}: name of game, availability of game's link, target audience, play time\\ 
\cite{clark_digital_2016, coenraad_experiencing_2020}: platform\\
New codes: number of games, number of mini-games or levels\\
\midrule

\textbf{Educational Factors}\\
\cite{coenraad_experiencing_2020}: cyber security concepts covered, framing (presentation) of cyber security, methods to deliver cyber security content, content alignment with standards\\
\cite{saglam_systematic_2023}: content development approach\\
New codes: proposed learning outcome, achieved learning outcome\\
\midrule

\textbf{Game Story, Players, and Characters}\\
\cite{coenraad_experiencing_2020}: game story motivation, player role, gender of characters, role of girls\\ 
\cite{clark_digital_2016, coenraad_experiencing_2020}: story relevance, story depth, anthropomorphism\\ 
New codes: number of players, interaction of players\\
\midrule

\textbf{Game Mechanics}\\
\cite{clark_digital_2016, coenraad_experiencing_2020}: player rewarding methods, variety of game actions, scaffolding ways, visual realism, camera view\\
New code: game genre\\
\midrule

\textbf{Design Variants}\\
\cite{clark_digital_2016}: enhanced scaffolding designs, collaborative social designs, competitive social designs, providing/situating context, interface enhancement designs, extended game play designs\\
\midrule

\textbf{Study Quality Variable}\\
\cite{clark_digital_2016}: media comparison (MC) condition availability, MC condition equivalence, sufficient condition reporting\\
New code: type of intervention, research method, statistical metrics, control group availability, consideration of ethics\\
\midrule

\textbf{Theories and Frameworks}\\
\cite{huang2022systematic}: theory referenced, theory name\\
New code: framework referenced, framework name\\
\bottomrule
\end{tabular}
\end{table}

\subsection{Analysis and Synthesis}

The analysis and synthesis were conducted mainly using qualitative approach, as most of the data extracted from the papers were qualitative. However, the frequency of the extracted data was quantitatively presented using numbers and percentages. Counts and percentages were generated for each code and category after all data were categorised and quantified. Thematic analysis was applied to present the results, with research questions used to identify overarching themes. Sub-themes were identified within these themes, primarily based on the categories used to group the codes.

\section{Results}
\label{sec:SLR_results}

\subsection{Searching and Screening Results}

The formulated query was used to conduct a search in each of the four databases mentioned above, and in total 6,400 data items (papers) were returned on 19 March 2025. The numbers of paper found in different databases are shown at the top of Figure~\ref{fig:prisma}. Among these papers, those eligible for the review were identified following the SLR process defined in the PRISMA 2020 guideline~\cite{prisma2020}, as mentioned in the previous section. There are 2,055 duplicate papers and one retracted. Therefore, only 4,344 papers were further screened and 3,986 of them were excluded based on the exclusion criteria. The remaining 358 papers were evaluated according to the inclusion criteria, and 68 were chosen for the review. The PRISMA flow diagram that describes the overall screening process is shown in Figure~\ref{fig:prisma}. 
Appendix~\ref{appendix_A} contains the list of papers included in the review.

\subsection{Trends and Distribution of Papers and Games}

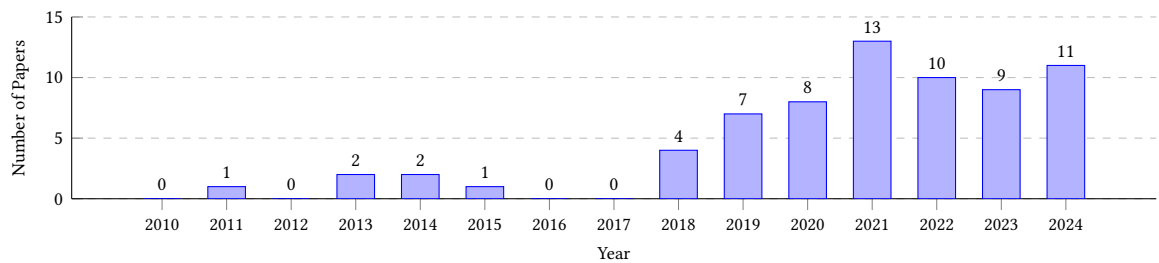
\begin{figure}[!htb]
\centering
\begin{tikzpicture}[font=\footnotesize]
\begin{axis}[
ybar,
bar width=.5cm,
width=16cm,
height=4cm, 
xlabel={Year},
ylabel={Number of Papers},
symbolic x coords={2010, 2011, 2012, 2013, 2014, 2015, 2016, 2017, 2018, 2019, 2020, 2021, 2022, 2023, 2024},
xtick=data,
nodes near coords,
ymin=0,
ymax=15,
ymajorgrids=true,
grid style=dashed,
enlarge x limits=0.1,
axis y line*=left,
axis x line*=bottom,
]
\addplot[
    fill=blue!30, 
    draw=blue    
] coordinates {
    (2010, 0)
    (2011, 1)
    (2012, 0)
    (2013, 2)
    (2014, 2)
    (2015, 1)
    (2016, 0)
    (2017, 0)
    (2018, 4)
    (2019, 7)
    (2020, 8)
    (2021, 13)
    (2022, 10)
    (2023, 9)
    (2024, 11)
};
\end{axis}
\end{tikzpicture}
\caption{Number of papers published from 2010 to 2024}
\label{fig:papers_per_year}
\end{figure}

We reviewed 68 papers published from 2010 to 2024. The majority of these papers were published in the latter half of the 15 years, with the highest number (13) appearing in 2021, as shown in Figure~\ref{fig:papers_per_year}.

A total of 91 games are reported across all papers. The number of games discussed in each reviewed paper ranges from one to eight. The majority of them (60) reported a single game. In contrast, only one paper~\cite[p.~75]{scholl_information_2020} described eight games designed to enhance ``information security'' awareness among students, teachers, and parents. In addition, one paper~\cite{videnovik_novel_2024} reported a game authoring approach in which students were involved in developing their own games. This initiative resulted in the creation of 18 different games by children. However, due to the lack of sufficient detail about each game, these student-created games were excluded from the final count of games in our analysis.

\subsection{RQ1: Design Elements of Cyber Games for Children}

Design elements of the games are grouped into two categories: educational and non-educational design elements.

\subsubsection{Educational Design Elements:}

This subsection presents the educational design elements used in the reported cyber games and how these design elements were incorporated.

\textbf{Proposed and achieved learning outcomes:}
Proposed learning outcomes are the designed/hoped educational goals of the games, which are usually defined during the game design process. In contrast, achieved learning outcomes refer to the actual observed improvements or benefits gained after playing the cyber security game, as reported in the corresponding paper. Note that different papers use different terms for these two concepts, but they can all be mapped to proposed and achieved learning outcomes. Nine categories of proposed and achieved learning outcomes have been identified across the 91 games, as shown in Table~\ref{tab:learning_outcomes}.

The most frequently proposed learning outcomes include enhanced cyber security awareness, cyber security knowledge, privacy awareness, and cyber security skills. Games designed to achieve cyber security awareness learning outcomes primarily focus on introducing cyber security concepts to children and enabling them to apply these concepts in real-world scenarios. This involves creating basic cyber security awareness, which provides children with a general understanding of fundamental concepts~\cite{alemany_assessing_2020}. Games aimed at building cyber security knowledge offer detailed and factual information about cyber security concepts, helping students gain a deeper and more comprehensive understanding of cyber security topics. Furthermore, games that focus on cyber security skills teach students how to apply their knowledge in practice by performing specific tasks, enabling them to effectively implement what they have learnt. Games designed to enhance privacy awareness aim to educate children about the importance of protecting their personal information while they are online and using digital media.

Although for all games the developers/researchers proposed one or more learning outcomes, many papers lack empirical evidence to demonstrate that the outcome or outcomes were actually achieved. For example, out of 32 games that aimed to build cyber security knowledge, only for 12 the corresponding papers provided evidence of achieving this learning outcome. Similarly, out of five games whose developers hoped to influence children's cyber security behaviour, only one game~\cite{maqsood_design_2021} reported a related achieved learning outcome, cyber security behavioural intent, which can only be a (not necessarily reliable) predictor of the actual behaviour in future. The overall findings on proposed versus achieved learning outcomes indicate that a significant number of games (44.8\%) fall short of meeting their intended educational goals, as shown in Table~\ref{tab:learning_outcomes}.

\begin{table}[!htb]
\centering
\caption{Proposed and Achieved Learning Outcomes of Games Reported in Papers Reviewed}
\label{tab:learning_outcomes}
\begin{tabular}{c cccc}
\toprule
\textbf{Category} & \multicolumn{2}{c}{\textbf{Proposed Learning Outcome}} & \multicolumn{2}{c}{\textbf{Achieved Learning Outcome}}\\
\cmidrule(lr){2-3} \cmidrule(lr){4-5}
& \textbf{\#(Games)} & \textbf{\#(Papers)} & \textbf{\#(Games)} & \textbf{\#(Papers)}\\
\midrule
Cyber security awareness & 37 (40.7\%) & 31 (45.6\%) & 27 (29.7\%) & 22 (32.4\%)\\
Cyber security knowledge & 32 (35.2\%) & 26 (38.2\%) & 12 (13.2\%) & 11 (16.2\%)\\
Privacy awareness & 19 (20.9\%) & 17 (25.0\%) & 14 (15.4\%) & 14 (20.6\%)\\
Cyber security skill & 14 (15.4\%) & 12 (17.6\%) & 6 (6.6\%) & 6 (8.8\%)\\
Online safety awareness & 10 (11.0\%) & 9 (13.2\%) & 5 (5.5\%) & 5 (7.4\%)\\
Cyber security behaviour & 5 (5.5\%) & 5 (7.4\%) & 1 (1.1\%) & 1 (1.5\%)\\
Privacy knowledge & 3 (3.3\%) & 3 (4.4\%) & 1 (1.1\%) & 1 (1.5\%)\\
Privacy behaviour & 3 (3.3\%) & 3 (4.4\%) & 3 (3.3\%) & 3 (4.4\%)\\
Privacy skill & 2 (2.2\%) & 2 (2.9\%) & 0 (0.0\%) & 0 (0.0\%)\\
\bottomrule
\end{tabular}
\end{table}

\textbf{Cyber security concepts:}
We identified a number of high-level cyber security concepts covered in the proposed learning outcomes of the games. These concepts, the number of games and the corresponding papers are shown in Table~\ref{tab:cyber_security_concepts}. \textit{Privacy} is the most covered concept, appearing in 30.8\% of the games, with core topics including privacy risk awareness, protective behaviours, and ethical considerations. \textit{Phishing} is the second popular concept covered, featured in 26.4\% of the games, often focusing on social engineering tactics and different types of phishing scams. Similarly, concepts related to \textit{passwords and multi-factor authentication (MFA)} are covered in 26.4\% of the games. \textit{Malware and anti-malware} and \textit{cryptography} also received significant attention compared to other concepts. \textit{Cyber security technologies and techniques} and \textit{technical cyber attack} gained considerable attention as well. Key cyber security technologies and techniques covered include application, mobile, web, and database security controls. On the other end of the spectrum, \textit{Information security management} appears in only one game's learning concepts~\cite{yamin_serious_2021}.

\textbf{Cyber security presentation method}:
Cyber security is presented in different ways within the games. Most of the games (74) embed cyber security topics covered in the formation of good cyber habits (cyber safety). This primarily looks at the actions children can / should take to defend against malicious actors such as those who try to steal their personal data or misuse the technology to cause harm~\cite{coenraad_experiencing_2020}. On the other hand, 19 games focus on advanced cyber security topics and present the `how' or `why' of cyber security concepts to deeply engage children with cyber security content.

\begin{table}[!htb]
\centering
\caption{Cyber Security Concepts Presented in Games and Corresponding Papers}
\label{tab:cyber_security_concepts}
\begin{tabularx}{\linewidth}{c cc Y}
\toprule
\textbf{Cyber Security Concepts} & \textbf{\#(Games)} & \textbf{\#(Papers)} & \textbf{Relevant Papers}\\
\midrule
Privacy & 28 (30.8\%) & 26 (38.2\%) & \cite{tjostheim_dark_2022, cardoso_playing_2022, neo_safe_2021, maqsood_design_2021, schuktomow_play_2020, hardin_digital_2020, alemany_assessing_2020, gioia_cyber_2019, giannakas_comprehensive_2019, ghazinour_digital-pass_2019, bioglio_social_2019, berger_privacity_2019, kumar_co-designing_2018, raynes-goldie_gaming_2014, christensen2023privacy, fountana_story_2011, giuseppe_blue_2024, denning_control-alt-hack_2013, doria_designing_2024, jaafar_empowering_2024, salazar_augmenting_2013, blinder_evaluating_2024, raihana_implementation_2024, yuan_redcapes_2024, chattopadhyay_vpet_2024, fujikawa_sns_2020}\\

Phishing & 24 (26.4\%) & 22 (32.4\%) & \cite{alsadhan_manar_2020, decusatis_cybersecurity_2022, drevin_story_2023, hodhod_cyberhero_2023, chiou_augmented_2021, jin_game_2018, neo_safe_2021, shen_work_progress_design_2021, sudha_impact_2023, tyner_framework_2022, zahed_play_2019, panga2022game, bassi2023serious, thomas2019educational, natalia2023gamification, del2023sec, alma2022soceng, christensen2023privacy, giuseppe_blue_2024, tseng_building_2024, visoottiviseth_pomega_2018, azzahra_socenggo_2024}\\

Password and MFA & 24 (26.4\%) & 22 (32.4\%) & \cite{hodhod_cyberhero_2023, tyner_framework_2022, decusatis_gamification_2022, decusatis_cybersecurity_2022, snyman_wolf_2021, qusa_cyber-hero_2021, neo_safe_2021, scholl_information_2020,alsadhan_manar_2020, zahed_play_2019, giannakas_comprehensive_2019, jin_game_2018, olano_securityempire_2014, del2023sec, giuseppe_blue_2024, tseng_building_2024, denning_control-alt-hack_2013, mohammed_designing_2024, salazar_augmenting_2013, blinder_evaluating_2024, visoottiviseth_pomega_2018, kim_security_2021}\\

Malware and anti-malware & 15 (16.5\%) & 13 (19.1\%) & \cite{sudha_impact_2023, tyner_framework_2022, alsadhan_manar_2020, hodhod_cyberhero_2023, jin_game_2018, zahed_play_2019, olano_securityempire_2014, bassi2023serious, thomas2019educational, del2023sec, tseng_building_2024, denning_control-alt-hack_2013, salazar_augmenting_2013}\\

Cryptography & 14 (15.4\%) & 14 (20.6\%) & \cite{pellicone_designing_2022, decusatis_gamification_2022, decusatis_cybersecurity_2022, yamin_serious_2021, shen_work_progress_design_2021, gonzalez-tablas_shuffle_2020, alsadhan_manar_2020, jin_game_2018, olano_securityempire_2014, thomas2019educational, del2023sec, denning_control-alt-hack_2013, kim_security_2021, chattopadhyay_vpet_2024}\\

Cyber security technologies and techniques & 12 (13.2\%) & 10 (14.7\%) & \cite{tyner_framework_2022, yamin_serious_2021, scholl_information_2020, giannakas_comprehensive_2019, jin_game_2018, olano_securityempire_2014, tseng_building_2024, denning_control-alt-hack_2013, jaafar_empowering_2024, sudha_impact_2023}\\

Technical cyber attack & 10 (11.0\%) & 10 (14.8\%) & \cite{costa_nerd_2020, bassi2023serious,  balakrishna_design_2021, thomas2019educational, jin_game_2018, yamin_serious_2021, tseng_building_2024, denning_control-alt-hack_2013, mnisi_digital_2024, kim_security_2021}\\

Cyber bullying & 9 (9.9\%) & 9 (13.2\%) & \cite{allers_childrens_2021, faith_intelligent_2022, lazarinis_raising_2015, maqsood_design_2021, shen_cyber_2021, alsadhan_manar_2020, fountana_story_2011, giuseppe_blue_2024, raihana_implementation_2024}\\

Network security & 8 (8.8\%) & 8 (11.8\%) & \cite{toledo_netdefense_2022, decusatis_cybersecurity_2022, hodhod_cyberhero_2023, jin_game_2018, yamin_serious_2021, shen_work_progress_design_2021, neo_safe_2021, thomas2019educational}\\

Social media security & 7 (7.7\%) & 7 (10.3\%) & \cite{faith_intelligent_2022, quayyum_collaboration_2023, scholl_information_2020, tuparova2021learning, jaafar_empowering_2024, visoottiviseth_pomega_2018, fujikawa_sns_2020}\\

Safe internet surfing \& downloading & 6 (6.6\%) & 6 (8.8\%) & \cite{mikka-muntuumo_designing_2021, olano_securityempire_2014, scholl_information_2020, 
shen_cyber_2021, tuparova2021learning, jaafar_empowering_2024}\\

Secure file \& information sharing & 6 (6.6\%) & 6 (8.8\%) & \cite{hodhod_cyberhero_2023, jin_game_2018, shen_cyber_2021, tuparova2021learning, salazar_augmenting_2013, blinder_evaluating_2024}\\

Hate speech and fake news & 4 (4.4\%) & 4 (5.9\%) & \cite{bassi2023serious, maqsood_design_2021, fujikawa_sns_2020, scholl_information_2020}\\

Physical and device security & 4 (4.4\%) & 4 (5.9\%) & \cite{hodhod_cyberhero_2023, schuktomow_play_2020, denning_control-alt-hack_2013, visoottiviseth_pomega_2018}\\

Digital forensic science & 3 (3.3\%) & 3 (4.4\%) & \cite{casey_motivating_2023, yamin_serious_2021, thomas2019educational}\\

Basic cyber security terms and concepts & 3 (3.3\%) & 2 (2.9\%) & \cite{schuktomow_play_2020, scholl_information_2020}\\

Cyber security incident report & 2 (2.2\%) & 2 (2.9\%) & \cite{yamin_serious_2021, thomas2019educational}\\

Information security management & 1 (1.1\%) & 1 (1.5\%) & \cite{yamin_serious_2021}\\
\bottomrule
\end{tabularx}
\end{table}

\textbf{Content development approaches:}
Game designers apply various approaches to develop the content of their games. These approaches can be bottom-up, top-down, or ad-hoc~\cite{saglam_systematic_2023}. The bottom-up approach involves intended game players in the content identification and development process, while the top-down approach utilises existing curricula, standards, and/or guidelines to develop the content. The ad-hoc approach lacks a predefined framework and is typically shaped by the game's designer in a less systematic manner. For the 91 cyber security games we reviewed, a majority (55 games, 60.4\%) were developed following an ad-hoc approach. The dominating adoption of an ad-hoc content development approach was surprising and raises concerns in areas such as justification of the cyber security concepts covered, generalisability of the game and any results obtained, and if such games can genuinely reflect children's needs and attract their interest in playing such games.

For the rest 36 games, 16 games (17.6\% of all games) were covered in 4 papers and there is no information about how the games were developed. This high percentage is even more surprising to us, indicating the lack of a reporting framework on such papers. For the final 20 games, their developers followed either a top-down (14 games, 15.4\%) approach or a bottom-up (6 games, 6.5\%) one, with none of them involving both approaches. Co-design workshops are the most popular research method for the bottom-up approach, and the frameworks used for top-down content development, spanning curricula, standards and guidelines, are highly diverse. Almost no games were developed following the same framework, with just one exception -- the Computer Science Teachers Association Standards (as a set of guidelines) was used for developing two games~\cite{shen_work_progress_design_2021, chiou_augmented_2021}. Moreover, the nature of these adopted frameworks varies significantly. While some game developers incorporated cyber security regulations, such as the European Union's General Data Protection Regulation (GDPR)~\cite{gioia_cyber_2019}, others relied on dedicated cyber security curricula and standards, such as the K-12 Cybersecurity Standards~\cite{toledo_netdefense_2022} in the United States. This variation underscores the lack of a unified, comprehensive framework that could serve as a common foundation for cyber security game design~\cite{saglam_systematic_2023}.

\textbf{Interactive content delivery methods:}
Cyber security content is delivered through six different methods as shown in Table~\ref{tab:cyber_security_content}. The primary interactive methods of content delivery are through gamified quizzes and arcade games, followed by puzzle games. Gamified quizzes are employed in 41.8\% (38) of all games. These quiz-based games use diverse types of questions, including multiple choice and matching, to deliver cyber security content or assess players' knowledge retention. Arcade games, which require rapid reflexes and twitch responses from players~\cite{coenraad_experiencing_2020}, make up 30.8\% (28) of the total game sample. Puzzle games, characterised by tasks that challenge players to solve problems, account for 17.6\% (16) of the total games. The prevalence of gamified quiz method indicates that many game designers preferred using simpler games to teach children cyber security concepts, supporting similar findings from a prior review of cyber security game products~\cite{coenraad_experiencing_2020}.

\begin{table}[!htb]
\centering
\caption{Content Related Design Elements Covered in Games}
\label{tab:cyber_security_content}
\begin{tabular}{ccc}
\toprule
\textbf{Design Elements} & \textbf{\#(Games)} & \textbf{\#(Papers)}\\
\midrule
\multicolumn{3}{l}{\textbf{Content Development Approaches}}\\
Ad-hoc approach & 55 (60.4\%) & 50 (73.5\%)\\
Top-down approach & 14 (15.4\%) & 12 (17.6\%)\\
Bottom-up approach & 6 (6.6\%) & 4 (5.9\%)\\
Not reported & 16 (17.6\%) & 4 (5.9\%)\\
\midrule
\multicolumn{3}{l}{\textbf{Interactive Content Delivery Methods}}\\
Gamified quiz & 38 (41.8\%) & 32 (47.1\%)\\
Arcade games & 28 (30.8\%) & 24 (35.3\%)\\
Puzzle games & 16 (17.6\%) & 15 (22.1\%)\\
Storytelling & 14 (15.4\%) & 14 (20.6\%)\\
Real world problem solving & 5 (5.5\%) & 5 (7.4\%)\\
Capture the flag competition & 2 (2.2\%) & 2 (2.9\%)\\
Not reported & 4 (4.4\%) & 3 (4.4\%)\\
\bottomrule
\end{tabular}
\end{table}

\textbf{Game story:} A story can better support the cyber security educational objectives of a game although it is not mandatory. It can provide a structured framework to present various scenarios and convey essential content in an engaging manner. However, only 45 (49.5\% of all) games incorporate one or more stories, with varying levels of motivation, relevance, and depth. Among these, 38 (41.8\% of all) games focus on defence-oriented narratives, two (2.2\%) adopt an attack-only perspective, and five (5.5\%) integrate both defence and attack motivations to offer a more comprehensive learning experience.

Games can include stories that are either relevant or irrelevant to the cyber security learning outcomes they aim to achieve. The relevance of a story is determined by its direct connection to the educational objectives rather than the game mechanics~\cite{clark_digital_2016}. For example, \textit{Safety Snail's Email}~\cite{drevin_story_2023}, an email security game designed for preschool children, presents a relevant story in which the main character, Snail, learns to identify and delete an email from a stranger. In contrast, \textit{SecurityEmpire}~\cite{olano_securityempire_2014}, where players take on the role of a green technology energy company owner, and construct green technology systems to compete with other companies, is considered to have an irrelevant story in terms of cyber security education. Notably, a majority (42) of the 45 games with one or more stories incorporate relevant stories, with only three games featuring irrelevant stories.

The depth of stories in the 45 games varies, with 31 games (34.1\%) featuring thin narratives, while only four games incorporate thick stories, and the rest 10 games cover stories with medium depth. Games with thin stories typically provide only a basic cyber security setting or context without extensive plot development. For example, \textit{CyberFamily} employs a thin story by presenting the following simple scenario and asking players to evaluate their actions in different scenarios: ``\textit{Suppose you have a Snapchat account. You have lots of friends on Snapchat. You regularly post pictures and videos on Snapchat and chat with your fiends.}''~\cite[p.~152]{quayyum_collaboration_2023}. This minimal storytelling approach establishes a context for players without a deeper narrative.

Games involving medium story depth present an emerging story, although the narrative does not deepen as the game evolves. In contrast, games with a thick story provide a narrative that evolves throughout the gameplay, offering new information to players as they progress~\cite{coenraad_experiencing_2020}. For instance, a game with a thick story presented an immersive and evolving storyline about a utility robot on a remote space station tasked with saving its creator's life~\cite{decusatis_cybersecurity_2022}.

The scarcity of games with thick stories highlights a gap in the use of evolving narratives within cyber security educational games. Games with thick story have the potential to create complex scenarios that closely mirror real-world situations. This depth not only enhances the learning experience but also provides a more immersive and engaging way to convey cyber security concepts.

\begin{table}[!htb]
\centering
\caption{Story Related Design Elements Covered in Games}
\label{tab:cyber_security_story}
\begin{tabular}{ccc}
\toprule
\textbf{Design Elements} & \textbf{\#(Games)} & \textbf{\#(Papers)}\\
\midrule
\multicolumn{3}{l}{\textbf{Story Motivation}}\\
Defence only & 38 (41.8\%) & 37 (54.4\%)\\
Attack only & 2 (2.2\%) & 2 (2.9\%)\\
Both defence and attack & 5 (5.5\%) & 5 (7.4\%)\\
Not reported & 46 (50.5\%) & 28 (41.2\%)\\
\midrule
\multicolumn{3}{l}{\textbf{Story Relevance}}\\
Relevant & 42 (46.2\%) & 41 (60.3\%)\\
Irrelevant & 3 (3.3\%) & 3 (4.4\%)\\
Not reported & 46 (50.5\%) & 28 (41.2\%)\\
\midrule
\multicolumn{3}{l}{\textbf{Story Depth}}\\
Thin & 31 (34.1\%) & 30 (44.1\%)\\
Medium & 10 (11.0\%) & 10 (14.7\%)\\
Thick & 4 (4.4\%) & 4 (5.9\%)\\
Not reported & 46 (50.5\%) & 28 (41.2\%)\\
\bottomrule
\end{tabular}
\end{table}

\textbf{Summary:}
The most frequently proposed learning outcomes across the games include cyber security awareness, cyber security knowledge, privacy awareness and cyber security skills. However, there is a significant discrepancy between proposed and achieved learning outcomes, indicating that these games may not have effectively achieved their goals. Only a small portion of games involve children in content development, with most relying on an ad-hoc approach that overlook engaging the target audience (children) or considering established standards and guidelines. The primary method for delivering cyber security content is through gamified quizzes. About half of the games incorporate storytelling, but most focus on relevant yet thin stories. There is lack of games with thick and engaging stories, which could provide more effective learning experiences.

\subsubsection{Non-Educational Design Elements}

Non-educational design elements of games can significantly affect the achievement of proposed learning outcomes by shaping the gameplay experience, although they are not directly tied to the educational content. This section highlights selected non-educational design elements of the games.

\textbf{Target audiences:}
The primary audiences for the games are young children and teenagers, as illustrated in Table~\ref{tab:games_papers_audience}. The number of games targeting young children is almost the same as those targeting teenagers. Games targeting young children range from those designed for pre-schoolers~\cite{drevin_story_2023, snyman_wolf_2021, mnisi_digital_2024} to those targeting children aged 11 to 13 years~\cite{maqsood_design_2021}. Most of these games focus on online safety issues. Conversely, a relatively higher number of games targeting teenagers focus on more advanced concepts such cryptography and technical cyber attack. In addition, some games include adults as their target audience, extending beyond their primary focus on young children and teenagers. Part of these games engage adults to create a conducive learning environment for children, enabling children to learn cyber security with the support of trusted adults such as their parents and teachers. For instance, \citet{quayyum_collaboration_2023} proposed a cyber security game that requires the engagement of parents and young children to raise cyber security awareness. This game facilitates discussions during gameplay, fostering an environment where both young children and parents can enhance their cyber security knowledge. There are also games that target a specific group of children such as girls~\cite{casey_motivating_2023}, autistic children~\cite{yuan_redcapes_2024} and minoritised groups~\cite{casey_motivating_2023, pellicone_designing_2022}. Overall, the findings show a balanced focus on young children and teenage audiences across the games.

\begin{table}[!htb]
\centering
\caption{Target Audiences of Games}
\label{tab:games_papers_audience} 
\begin{tabular}{ccc}
\toprule
\textbf{Audiences} & \textbf{\#(Games)} & \textbf{\#(Papers)}\\
\midrule
Young children & 55 (60.4\%) & 39 (57.4\%)\\
Teenagers & 54 (59.3\%) & 37 (54.4\%) \\
Teenagers \& adults & 21 (23.1\%) & 15 (22.1\%)\\
Young children \& teenagers & 21 (23.1\%) & 11 (16.2\%)\\
Young children \& adults & 8 (8.8\%) & 8 (11.8\%) \\
Young children, teenagers \& adults & 7 (7.7\%) & 5 (7.4\%)\\
Minoritised groups & 2 (2.2\%) & 2 (2.9\%)\\
Girls only & 1 (1.1\%) & 1 (1.5\%)\\
Autistic children & 1 (1.1\%) & 1 (1.5\%)\\
Not reported & 3 (3.3\%) & 3 (4.4\%)\\
\bottomrule
\end{tabular}
\end{table}

\textbf{Players and characters of the games:}
Games assign different roles to players based on the cyber security concepts they aim to teach and the story motivation they have. A majority of games (75, 82.4\%) focus solely on defence, allowing players to take on protective roles. In contrast, only two games (2.2\%) provide an attacker role, offering a perspective on offensive strategies. Additionally, seven games (7.7\%) incorporate both attack and defence roles, giving players a more comprehensive experience. The remaining seven games (7.7\%) do not specify player roles. The number of participating players also varies across these games. Most games (55, 60.4\%) were designed for single players, while ten games (11.0\%) support two-player gameplay. A total of 26 games (28.6\%) accommodate three or more players. Among the games that support two or more players, five also offer both single- and multi-player modes. The number of players is not specified for the remaining five games.

Game developers and designers incorporate different character genders, although gender remains unreported in a majority of games (66, 72.5\%). Three games (3.3\%) allow players to choose between male and female characters, while six games (6.6\%) feature an equal representation of both genders. Additionally, six games (6.6\%) include only female characters, comprising five human characters (girls) and one non-human character (a snail)~\cite{drevin_story_2023}, whereas three game (3.3\%) has an exclusively male character. Another two games (2.2\%) use generic characters that are not gender-specific and can be interpreted as either. The representation of female characters also varies across the games that include them. In five games (5.5\%), girls take on main character roles, while they appear as secondary character in one game.

Beyond human representation, ten games (11.0\%) feature non-human characters, e.g., animals that exhibit human-like traits (anthropomorphism). The degree of anthropomorphism differs among these games: five games depict highly human-like non-human characters, four games use a medium level of anthropomorphism, and one game employs a low level, where human-like traits are minimal.

\textbf{Game play time and mini-games:}
The time required for the gameplay varies across different games, ranging from 1.5 to 120 minutes, as shown in Table~\ref{tab:play_time}. In most cases (72 games, 79.1\%), the gameplay time is not reported. Four games, with ``Not Applicable'' playtime, feature playable game designs but have not yet been implemented or tested. The number of mini-games or levels in each game ranges from one to nine, as detailed in Table~\ref{tab:mini-games}. The majority of games which reported their number of mini-games or levels feature one to three mini-games or levels. Some games present distinct mini-games that focus on different concepts or incorporate various game mechanics, while others offer levels that vary in difficulty, such as easy, medium, or hard~\cite{toledo_netdefense_2022}.

\begin{table}[!htb]
\centering
\small
\caption{Play Time (NR = not reported, NA = not applicable)}
\label{tab:play_time}
\setlength{\tabcolsep}{3.5pt}
\begin{tabular}{lcccccccc}
\toprule
\multicolumn{9}{c}{\textbf{Play Time (minutes)}}\\
\cmidrule(lr){2-9}
\textbf{Category} & 1-15 & 16-30 & 31-45 & 46-60 & 61-90 & 91-120 & NR & NA\\
\midrule
\textbf{\#(Games)} & 2 (2.2\%) & 5 (5.5\%) & 2 (2.2\%) & 2 (2.2\%) & 3 (3.3\%) & 1 (1.1\%) & 72 (79.1\%) & 4 (4.4\%)\\
\bottomrule
\end{tabular}
\end{table}

\begin{table}[!htb]
\centering
\small
\caption{Number of Mini-games or Levels (NR = not reported, NA = not applicable)}
\label{tab:mini-games}
\setlength{\tabcolsep}{3.5pt}
\begin{tabular}{lcccccccccccc}
\toprule
\multicolumn{13}{c}{\textbf{Number of Mini-games or Levels}}\\
\cmidrule(lr){2-13}
\textbf{Category} & 1 & 2 & 3 & 4 & 5 & 6 & 7 & 8 & 9 & NR & NA &\\
\midrule
\textbf{\#(Games)} & 24 (26.4\%) & 5 (5.5\%) & 15 (16.5\%) & 9 (9.9\%) & 7 (7.7\%) & 0 (0.0\%) & 3 (3.3\%) & 1 (1.1\%) & 1 (1.1\%) & 23 (25.3\%) & 3 (3.3\%) &\\
\bottomrule
\end{tabular}
\end{table}

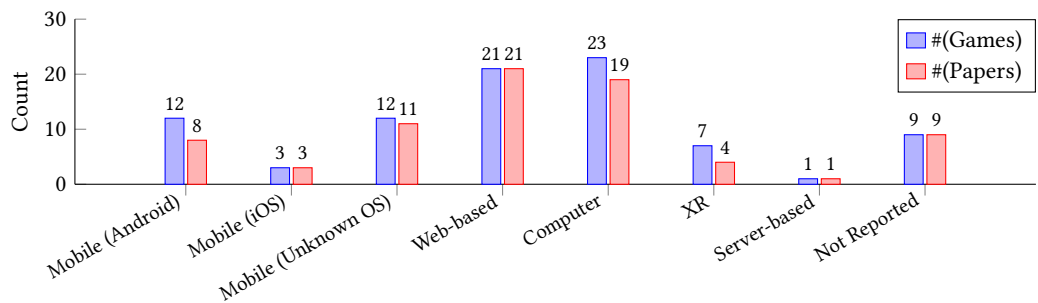
\begin{figure}[!htb]
\centering
\begin{tikzpicture}
\begin{axis}[
ybar,
bar width=0.25cm,
width=0.95\textwidth,
height=0.25\textwidth,
enlarge x limits=0.15,
axis y line*=left,
axis x line*=bottom,
ylabel={Count},
symbolic x coords={Mobile (Android), Mobile (iOS), Mobile (Unknown OS), Web-based, Computer, XR, Server-based, Not Reported},
xtick=data,
x tick label style={rotate=30, anchor=east, font=\small},
ymin=0,
ymax=30,
nodes near coords,
nodes near coords align={vertical},
every node near coord/.append style={font=\small},
legend style={at={(1,1)}, anchor=north east, legend columns=1},
legend image code/.code={%
    \draw[#1] (0cm,-0.1cm) rectangle (0.25cm,0.1cm);
},
cycle list={{blue, fill=blue!30}, {red, fill=red!30}},
]

\addplot+[
ybar,
bar shift=-0.15cm,
nodes near coords,
every node near coord/.append style={text=black},
] coordinates {
(Mobile (Android),12) (Mobile (iOS),3)
(Mobile (Unknown OS),12) (Web-based,21) (Computer,23) (XR,7) (Server-based,1)
(Not Reported, 9)
};

\addplot+[
ybar,
bar shift=0.15cm,
nodes near coords,
every node near coord/.append style={text=black},
] coordinates {
(Mobile (Android),8) (Mobile (iOS),3)
(Mobile (Unknown OS),11) (Web-based,21) (Computer,19) (XR,4) (Server-based,1)
(Not Reported, 9)
};
\legend{\#(Games), \#(Papers)}
\end{axis}
\end{tikzpicture}
\caption{Platforms Used to Present the Games}
\label{fig:games_papers_platform}
\end{figure}

\textbf{Platforms:}
``Platforms are technical objects that facilitate digital production'', primarily functioning as layered abstractions that enable digital creation~\cite{murray2020more}. Therefore, the platforms identified in this section address only digital games reported in the papers. The choice of the platform used to deliver a game is influenced by target audiences, as different age groups prefer different platforms based on their interest and accessibility of platforms~\cite{coenraad_experiencing_2020}. The platforms used to present cyber security games include mobile devices, laptop and desktop computers, web browsers, servers and XR, as illustrated in Figure~\ref{fig:games_papers_platform}.\footnote{Game consoles, such as Microsoft Xbox and Sony PlayStation, are also recognised as gaming platforms; however, none of the reviewed games utilised these platforms.}

Mobile games constitute the largest category, followed by computer and Web-based games. However, the operating system for nearly half of mobile platforms was not reported. Game platforms with limited presence include server-based (1) and XR (7). The reported XR games consist of two AR, two VR, and three MR games.

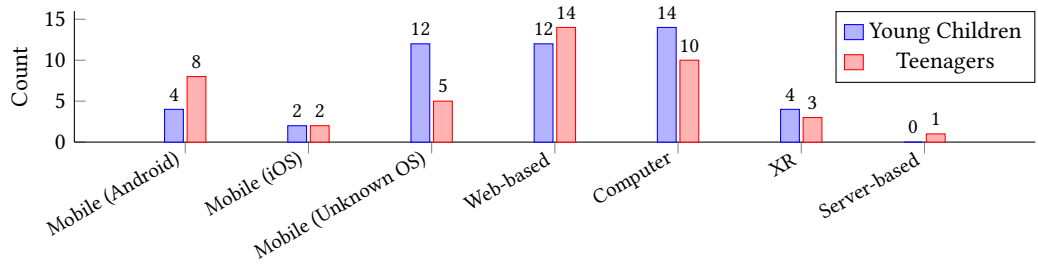
\begin{figure}[!htb]
\centering
\begin{tikzpicture}
\begin{axis}[
ybar,
bar width=0.25cm,
width=0.95\textwidth,
height=0.22\textwidth,
enlarge x limits=0.15,
axis y line*=left,
axis x line*=bottom,
ylabel={Count},
symbolic x coords={Mobile (Android), Mobile (iOS), Mobile (Unknown OS), Web-based, Computer, XR, Server-based},
xtick=data,
x tick label style={rotate=30, anchor=east, font=\small},
ymin=0,
ymax=16,
nodes near coords,
nodes near coords align={vertical},
every node near coord/.append style={font=\small},
legend style={at={(1,1)}, anchor=north east, legend columns=1},
legend image code/.code={%
\draw[#1] (0cm,-0.1cm) rectangle (0.25cm,0.1cm);
},
cycle list={{blue, fill=blue!30}, {red, fill=red!30}},
]

\addplot+[
ybar,
bar shift=-0.15cm,
nodes near coords,
every node near coord/.append style={text=black},
] coordinates {
(Mobile (Android),4) (Mobile (iOS),2) 
(Mobile (Unknown OS),12) (Web-based,12) (Computer,14) (XR,4) (Server-based,0)
};

\addplot+[
ybar,
bar shift=0.15cm,
nodes near coords,
every node near coord/.append style={text=black},
] coordinates {
(Mobile (Android),8) (Mobile (iOS),2)
(Mobile (Unknown OS),5) (Web-based,14) (Computer,10) (XR,3) (Server-based,1)
};
\legend{Young Children, Teenagers}
\end{axis}
\end{tikzpicture}
\caption{Distribution of target audience across platforms}
\label{fig:audience_across_platform}
\end{figure}

Platforms used for different audiences have slight variation, as shown in Figure~\ref{fig:audience_across_platform}. Mobile, web and computer are the leading platforms that serve games for young children and teenagers. Games developed for young children are predominantly run on computer and mobile platforms. On the other hand, web browsers are the main platforms for games targeting teenagers. 

\textbf{Types of non-digital games:}
The main types of non-digital games identified in the reviewed papers include card games, board games, toys, and other games involving physical environments, as shown in Table~\ref{tab:non-digigal_game_types}. Card (9) and board (8) games are the most commonly reported non-digital games. Of the two other games that involve a physical environment, one is a physical escape the room game that integrates both virtual and physical activity~\cite{costa_nerd_2020}, while the other uses ``Would You Rather'' scenarios to prompt discussions on privacy and security concepts with children in a virtual meeting setting~\cite{blinder_evaluating_2024}. Both of these games incorporate digital components to some extent; however, their core mechanics rely primarily on physical environments and interactions. Therefore, they are classified as non-digital games based on the dominant modality of gameplay.

\begin{table}[!htb]
\centering
\caption{Types of Non-digital Games}
\label{tab:non-digigal_game_types}
\begin{tabular}{ccc}
\toprule
\textbf{Types} & \textbf{\#(Games)} & \textbf{\#(Papers)}\\
\midrule
Card & 9 (9.9\%) & 5 (7.4\%)\\
Board & 8 (8.8\%) & 5 (7.4\%)\\
Physical environment & 2 (2.2\%) & 2 (2.9\%)\\
Toy & 1 (1.1\%) & 1 (1.5\%)\\
\bottomrule
\end{tabular}
\end{table}

\textbf{Game genres:}
On their study of game genres, \citet[p.175]{heintz2015game} noted that ``the existing common video game genres lack clarity as well as consistency and thus cannot serve as a solid reference to inform the research on digital educational games.'' Despite this limitation, commonly recognised game genres reported in prior studies~\cite{heintz2015game, silva2019practical, apperley2006genre} were adopted to identify and categorise the game genres applicable to the educational games reviewed in this study.

The reported games include a diverse range of genres, as presented in the Table~\ref{tab:game_genres}, despite all being classified as educational. Among these, quiz or trivia games are the most prevalent, accounting for 38 games (41.8\%), followed by simulations (18 games, 19.8\%) and puzzle games (16 games, 17.6\%). The predominance of quiz or trivia games suggests that a significant number of games deliver their educational content through multiple-choice and matching questions. This is consistent with the observation that gamified quizzes are the leading approach to interactive content delivery methods, as illustrated in Table~\ref{tab:cyber_security_content}.

The terms `quiz' and `puzzle' are used in this study to describe both interactive content delivery methods and game genres. This dual usage was found to be the most appropriate way to represent these elements, aligning with previous studies that employed similar terminology to describe the content delivery mechanisms in cyber security educational games~\cite{coenraad_experiencing_2020} and the genres of educational serious games~\cite{silva2019practical}.

\begin{table}[!htb]
\centering
\caption{Genres of the Games}
\label{tab:game_genres}
\begin{tabular}{ccc}
\toprule
\textbf{Genres} & \textbf{\#(Games)} & \textbf{\#(Papers)}\\
\midrule
Educational & 91 (100.0\%) & 68 (100.0\%)\\
Quiz or Trivia & 38 (41.8\%) & 32 (47.1\%)\\
Simulation & 18 (19.8\%) & 14 (20.6\%)\\
Puzzle & 16 (17.6\%) & 15 (22.1\%)\\
Strategy & 12 (13.2\%) & 11 (16.2\%)\\
Role-playing & 9 (9.9\%) & 8 (11.8\%)\\
Adventure & 6 (6.6\%) & 6 (8.8\%)\\
Point and Click & 2 (2.2\%) & 2 (2.9\%)\\
Shooter & 2 (2.2\%) & 2 (2.9\%)\\
Action & 1 (1.1\%) & 1 (1.5\%)\\
Detective & 1 (1.1\%) & 1 (1.5\%)\\
Exploration & 1 (1.1\%) & 1 (1.5\%)\\
Racing & 1 (1.1\%) & 1 (1.5\%)\\
\bottomrule
\end{tabular}
\end{table}

\textbf{Visual realism and camera view:}
Visual realism in games refers to how realistic the characters and environments of the games appear. A majority of games (58 games, 63.7\%) fall into the category of cartoon games, featuring non-realistic characters or objects and typically presented in a two-dimensional format (see Table~\ref{tab:Visual_realism_camera_view}). Games characterised by schematic visual realism, that are schematic, symbolic, or text-based with minimalistic graphics, constitute 12 games (13.2\%). These are followed by eight realistic games (8.8\%) that incorporate realistic shading or real pictures. The visual realism of seven games (7.7\%) is not reported in the corresponding papers.

The camera view in a game determines how players see the game world. The side-scrolling camera view is noted as the most prevalent among the 91 games, accounting for 23 games (25.3\%), with a fixed camera view coming next at 18.7\% (17 games). Games utilising third-person and first-person shooter camera views are much less: 8 (8.8\%) and 3 (3.3\%) games, respectively. Moreover, the camera views for 40 games (44.0\%) are either not reported in the corresponding paper or not applicable. Most of the games in the ``not applicable'' category are non-digital such as card and board games.

\begin{table}[!htb]
\centering
\caption{Visual Realism and Camera View of the Games}
\label{tab:Visual_realism_camera_view}
\begin{tabular}{ccc}
\toprule
\textbf{Design Elements} & \textbf{\#(Games)} & \textbf{\#(Papers)}\\
\midrule
\multicolumn{3}{l}{\textbf{Visual Realism}}\\
Cartoon & 58 (63.7\%) & 48 (70.6\%)\\
Schematic & 12 (13.2\%) & 8 (11.8\%)\\
Realistic & 8 (8.8\%) & 6 (8.8\%)\\
Not reported & 7 (7.7\%) & 7 (10.3\%)\\
\midrule
\multicolumn{3}{l}{\textbf{Camera View}}\\
Side-scrolling view & 23 (25.3\%) & 21 (30.9\%)\\
Fixed camera view & 17 (18.7\%) & 12 (17.7\%)\\
Third person & 8 (8.8\%) & 7 (10.3\%)\\
First person shooter & 3 (3.3\%) & 3 (4.4\%)\\
Not reported & 17 (18.7\%) & 15 (22.1\%)\\
Not applicable & 23 (25.3\%) & 16 (23.5\%)\\
\bottomrule
\end{tabular}
\end{table}

\textbf{Variety of game actions:}
Games encompass a range of actions to engage players and effectively convey educational content. These actions can vary in scope and can be small, medium, or large. Games with small variety of actions require players to perform one or two simple tasks such as clicking on answers or dragging an item~\cite{clark_digital_2016}. The complexity of the graphics or the platform does not affect the variety of game actions~\cite{clark_digital_2016}. For instance, a VR game involving object rotation or answering different types of questions is considered to have a small variety of game actions. Games with medium variety of actions offer several ways to interacting with the environment, while games with large variety of actions provide numerous methods for player interaction~\cite{clark_digital_2016}. According to the findings, 59 games (64.8\%) feature a small variety of actions. Games with medium and large varieties of actions represent 21 (23.1\%) and 6 (6.6\%) games, respectively. This limited diversity in variety of actions across games restricts children's interactions and can diminish the entertainment value of the games, potentially impacting the achieved learning outcomes.

\textbf{Scaffolding methods:}
Different scaffolding methods are used in the reported games to offer support for players, and hence promote deeper engagement. The primary three methods of scaffolding encompass success or failure notifications (26 games, 28.6\%), adult-provided support (24 games, 26.4\%), and revealing the answer for the player (13 games, 14.3\%). The games with adult-provided scaffolding involve some form of supervision and assistance from teachers, parents or other adults. Enhanced scaffolding, which includes providing personalised support, intelligent agents, and tailoring the game to the player's needs~\cite{clark_digital_2016}, was observed in only four games (4.4\%). The clear lack of games that offer enhanced scaffolding indicates a gap, given the wide range of interests and different comprehension levels among children. Games offering customised support and experiences adapted to individual requirements can improve the learning outcomes and player engagement.

\textbf{Design variants:}
Games often incorporate various design variants that enhance their effectiveness and improve the overall player experience. These variants include providing or situating context, interface enhancement designs, competitive social designs, collaborative social designs, and enhanced scaffolding designs (see Table~\ref{tab:design_variants}).

Providing or situating context, which involves setting the stage through introductions, background narratives, character backstories and environmental cues that convey contextual information, is reported in 32 games (35.2\%). Interface enhancement designs are presented in 30 games (33.0\%). These games incorporate features such as intuitive navigation for ease of use, visual consistency across different elements, feedback mechanisms that respond to user actions, and customisation options that allow players to personalise aspects of the interface. Collaborative and competitive social designs are implemented in varying ways, with competitive designs being more prevalent than collaborative ones. Collaborative social designs involve multiple players working towards shared goals, whereas competitive social designs enable individual or team-based competition. Out of the 91 games, 24 (26.4\%) incorporate competitive designs, while 22 (24.2\%) focus on collaborative elements, and eight games (8.8\%) integrate both collaborative and competitive social designs, therefore blending cooperative and competitive interactions within gameplay. Extended gameplay design, observed in 12 games (13.2\%), offers players opportunities to explore multiple endings, featuring branching storylines and alternative outcomes. In addition, it includes progression-based rewards, such as unlocking new levels or features.

\begin{table}[!htb]
\centering
\caption{Design Variants Applied in the Games Design}
\label{tab:design_variants}
\begin{tabular}{ccc}
\toprule
\textbf{Design variants} & \textbf{\#(Games)} & \textbf{\#(Papers)}\\
\midrule
Providing/situating context & 32 (35.2\%) & 31 (45.6\%)\\
Interface enhancement designs & 30 (33.0\%) & 29 (42.7\%)\\
Competitive social designs & 24 (26.4\%) & 17 (25.0\%)\\
Collaborative social designs & 22 (24.2\%) & 14 (20.6\%)\\
Collaborative and competitive social designs & 8 (8.8\%) & 5 (7.4\%)\\
Extended game play designs & 12 (13.2\%) & 12 (17.7\%)\\
Enhanced scaffolding designs & 4 (4.4\%) & 4 (5.9\%)\\
\bottomrule
\end{tabular}
\end{table}

\textbf{Availability of the games:}
Availability of games involves providing URLs, physical addresses, and/or names of platforms where interested individuals can find, purchase, or download the games. Among the total games, 73.6\% (67) do not have associated URLs reported in their corresponding paper. Only 24 games (26.3\%) provide a URL to one or more essential elements such as source code, executable files, websites, and videos. This lack of availability limits both the usability and potential impact of most games. Furthermore, the absence of URLs in the majority of papers makes many of the game products inaccessible for review. This highlights a notable gap in the literature, where limited access to game products and supporting materials hinders transparency and reproducibility. Such gaps may lead to discrepancies between the reported and actual educational or non-educational design elements of the games.

The findings on unavailability of game products mirrors the observations from a previous study~\cite{maqsood_design_2021}, indicating that the majority of these games were developed by academic researchers and had not been made widely available, thereby limiting their real-world impact and availability for further study. This lack of availability underscores a similar concern highlighted by \citet{roepke_teaching_2020}, where game-based learning applications and educational games often do not progress beyond the prototype phase and remain unavailable to the public.

\textbf{Summary:}
The primary audiences for the reported games are young children and teenagers, demonstrating a balanced focus on these audiences. Mobile games form the largest category of platforms, followed by computer and web-based games. In contrast, server-based and XR-based games have minimal representation in the digital game landscape. Most games offer a limited variety of game actions, which can restrict children's interactions and reduce entertainment value of the games. This, in turn, could impact their educational effectiveness. There is a notable lack of games that provide enhanced scaffolding. This highlights a gap in delivering tailored support and experiences adapted to individual needs, which are crucial for improving learning outcomes and player engagement. Most of the games lack URLs of game products, therefore affecting the broader dissemination and potential impact of these cyber security games.

\subsection{RQ2: Evaluation of the Games}

The effectiveness of the cyber security games reported in the papers was evaluated using various methods. This section addresses how these games were assessed and analyses their effectiveness based on the reported results, thus addressing RQ2. It presents findings on types of intervention, research design, participant demographics, media comparison conditions, and ethical considerations. These findings rely primarily on the experiments reported in the papers and aim to indicate the quality of the studies and the reliability of the reported evaluation results.

Although most of the papers reported the evaluation process and findings of the games, some papers lacked detailed evaluations, limiting the analysis on the effectiveness of the games. For example, \citet{scholl_information_2020} described eight games but omitted details of the experiments used to evaluate them. Such omission makes it challenging for readers to understand the effectiveness of these games and reproduce the evaluation process.

\textbf{Types of intervention:}
The game-based interventions reported in the papers followed three distinct approaches. The first and most common approach involves engaging children in gameplay and evaluating the educational impact of the experience. This method focuses on assessing the learning outcomes of games that were either designed by the study authors or sourced from other developers. The majority of the studies (60 papers, 88.2\%) employed this intervention strategy.

The second type of intervention focuses on involving children as co-designers of cyber security educational games, working collaboratively with adult design partners. This participatory design approach aims to assess the educational effectiveness of children's involvement in the game design process. It was adopted in six studies, accounting for 10.2\% of the total papers.

The third and least common intervention involves enabling children to develop cyber security educational games themselves, followed by an evaluation of the outcomes. This approach was found in only one study~\cite{videnovik_novel_2024}, which involved upper-grade students in creating educational materials, such as games and lectures, for use in teaching younger peers. As a result of this initiative, the students developed 18 interactive Scratch-based games addressing a variety of cyber security concepts.

\textbf{Research design:}
Most of the papers reviewed employ research designs that do not incorporate a control group, which provides a baseline to compare the observed changes in the experimental group(s) (see Table~\ref{tab:research_design}). Among 57 papers that report experiments conducted to evaluate the proposed games, only five (8.8\%) reported the use of a control group. This lack of a control group can introduce methodological weakness in terms of validity and reliability of the findings.

There are studies which applied pre-post testing instead of a control group. However, the time between the pre- and post-evaluation is short in most of the studies. Among 23 studies that reported the timing information, 14 of them conducted their post-test(s) on the same day while five studies conducted such test(s) within a week (1-7 days). Such port-test experiments are unable to show the long-term effects of the reported games, which may cast doubts about the achieved learning outcomes.

The reviewed papers employed a variety of research methods in conducting their experiments, including qualitative, quantitative, and mixed-methods approaches, as shown in Table~\ref{tab:research_design}. The majority of studies (29 papers, 42.6\%) adopted only quantitative method, while 11 papers (16.2\%) employed exclusively qualitative method. In addition, 18 papers (26.5\%) utilised a mixed methods approach, incorporating both qualitative and quantitative elements. Five papers did not specify the research method used, and an additional five did not involve any form of experimental study.

In papers that employed quantitative only or mixed-methods approaches, statistical metrics were used to analyse and report findings. These metrics can be categorised into descriptive and inferential statistics (see Table~\ref{tab:research_design}). Descriptive only statistical metrics, such as mean, median, standard deviation, and percentages, were used in 32 papers (47.1\%) to present their findings. In addition to descriptive statistical metrics, inferential statistical metrics, which help to assess the statistical significance of findings, were reported in 16 papers (23.5\%). All of these studies reported p-values which indicate whether the results are statistically significant. However, only six of them reported effect sizes to indicate the magnitude of the observed effects. The notable absence of experiments reporting effect sizes shows a lack of rigorous quantitative approaches in these studies. The omission of such quantitative measures of achieved learning outcomes can introduce subjectivity into the analysis and undermine the credibility of the findings.

\begin{table}[!htb]
\centering
\caption{Research Design}
\begin{tabular}{cc}
\toprule
\textbf{Categories} & \textbf{\#(Papers)}\\
\midrule
\multicolumn{2}{l}{\textbf{Control group}}\\
Without control group & 52 (76.5\%)\\
With control group & 5 (7.4\%)\\
Not reported & 6 (8.8\%)\\
Not applicable & 5 (7.4\%)\\
\midrule

\multicolumn{2}{l}{\textbf{Research method}}\\
Quantitative only & 29 (42.6\%)\\
Mixed (quantitative \& qualitative) & 18 (26.5\%)\\
Qualitative only & 11 (16.2\%)\\
Not reported & 5 (7.4\%)\\
Not applicable & 5 (7.4\%)\\
\midrule

\multicolumn{2}{l}{\textbf{Statistical metrics}}\\
Descriptive only & 32 (47.1\%)\\
Inferential \& descriptive & 16 (23.5\%)\\
Not applicable & 20 (29.4\%)\\
\bottomrule
\end{tabular}
\label{tab:research_design}
\end{table}

\textbf{Participant demographics:}
Among the 68 papers, 63 reported either one or two studies, while the remaining five did not include any experimental component, despite reporting one or more games. Out of the 63 papers, four reported two studies each, bringing the total number of studies across all papers to 67.

Experiments of game evaluation need to involve participants of varying genders, age groups, roles, and educational levels. However, out of the 67 studies, 35 (52.2\%) did not report participant demographics by gender. This highlights a significant gap in accurately and transparently reporting participant characteristics. Across the 32 studies that reported participant gender, a total of 1,499 male and 1,252 female participants were recorded, with an additional 30 participants whose gender was not specified. The gender distribution varies across studies: 15 reported a predominance of male participants, 13 indicated a predominance of female participants, and only four reported an equal number of male and female participants. The highest numbers of female and male participants are 196 and 191, respectively. Both of them were recorded in a gamified social network study involving both genders~\cite{alemany_assessing_2020}. \citet{yuan_redcapes_2024} recruited the fewest female participants, with just two individuals involved. Likewise, \citet{blinder_evaluating_2024} reported the minimal male participation, with only a single male participant in one of their two experiments. Overall, our findings show a predominance of male participants in the studies.

The age groups of participants involved in the studies vary, although the primary audience for most of the games are teenagers and young children (see Table~\ref{tab:participant_characteristics}). Teenagers make up the highest group of participants (55.2\% of total number of participant groups). Most of these participants are high school or secondary school students, as they typically fall within this educational level. Young children account only for 38.8\% of the total studies although they are the main target audience for many of the games (60.4\%) evaluated in these studies (see Table~\ref{tab:games_papers_audience}). In contrast, adults participated in 46.3\% of the studies although they are not the main target audiences of the games. The relatively significant participation of adults in the evaluation of these games is often due to the inclusion of adult audiences in some games or the involvement of experts assessing games designed for children. The relatively low participation of young children highlights a notable gap in ensuring the games' impact on this key demographic. It overlooked the need for including young children in the evaluation of cyber security educational games designed specifically for them~\cite{hartikainen_childrens_2019}.

The participants involved in the evaluation of cyber security games include students, teachers, parents, and different types of experts, as detailed in Table~\ref{tab:participant_characteristics}. Students represent the largest group of participants, which is expected since most of the games are developed for children. The expert groups comprise various types of professionals with experience in the field. Nine games were evaluated by these groups, which included researcher groups~\cite{shen_cyber_2021, costa_nerd_2020, kumar_co-designing_2018}, a consultant group~\cite{shen_cyber_2021}, experts working with children in schools~\cite{drevin_story_2023, allers_childrens_2021, snyman_wolf_2021, mnisi_digital_2024}, and IT professionals~\cite{costa_nerd_2020}. In contrast, parents and caregivers participated in the evaluation of only four games~\cite{quayyum_collaboration_2023, mikka-muntuumo_designing_2021, raihana_implementation_2024, yuan_redcapes_2024}. This limited involvement of parents in the evaluation of cyber security games requires attention, given their crucial role in shaping their children's cyber security practices. \citet{quayyum_collaboration_2023} argued that the role of parents and their participation should be considered during the development of cyber security games, as it is valuable for educating both children and parents.

\begin{table}[!htb]
\centering
\caption{Demographics of Participants Involved in the Evaluation of Games}
\begin{tabular}{ccc}
\toprule
\textbf{Participant demographics} & \textbf{\#(Participant groups)} & \textbf{\#(Papers)}\\
\midrule
\multicolumn{3}{l}{\textbf{Participant age group}}\\
Teenagers & 37 (55.2\%) & 35 (51.5\%)\\
Young children & 26 (38.8\%) & 26 (38.2\%)\\
Adults & 31 (46.3\%) & 30 (44.1\%)\\
Not reported & 5 (7.5\%) & 5 (7.4\%)\\
Not applicable & 5 (7.5\%) & 5 (7.4\%)\\
\midrule

\multicolumn{3}{l}{\textbf{Participant type}}\\
Students & 54 (80.6\%) & 51 (75.0\%)\\
Teachers & 9 (13.4\%) & 9 (13.2\%)\\
Experts \& professionals  & 9 (13.4\%) & 9 (13.2\%)\\
Parents \& caregivers & 4 (6.0\%) & 4 (5.9\%)\\
Not reported & 7 (10.4\%) & 7 (10.3\%)\\
Not applicable & 5 (7.5\%) & 5 (7.4\%)\\
\midrule

\multicolumn{3}{l}{\textbf{Participant education level}}\\
High or secondary school & 31 (46.3\%) & 29 (41.6\%)\\
Higher education & 13 (19.4\%) & 13 (19.1\%)\\
Primary or elementary school & 11 (16.4\%) & 11 (16.2\%)\\
Medium school & 4 (6.0\%) & 4 (5.9\%)\\
Pre-college & 2 (3.0\%) & 2 (2.9\%)\\
Not reported & 24 (35.8\%) & 23 (35.8\%)\\
Not applicable & 5 (7.5\%) & 5 (7.4\%)\\
\bottomrule
\end{tabular}
\label{tab:participant_characteristics}
\end{table}

\textbf{Media comparison condition:}
Media comparison conditions involve comparing the proposed game with another medium that could be used for delivering the same cyber security education intended to be conducted using the game~\cite{clark_digital_2016}. The media can be deemed equivalent if it provides an experience comparable to the proposed game. In contrast, a non-equivalent media comparison involves comparing two media conditions that differ significantly, making a direct `apple-to-apple' comparison impossible. For example, a VR-based cyber security game can be considered equivalent media to an AR-based game with the same content, whereas a card game would be non-equivalent media.

Among the reviewed papers, only seven included media comparison conditions. One of these studies~\cite{zahed_play_2019} reported both equivalent and non-equivalent conditions as the evaluation experiments involved two media comparison conditions: comparing a computer game with an interactive multimedia-based e-learning condition (equivalent) and a text-based handout (non-equivalent). Two other studies~\cite{gonzalez-tablas_shuffle_2020, alemany_assessing_2020} involved equivalent media conditions, while the remaining four studies~\cite{giannakas_comprehensive_2019, panga2022game, salazar_augmenting_2013, chattopadhyay_vpet_2024} utilised a non-equivalent condition. The limited number of studies incorporating media comparison undermines the credibility of the claimed effectiveness of game-based conditions when compared to other media.

\textbf{Sufficient condition reporting:}
Game conditions are reported in varying ways across different games. A game's condition is considered sufficiently reported if it includes at least one image that visually represents the game~\cite{clark_digital_2016}. Among all the games analysed, 84 games were reported sufficiently across 61 research papers. However, the remaining seven games in seven papers lacked adequate reporting in their respective paper.

\textbf{Ethical consideration:}
Ethical considerations within the studies were assessed in each paper. A notable and hugely concerning gap was identified, with a majority of papers (43, 63.2\%) failing to report ethical considerations. Only 20 papers (29.4\%) reported ethical considerations while conducting their studies, and the remaining five papers did not report experimental studies. Out of these 20 papers that reported ethical considerations, 11 reported obtaining an approval from their Institutional Review Board (IRB) or an equivalent body, while nine papers indicated getting consent from their participants. The remaining five papers did not report experiments. Given the significant involvement of children in these experiments, the omission of ethical reporting in a substantial number of papers raises substantial concerns regarding the ethical standards and overall quality of these studies.

\textbf{Summary:}
Effectiveness of the cyber security games was assessed through various experiments. However, most of the studies do not incorporate a control group in their experiments, which represents a gap in methodological rigour. The combination of participants in the experiments also show a predominance of male participants across the studies. Most participants were teenagers, with notably low involvement of young children despite being the primary target audience for many games. This highlights the necessity of including young children in the evaluation of cyber security educational games designed specifically for them. The limited participation of parents and caregivers in evaluating cyber security games targeting children also requires attention, given their pivotal role in shaping their children’s cyber security practices.

 While most of the studies applied quantitative or mixed methods to evaluate the effectiveness of the games, the notable absence of reported effect sizes highlights a lack of methodological rigour in the quantitative approaches employed. The scarcity of studies which involve equivalent media comparison conditions undermines the credibility of claims regarding the effectiveness of game-based conditions compared to other media. Furthermore, most of the papers failed to report ethical considerations, raising critical concerns about the ethical standards and overall quality of these studies.

\subsection{RQ3: Theories and Frameworks Referred in the Papers}

Applying relevant theories and frameworks in the design, implementation, and evaluation of cyber security games provides a theoretical foundation that supports both effectiveness and pedagogical soundness. Some of the games reviewed in this SLR incorporated theories and frameworks, particularly from social science disciplines, into their design, implementation, and evaluation. This section presents the theories and frameworks referenced in these cyber security games, thereby addressing RQ3.

\textbf{Theories referred:}
Among the 68 reviewed papers, only six (8.8\%) explicitly referenced the use of theories in the design, implementation, or evaluation of their cyber security games. Five of them clearly identified the specific theories applied. These theories include constructivism~\cite{hodhod_cyberhero_2023}, constructivist grounded theory~\cite{casey_motivating_2023}, interpretive grounded theory~\cite{pellicone_designing_2022}, theory of planned behaviour~\cite{maqsood_design_2021}, theory of procedural rhetoric~\cite{maqsood_design_2021} and motivational theories~\cite{giannakas_comprehensive_2019}. \citet[p.~92]{giannakas_comprehensive_2019} presented the motivational theories as part of the Attention, Relevance, Confidence, and Satisfaction (ARCS) model of motivation, which is claimed to comprise different motivational theories that include ``skills and knowledge, cognitive accounting of individual abilities, behavioral contingency design and management, and expectancy-value theory''. The sixth paper\cite{del2023sec}, while acknowledging the use of theoretical principles, did not specify the name of the theory applied. Instead, it described a process where game design theory was integrated with gamification techniques during development.

Each of the reported theories are notably distinct. This diversity indicates that there has not been a single, dominant theory guiding the design of cyber security educational games. The limited application of theories in the reviewed papers represents a significant gap, particularly given the well-documented advantages of grounding educational game design in social science theories~\cite{giannakas_comprehensive_2019, maqsood_design_2021, clark_digital_2016}. This scarcity suggests an opportunity for more systematic incorporation of relevant theories in future research to enhance the educational effectiveness of games.

\textbf{Frameworks referred:}
A total of 20 papers (29.4\%) mentioned the use of one or more frameworks in their work on educational games. 14 of them explicitly identified the frameworks applied. The reported frameworks are Stenmap Framework~\cite{hodhod_cyberhero_2023}, Bloom's Taxonomy~\cite{hodhod_cyberhero_2023, kim_security_2021}, Investigate and Decide Learning Environment (IDLE)~\cite{casey_motivating_2023}, Octalysis Gamification Framework~\cite{decusatis_gamification_2022, decusatis_cybersecurity_2022}, Procedural Rhetoric Guidelines~\cite{maqsood_design_2021}, ARCS (Attention, Relevance, Confidence, and Satisfaction) model of motivation~\cite{giannakas_comprehensive_2019}, Framework for Developing a Moral Decision-making Curriculum~\cite{raynes-goldie_gaming_2014}, EDR (Educational Design Research) Framework~\cite{thomas2019educational}, MDA (Mechanics, Dynamics, Aesthetics) Framework~\cite{natalia2023gamification}, 5Ds of Privacy Literacy~\cite{yuan_redcapes_2024}, EDA (Experience, Dynamics and Artifacts)~\cite{jaafar_empowering_2024}, and CI (Cooperative Inquiry) Framework~\cite{blinder_evaluating_2024}, Digital Competence framework (DigComp)~\cite{gioia_cyber_2019}. The remaining six papers acknowledged the use of one or more frameworks but did not provide specific names.

Most of the frameworks reported across the papers are distinct, except Bloom's Taxonomy and Octalysis Gamification Framework, which are referred in two papers each. This diversity underscores the absence of a widely accepted or standardised framework in the field. The lack of uniformity in framework adoption highlights a challenge for researchers and practitioners seeking to build upon prior work or establish consistent methodologies in educational game development. This fragmentation indicates the need for more efforts to develop and promote adaptable frameworks that can serve as a foundation for future research and application.

% \begin{table}[!htb]
% \centering
% \caption{Theories and Frameworks Referred in the Papers}
% \begin{tabular}{lcc}
% \toprule
% \textbf{Category} & \textbf{Status} & \textbf{\#(Papers)}\\
% \midrule
% Theory referred & Yes & 6 (8.8\%)\\
% & No & 62 (91.2\%)\\
% \midrule
% Theory name listed & Yes & 5 (7.4\%)\\
% & No & 1 (1.5\%)\\
% & Not Applicable & 62 (91.2\%)\\
% \midrule
% Framework referred & Yes & 20 (29.4\%)\\
% & No & 48 (70.6\%)\\
% \midrule
% Framework name listed & Yes & 14 (20.6\%)\\
% & No & 6 (8.8\%)\\
% & Not Applicable & 48 (70.6\%)\\
% \bottomrule
% \end{tabular}
% \label{tab:theories_and_frameworks}
% \end{table}

\textbf{Summary:}
The reviewed papers reveal notable gaps in the systematic application of theories and frameworks in the design, implementation, and evaluation of educational games. Only six papers explicitly referenced theories, and five reported the distinct theories they applied. Similarly, while 20 papers referred frameworks, only 14 provided specific names, most of which were distinct. The lack of commonly adopted theories and frameworks highlights the absence of widely accepted dominant theories and frameworks.

\section{Further Discussions}
\label{sec:further_discussions}

Serious games educating children cyber security should be designed to influence children's cyber security practices in real-world contexts~\cite{drevin_story_2023}. This can be achieved by considering appropriate educational and non-educational design elements. In addition, the effectiveness of these games should be rigorously evaluated, and the achieved learning outcomes should be clearly demonstrated. This section will focus on the principal aspects with respect to the design elements and effectiveness of the games, corresponding to the main themes used to present the results in the preceding section. The research gaps and future directions regarding these aspects will be discussed, with the goal of addressing RQ4.

\subsection{Educational Design Elements of the Games}

The cyber security games reported in the papers were designed to address specific cyber security learning outcomes. They address these outcomes by presenting a content related to the cyber security concepts they intend to address. This content is essential to effectively achieve the learning outcomes. The content presented throughout the games addressed both technical and non-technical concepts of cyber security, in contrast to gaming products examined in an earlier study~\cite{coenraad_experiencing_2020}, which were criticised for overly concentrating on technical elements and overlooking the social dimensions of cyber security.

The findings revealed a significant gap between the proposed and achieved learning outcomes. This significant discrepancy necessitates a discussion on how cyber security content should be developed and embedded within the games. Therefore, the following three main recommendations are given as potential solutions to address this discrepancy.

\textbf{Adoption of hybrid (combining bottom-up and top-down) content development approaches:}
More than half of the total games reviewed in this paper used an ad-hoc content development approach when identifying security concepts and developing the required content. Only a small portion of the total games followed either a bottom-up or top-down approach, and none of them applied both approaches.

The bottom-up approach requires the participation of target audiences, such as children, in the content creation process, and is one of the recommended approaches to content development. More than 10 years ago, \citet{raynes-goldie_gaming_2014} suggested that instructional methods that transfer knowledge from authorised adults to children can hinder the development of children's cognitive and social skills, which is essential for effective privacy management. They highlighted the benefits of allowing children to act independently, make suitable choices, and rely on their own resources instead of depending on adults for guidance and materials. \citet{hartikainen_childrens_2019} also argued that children should be involved in the development and evaluation of their own cyber security education. However, they acknowledged the importance of adult participation in the development of cyber security content for children. In this regard, the bottom-up approach provides an opportunity for children who co-design games with adult partners to determine or propose the content of their games based on the problems they experience and visualise, which can lead to better outcomes.

On the other hand, the top-down approach to content development relies on curricula, standards, guidelines, and other frameworks developed by governmental and non-governmental bodies to define what children should or need to learn through games~\cite{saglam_systematic_2023}. However, these frameworks are quite varied, and there is a significant difference in their adoption across different games. In their SLR on this topic, \citet{saglam_systematic_2023} found that the existing literature on cyber security curriculum topics is highly fragmented, and there is no recommended comprehensive curriculum for schools. However, multiple context-relevant frameworks can be used to create cyber security content for games, as demonstrated in the development of a game for young children reported in~\cite{giannakas_comprehensive_2019}.

These two approaches to content development often lead to different conclusions about what content is needed most~\cite{saglam_systematic_2023}. Therefore, it is recommended to consider both approaches based on the objectives and related values of children and educators when developing cyber security content for games~\cite{hartikainen_childrens_2019}. Both approaches are useful to develop effective cyber security content that can engage and benefit children~\cite{saglam_systematic_2023}, and hence a \textbf{\textit{hybrid content development approach}} is recommended.

In the hybrid approach, relevant content is identified based on the selected curriculum or standard. This content then guides discussions with children in bottom-up content creation sessions, providing a framework to structure the ideas raised by the children. Conversely, the original content is refined based on the children's input. This helps to incorporate their needs as well as address the requirements of the curriculum or standard.

\textbf{Using theory-informed and hybrid content development approach:}
Socio-technical aspects of cyber security require interdisciplinary approaches that demand consideration of people and technology~\cite{pellicone_designing_2022}. Therefore, designing an effective educational cyber security game for children requires a theory-informed interdisciplinary approach~\cite{maqsood_design_2021}. Previous studies on game-based learning emphasised the demand for theory-driven game designs. This has been advocated by some researchers for many years, e.g., almost ten years ago, \citet[p.~116]{clark_digital_2016} urged researchers to shift emphasis from studies that ask ``Can games support learning?'' and ``Are games better or worse than other media for learning?'' to ``exploring how theoretically driven design decisions influence situated learning outcomes for the broad diversity of learners''.

Utilising social science theories, such as psychological and educational theories, can help cyber security games to be more effective, engaging, interactive and personalised to the target audiences~\cite{hodhod_cyberhero_2023, maqsood_design_2021}. These theories can help game designers/developers to understand the necessary requirements and factors more systematically, informing the design of effective educational games. The effectiveness of this approach was reported in the development and evaluation of games such as \textit{CyberAware}~\cite{giannakas_comprehensive_2019} and ``\textit{A Day in the Life of the JOs}''~\cite{maqsood_design_2021}. The authors applied relevant psychological theories and learning strategies while developing these games. The effectiveness of this approach was demonstrated by significantly achieving their learning outcomes.

Social science theories can be utilised in a hybrid content development approach. Therefore, we recommend a \textbf{\textit{theory-informed hybrid content development approach}} which involve using relevant theories in the top-down and bottom-up stages of the hybrid approach. These theories can be applied to cascade curriculum or standard into content that is relevant to children's contexts. This approach facilitates adapting curriculum content according to children's developmental stages, ensuring alignment with their age and comprehension levels. Moreover, these theories assist in understanding children's contexts and requirements during the bottom-up content development, thereby enhancing interaction during content creation and fostering conditions that increase their engagement in the process. 

In addition, these theories help to design effective methods to share the content. They can help to identify more effective intervention methods to positively influence children towards desired cyber security outcomes. This understanding ultimately supports the effective dissemination of the developed content through games. This can be achieved by following a theory-informed approach while designing game scenarios and stories, as discussed below.

\textbf{Designing theory-informed and hybrid scenarios and game stories:}
When developing a cyber security educational game, the development of cyber security content will be followed by embedding the content into the game. This involves designing relevant scenarios and effective game stories that can help to achieve the desired learning outcomes. Game stories are essential for embedding effective, engaging, and enduring educational content into the game~\cite{casey_motivating_2023}. However, most of the games reviewed have overlooked designing stories and scenarios that effectively integrate cyber security content into engaging gameplay. Games with rich or moderately detailed stories (thick and medium depth stories) capable of accommodating such scenarios are notably rare.

Game scenarios should be realistic and reflect real-world situations to enable players to apply the learned behaviours or skills in their daily lives~\cite{maqsood_design_2021}. Games that simulate real-world scenarios provide children with practical experience in identifying and responding to cyber threats~\cite{hodhod_cyberhero_2023}. Overlooking this fact can result in games that do not address real-world cyber security problems of children beyond entertaining them. \citet{coenraad_experiencing_2020} emphasised this concern by criticising games trading simply by using `cyber' as an aesthetic hook to attract players without engaging deeply with cyber security content and addressing practical learning outcomes.

A good example that demonstrate utilisation of real-world scenarios is the ``\textit{A Day in the Life of the JOs}'' game~\cite{maqsood_design_2021}. This game has evolving (thick) story and involves realistic scenarios that can help to teach 11-13-year-olds cyber security, privacy, and digital literacy issues based on scenarios they encounter in their day-to-day activity. The game demonstrates the importance of considering social science theories and creating realistic scenarios to increase cyber security awareness of children and positively influence their behavioural intent. This indicates a \textbf{\textit{theory-informed hybrid scenario and story design}} approach is relevant to address learning outcomes and consider the context of children in the game design.

Hybrid scenario design incorporates involving children in scenario design process (bottom-up), and designing scenarios based on the previously developed content (top-down). Children will actively engage in scenario design discussions and be encouraged to suggest scenarios relevant to them. The content will be used to guide the scenario design session although children have the freedom to deviate from it. Based on the output of the children, adults will design additional scenarios that are not covered by children and required to address the content. These scenarios will then be woven into a unified overarching narrative for the game's storyline. The overall design process will be driven by appropriate social science theories.

In conclusion, cyber security content constitutes the primary component of cyber security educational games targeting children. Therefore, a more systematic approach should be adopted when developing and integrating it into games as recommended in this section. The three recommendations discussed in this subsection will be summarised by proposing a \textbf{\textit{theory-informed and hybrid game design approach}}. The following main activities will help to implement this approach:

\begin{enumerate}
\item \textbf{\textit{Developing content based on frameworks:}} This initial step entails selecting applicable cyber security education frameworks such as curricula, standards, and guidelines, and identifying content aligned with these frameworks. Theories that aid in understanding children's developmental stages and comprehension levels should guide this content development.

\item \textbf{\textit{Engaging children in content development:}} Children will actively participate in discussions to identify content relevant to their needs and contexts. Their ideas will be valued throughout the process. Utilising theories that enhance understanding of children's contexts and perspectives can improve interaction during content development and create conditions that increase their engagement in the process.

\item \textbf{\textit{Harmonising derived contents:}} This involves harmonising the contents developed in the above two approaches. There can be variations between the content identified from frameworks and identified by children. These distinct contents should be harmonised to create the final content that will guide scenario development. 

\item \textbf{\textit{Involving children in scenario design:}} Children will participate in engaging scenario design sessions, which involve adult design partners, to identify scenarios that meet their needs. While the developed content will guide scenario design, children will have the freedom to deviate from it. They should be encouraged to create their own scenarios. At this stage, it is crucial to utilise relevant social science theories to empower children in developing scenarios that authentically reflect their daily interactions and real-world situations.

\item \textbf{\textit{Designing supplementary scenarios:}} Scenarios derived from children's engagement should be supplemented with additional scenarios if those designed by the children are inadequate to address the identified content. This ensures comprehensive coverage of proposed learning outcomes. Adult game designers should engage in discussions with children to gather their input on the scenarios they design, ensuring the scenarios are tailored to align with children's interests and comprehension levels.

\item \textbf{\textit{Designing one or more game stories:}} The scenarios should be integrated into a cohesive overarching narrative for the game's storyline. This narrative helps to seamlessly incorporate scenarios, enhancing continuity and engagement throughout gameplay. This step should consider relevant theories to ensure that the overall game story effectively considers children's context and addresses the proposed learning outcomes.
\end{enumerate}

\subsection{Non-Educational Design Elements of the Games}

Various non-educational design elements are identified, along with associated findings and research gaps, in Section~\ref{sec:SLR_results}. Among these elements, this further discussion focuses specifically on platforms and related design elements. Variations in platforms significantly influence the types of games played by people in different age groups~\cite{coenraad_experiencing_2020} and, consequently, the cyber security outcomes achieved by these groups. Most of the games reported in the literature utilised traditional platforms (i.e., computers, mobile devices, and web browsers) to address their audiences. In contrast, emerging platforms are underused despite their potential impact on cyber security education. In addition, physical games are under-represented in the reported games.

XR platforms are among the underutilised platforms. These platforms were used only in seven games, and there was no reported evidence demonstrating the effectiveness of most of these games. There are promising findings that showed XR platforms have the potential to be an effective medium for cyber security education targeting adults~\cite{ulsamer2021immersive}. The key features that make them attractive for cyber security education are immersion, navigation, and interaction~\cite{gibson2018virtual}. These immersive and interactive experiences can engage children, and will have the potential to convey abstract cyber security concepts, making learning both effective and enjoyable~\cite{chiou_augmented_2021}. These features of AR, VR and MR platforms make them potentially one of the better platforms for game-based cyber security education targeting children.

AR platforms have the ability to overlay digital content onto the real-world environment in real time, improving children's perception of reality by adding digital elements that complement or augment the physical world~\cite{chiou_augmented_2021}. This coexistence of virtual and real information in AR enables children to visualise and interact with abstract cyber security concepts in an engaging and intuitive way~\cite{chiou_augmented_2021}. AR games can be played using smart mobile devices or AR glasses designed specifically for AR applications. While there is no age limit for mobile device-based AR games, AR glasses are not recommended for children below 13 years old~\cite{hololens2_age_limit}. Consequently, mobile device-based AR games are a more suitable option for educational cyber security games targeting young children. This platform has been used by a collaborative AR game aimed at teaching abstract mathematics concepts to young children~\cite{stappen2019mathbuilder}.

The immersive nature of VR isolates users from the real world and immerses them into a fully virtual space where they can interact with the digital environment in a way that feels realistic~\cite{gibson2018virtual}. VR platforms have promising potential in cyber security education~\cite{ulsamer2021immersive} and may improve learning outcomes for children. However, this platform is recommended for children aged 13 and older, with children between 10 and 12 years of age allowed to use it under parental controls and additional protections~\cite{metaquest_age_limit}. This limitation shows that VR is not suitable for cyber security education targeting young children.

MR combines elements from both AR and VR, using AR's overlay capabilities and VR's immersive simulation to blend the real and virtual worlds, making it a versatile tool for cyber security education~\cite{shen_work_progress_design_2021}. However, similar age restrictions apply to MR as well, making it unsuitable for young children. The promising potential of AR, VR and MR, combined with the established success of game-based learning, could present a promising alternative to teach cyber security for children. However, research on applying these platforms specifically for cyber security education targeting children remains limited, highlighting the need for further studies.

Physical games that involve physical bodily movement are relatively scarce, despite their important role in promoting physical activity among children. One of the two games that involve a physical environment, `A NERD DOGMA'~\cite{costa_nerd_2020}, integrates both virtual and physical activity. This game combines a quest-style capture the flag (CTF) challenge with a physical escape room, incorporating four entry-level cyber security tasks. Completion of the game requires players to engage in physical movement, effectively blending digital gameplay with real-world activity.

Active video games that require physical interaction can serve as viable alternatives that leverage the benefits of both digital and physical games. A 2010 literature review on such games~\cite{foley2010use} highlighted that active video games, where players physically engage with on-screen content, can result in increased energy expenditure and improved body composition in children when compared with non-active video games. Although not called physical games, such digital games still allow children to physically interact, using arm, leg, or whole-body movement, with images on screen in a variety of activities, and are equivalent to mild to moderate-intensity physical activity. AR, VR, and MR-based cyber security games can be considered as typical examples of such active video games due to their interactive features that involve physical movement. Among these, AR games -- particularly those developed for mobile platforms -- show strong potential for engaging children in a wider variety of physical activities, extending beyond the limited movement patterns typically associated with VR-based games.

Other emerging platforms, such as social robots~\cite{chiou2020social}, the metaverse~\cite{yeoh2025immersive}, artificial intelligence (AI) and chatbots~\cite{piccolo2021chatbots}, also demonstrate significant potential for delivering cyber security education to children in more engaging and interactive ways. The combination of physical interaction and conversational capabilities makes social robots promising alternatives in capturing children's attention and enhancing learning outcomes. However, the high costs incurred by the use of robots may limit widespread adoption in educational settings. The immersive virtual environment provided by the metaverse has the potential to offer engaging cyber security educational experiences for children. The increasing integration of AI, along with its capacity for adaptive learning and personalised support, positions AI as a promising platform for cyber security education. Similarly, chatbots offer interactive, dialogue-based learning experiences and provide a sense of anonymity, creating a safe space for children to ask questions they may feel uncomfortable raising in person.

\subsection{Effectiveness of the Games}

The effectiveness of cyber security games was evaluated through experiments, with or without control groups, and the achieved learning outcomes were reported. Some games, such as~\cite{maqsood_design_2021, giannakas_comprehensive_2019}, provided sufficient evidence to support their outcomes. However, in many cases, the achieved learning outcomes fell short of those proposed, raising concerns about the overall effectiveness of the games. Some of the reported findings are also questionable. For example, \citet{jin_game_2018} presented evaluation results for four games without providing specific evidence on the achieved learning outcomes of each game. Furthermore, some games have not been evaluated at all, e.g., four XR-based games targeting young children reported in~\cite{shen_work_progress_design_2021, chiou_augmented_2021}, and eight games that utilise different platforms reported in~\cite{scholl_information_2020}. These notable gaps cast uncertainty on the true effectiveness of these games.

The number of games with achieved learning outcomes related to cyber security and privacy behaviours is low. The minimal reporting on these learning outcomes could be attributed to the difficulties in measuring these effects. \citet{maqsood_design_2021} argued that evaluating real behaviours is impractical because it is unethical to place experiment participants in dangerous situations to assess their behaviours. As an alternative approach, they recommended measuring behavioural intent which is ``a predictor of whether users will take part in certain behaviours in the future''~\cite[p.~28:15]{maqsood_design_2021}.

In more than three quarter of papers (76.5\%), control groups were not employed when evaluating the achieved learning outcomes of the games. Furthermore, although quantitative or mixed research methods were applied by authors of 47 papers (69.1\%), only 16 papers (23.5\%) report evaluation results using both inferential and descriptive statistical metrics. This highlights a gap in the research methodology and adherence to a rigorous quantitative research design. Moreover, across the studies conducted with quantitative approaches, the critical measure of the effect size was largely overlooked, with the exception of six studies. This omission undermines the interpretation, credibility, and practical applicability of the reported findings. Without sufficient statistical evidence to support the claims, it can be difficult to conclusively establish the effectiveness of these games.

Concerns regarding the efficacy of the reported cyber security educational games arise due to the gaps in assessing their effectiveness and the absence of methodological rigour~\cite{maqsood_design_2021, hodhod_cyberhero_2023}. Most of the studies relied on pre- and post-surveys to assess the effectiveness of the games. However, this method has faced criticism due to the potential gap in demonstrating the game's effectiveness, as surveys might not gather accurate and honest responses from participants~\cite{hodhod_cyberhero_2023}. Furthermore, some games are assessed using only a post-test evaluation, without establishing a baseline through a pre-test. This results in largely untested empirical effectiveness of the games in meeting their proposed learning outcomes~\cite{maqsood_design_2021}.

The identified gaps underscore the need for a dedicated framework that specifically evaluates the effectiveness of cyber security games designed for children. While existing frameworks~\cite{all2015towards, serrano2012framework} proposed to evaluate various aspects of games can provide valuable inputs, they do not align with the unique context of cyber security games targeting children. Therefore, developing a tailored framework that considers both key educational and non-educational design elements is essential for effectively evaluating cyber security games targeting children.

\subsection{Research Gaps and Future Work}

The research gaps were presented in Section~\ref{sec:SLR_results} and the preceding three subsections of this section. Moreover, no review was found on existing cyber security educational games targeting children that have been developed by stakeholders beyond academia and distributed through various application stores and websites. A related study was conducted by \citet{coenraad_experiencing_2020} five years ago; however, it does not reflect the current landscape and its focus is on general cyber security games rather than those specifically designed for children.

The following future research directions are proposed in light of these gaps.

\begin{enumerate}
\item Exploring the key areas of cyber security education that are most relevant and appropriate for children by involving children, parents, teachers and other school staff members.

\item Analysing relevant social science theories that can inform the design of cyber security games for children.

\item Developing frameworks for evaluating the effectiveness of cyber security games designed for children.

\item Analysing real-world cyber security scenarios in existing games, and systematically classifying them to support development of more effective games for children.

\item Designing games that utilise emerging platforms such as XR, AI, chatbots, robots and metaverse.

\item Following a theory-informed and hybrid (combining bottom-up and top-down) game design approach.
\end{enumerate}

\section{Conclusion}
\label{sec:conclusion}

This SLR analysed educational and non-educational design elements of 91 games reported in 68 papers published in the past 15 years. These games are primarily designed for young children and teenagers to achieve various cyber security learning outcomes. Significant learning outcomes were achieved in some of the games. A range of technical and non-technical cyber security concepts are covered across the games.

Our findings revealed that the achieved learning outcomes often fall short of the proposed ones in many games, raising doubts on their effectiveness. Furthermore, there are gaps in the effective evaluation of the games, and a lack of methodological rigour undermines the efficacy of the proposed games. Most of the games lack rich and engaging stories that include scenarios relevant to the real-world context of children. The systematic application of relevant social science theories in the design process of the games is rare. A theory-informed and hybrid (combining bottom-up and top-down) content development and scenario design approach is recommended to enhance the effectiveness of educational games and the rigour of the design methods.

Delivering the learning content embedded in the game story requires selecting an appropriate game platform that caters to the target audience, as platform preferences vary based on needs and platform availability. Most of the games utilise traditional platforms such as mobile devices, computers, and the web. Emerging platforms, such as XR and robots, are seldom used despite their potential to deliver cyber security education.

\bibliographystyle{ACM-Reference-Format}
\bibliography{main}


\begin{thebibliography}{108}

%%% the Web address.

\ifx \showCODEN    \undefined \def \showCODEN     #1{\unskip}     \fi
\ifx \showISBNx    \undefined \def \showISBNx     #1{\unskip}     \fi
\ifx \showISBNxiii \undefined \def \showISBNxiii  #1{\unskip}     \fi
\ifx \showISSN     \undefined \def \showISSN      #1{\unskip}     \fi
\ifx \showLCCN     \undefined \def \showLCCN      #1{\unskip}     \fi
\ifx \shownote     \undefined \def \shownote      #1{#1}          \fi
\ifx \showarticletitle \undefined \def \showarticletitle #1{#1}   \fi
\ifx \showURL      \undefined \def \showURL       {\relax}        \fi
% invisible to TeX
\providecommand\bibfield[2]{#2}
\providecommand\bibinfo[2]{#2}
\providecommand\natexlab[1]{#1}
\providecommand\showeprint[2][]{arXiv:#2}

\bibitem[Alahmari et~al\mbox{.}(2022)]%
        {alahmari2022moving}
\bibfield{author}{\bibinfo{person}{Salwa Alahmari}, \bibinfo{person}{Karen Renaud}, {and} \bibinfo{person}{Isaac Omoronyia}.} \bibinfo{year}{2022}\natexlab{}.
\newblock \showarticletitle{Moving beyond cyber security awareness and training to engendering security knowledge sharing}.
\newblock \bibinfo{journal}{\emph{Information Systems and e-Business Management}}  \bibinfo{volume}{21} (\bibinfo{year}{2022}), \bibinfo{pages}{123--158}.
\newblock
\href{https://doi.org/10.1007/s10257-022-00575-2}{doi:\nolinkurl{10.1007/s10257-022-00575-2}}


\bibitem[Alemanno et~al\mbox{.}(2024)]%
        {giuseppe_blue_2024}
\bibfield{author}{\bibinfo{person}{Giuseppe Alemanno}, \bibinfo{person}{Daniele Semeraro}, {and} \bibinfo{person}{Veronica Rossano}.} \bibinfo{year}{2024}\natexlab{}.
\newblock \showarticletitle{Blue and Red Team Quiz Game to Train High School Students}. In \bibinfo{booktitle}{\emph{Proceedings of the 2nd International Workshop on CyberSecurity Education for Industry and Academia ({CSE4IA} 2024)}}.
\newblock
\urldef\tempurl%
\url{https://ceur-ws.org/Vol-3700/paper2.pdf}
\showURL{%
\tempurl}


\bibitem[Alemany et~al\mbox{.}(2020)]%
        {alemany_assessing_2020}
\bibfield{author}{\bibinfo{person}{Jose Alemany}, \bibinfo{person}{Elena Del~Val}, {and} \bibinfo{person}{Ana Garcia-Fornes}.} \bibinfo{year}{2020}\natexlab{}.
\newblock \showarticletitle{Assessing the Effectiveness of a Gamified Social Network for Applying Privacy Concepts: An Empirical Study With Teens}.
\newblock \bibinfo{journal}{\emph{IEEE Transactions on Learning Technologies}} \bibinfo{volume}{13}, \bibinfo{number}{4} (\bibinfo{year}{2020}), \bibinfo{pages}{777--789}.
\newblock
\href{https://doi.org/10.1109/TLT.2020.3026584}{doi:\nolinkurl{10.1109/TLT.2020.3026584}}


\bibitem[All et~al\mbox{.}(2015)]%
        {all2015towards}
\bibfield{author}{\bibinfo{person}{Anissa All}, \bibinfo{person}{Elena Patricia~Nu{\~n}ez Castellar}, {and} \bibinfo{person}{Jan Van~Looy}.} \bibinfo{year}{2015}\natexlab{}.
\newblock \showarticletitle{Towards a conceptual framework for assessing the effectiveness of digital game-based learning}.
\newblock \bibinfo{journal}{\emph{Computers \& Education}}  \bibinfo{volume}{88} (\bibinfo{year}{2015}), \bibinfo{pages}{29--37}.
\newblock
\href{https://doi.org/10.1016/j.compedu.2015.04.012}{doi:\nolinkurl{10.1016/j.compedu.2015.04.012}}


\bibitem[Allers et~al\mbox{.}(2021)]%
        {allers_childrens_2021}
\bibfield{author}{\bibinfo{person}{J. Allers}, \bibinfo{person}{G.~R. Drevin}, \bibinfo{person}{D.~P. Snyman}, \bibinfo{person}{H.~A. Kruger}, {and} \bibinfo{person}{L. Drevin}.} \bibinfo{year}{2021}\natexlab{}.
\newblock \showarticletitle{Children’s Awareness of Digital Wellness: A Serious Games Approach}.
\newblock In \bibinfo{booktitle}{\emph{Information Security Education for Cyber Resilience: 14th IFIP WG 11.8 World Conference, WISE 2021, Virtual Event, June 22–24, 2021, Proceedings}}. \bibinfo{series}{IFIP Advances in Information and Communication Technology}, Vol.~\bibinfo{volume}{615}. \bibinfo{publisher}{Springer}, \bibinfo{pages}{95--110}.
\newblock
\href{https://doi.org/10.1007/978-3-030-80865-5_7}{doi:\nolinkurl{10.1007/978-3-030-80865-5_7}}


\bibitem[Alma'ariz et~al\mbox{.}(2022)]%
        {alma2022soceng}
\bibfield{author}{\bibinfo{person}{Salsa Alma'ariz}, \bibinfo{person}{Raden~Budiarto Hadiprakoso}, \bibinfo{person}{Nurul Qomariasih}, {et~al\mbox{.}}} \bibinfo{year}{2022}\natexlab{}.
\newblock \showarticletitle{{Soceng Warriors}: Game-Based Learning to Increase Security Awareness Against Social Engineering Attacks}. In \bibinfo{booktitle}{\emph{Proceedings of the 2022 {IEEE} 8th Information Technology International Seminar}}. \bibinfo{publisher}{IEEE}, \bibinfo{pages}{124--129}.
\newblock
\href{https://doi.org/10.1109/ITIS57155.2022.10009041}{doi:\nolinkurl{10.1109/ITIS57155.2022.10009041}}


\bibitem[Alsadhan et~al\mbox{.}(2020)]%
        {alsadhan_manar_2020}
\bibfield{author}{\bibinfo{person}{Afnan Alsadhan}, \bibinfo{person}{Asma Alotaibi}, \bibinfo{person}{Lulu Altamran}, \bibinfo{person}{Majd Almalki}, \bibinfo{person}{Moneera Alfulaij}, {and} \bibinfo{person}{Tarfa Almoneef}.} \bibinfo{year}{2020}\natexlab{}.
\newblock \showarticletitle{{Manar}: An Arabic Game-based Application Aimed for Teaching Cybersecurity using Image Processing}.
\newblock \bibinfo{journal}{\emph{International Journal of Advanced Computer Science and Applications}} \bibinfo{volume}{11}, \bibinfo{number}{10} (\bibinfo{year}{2020}), \bibinfo{pages}{410--416}.
\newblock
\href{https://doi.org/10.14569/IJACSA.2020.0111051}{doi:\nolinkurl{10.14569/IJACSA.2020.0111051}}


\bibitem[Apperley(2006)]%
        {apperley2006genre}
\bibfield{author}{\bibinfo{person}{Thomas~H. Apperley}.} \bibinfo{year}{2006}\natexlab{}.
\newblock \showarticletitle{Genre and game studies: Toward a critical approach to video game genres}.
\newblock \bibinfo{journal}{\emph{Simulation \& Gaming}} \bibinfo{volume}{37}, \bibinfo{number}{1} (\bibinfo{year}{2006}), \bibinfo{pages}{6--23}.
\newblock
\href{https://doi.org/10.1177/1046878105282278}{doi:\nolinkurl{10.1177/1046878105282278}}


\bibitem[Azzahra et~al\mbox{.}(2024)]%
        {azzahra_socenggo_2024}
\bibfield{author}{\bibinfo{person}{Fadel Azzahra}, \bibinfo{person}{Nurul Qomariasih}, \bibinfo{person}{Herman Kabetta}, \bibinfo{person}{Hermawan Setiawan}, \bibinfo{person}{Rheva~Anindya Wijayanti}, {and} \bibinfo{person}{Taqiya Nabilla~Nathania Afnani}.} \bibinfo{year}{2024}\natexlab{}.
\newblock \showarticletitle{{SocengGo}: Social Engineering Educational Application Based on Attack--Defense Multiplayer Card Game}. In \bibinfo{booktitle}{\emph{Proceedings of the 2024 10th International Conference on Education Technology, Education Innovation and Society ({ICET})}}. \bibinfo{publisher}{IEEE}, \bibinfo{pages}{118--123}.
\newblock
\href{https://doi.org/10.1109/ICET64717.2024.10778458}{doi:\nolinkurl{10.1109/ICET64717.2024.10778458}}


\bibitem[Balakrishna(2021)]%
        {balakrishna_design_2021}
\bibfield{author}{\bibinfo{person}{Chitra Balakrishna}.} \bibinfo{year}{2021}\natexlab{}.
\newblock \showarticletitle{Design considerations for developing a game-based learning resource for cyber security education}.
\newblock \bibinfo{journal}{\emph{Proceedings of the 2021 European Conference on Games-based Learning}}  \bibinfo{volume}{2021} (\bibinfo{year}{2021}), \bibinfo{pages}{80--89}.
\newblock
\urldef\tempurl%
\url{https://oro.open.ac.uk/81179/}
\showURL{%
\tempurl}


\bibitem[Bassi et~al\mbox{.}(2023)]%
        {bassi2023serious}
\bibfield{author}{\bibinfo{person}{Giorgia Bassi}, \bibinfo{person}{Stefania Fabbri}, {and} \bibinfo{person}{Anna Vaccarelli}.} \bibinfo{year}{2023}\natexlab{}.
\newblock \showarticletitle{A Serious Video Game on Cybersecurity}. In \bibinfo{booktitle}{\emph{Entertainment Computing -- ICEC 2023: 22nd IFIP TC 14 International Conference, ICEC 2023, Bologna, Italy, November 15--17, 2023, Proceedings}}. \bibinfo{publisher}{Springer}, \bibinfo{pages}{341--345}.
\newblock
\href{https://doi.org/10.1007/978-981-99-8248-6_29}{doi:\nolinkurl{10.1007/978-981-99-8248-6_29}}


\bibitem[Berger et~al\mbox{.}(2019)]%
        {berger_privacity_2019}
\bibfield{author}{\bibinfo{person}{Erlend Berger}, \bibinfo{person}{Torjus~H. Sæthre}, {and} \bibinfo{person}{Monica Divitini}.} \bibinfo{year}{2019}\natexlab{}.
\newblock \showarticletitle{{PrivaCity}: A Chatbot Game to Raise Privacy Awareness Among Teenagers}.
\newblock In \bibinfo{booktitle}{\emph{Informatics in Schools. New Ideas in School Informatics: 12th International Conference on Informatics in Schools: Situation, Evolution, and Perspectives, ISSEP 2019, Larnaca, Cyprus, November 18–20, 2019, Proceedings}}. \bibinfo{series}{Lecture Notes in Computer Science}, Vol.~\bibinfo{volume}{11913}. \bibinfo{publisher}{Springer}, \bibinfo{pages}{293--304}.
\newblock
\href{https://doi.org/10.1007/978-3-030-33759-9_23}{doi:\nolinkurl{10.1007/978-3-030-33759-9_23}}


\bibitem[Bioglio et~al\mbox{.}(2019)]%
        {bioglio_social_2019}
\bibfield{author}{\bibinfo{person}{Livio Bioglio}, \bibinfo{person}{Sara Capecchi}, \bibinfo{person}{Federico Peiretti}, \bibinfo{person}{Dennis Sayed}, \bibinfo{person}{Antonella Torasso}, {and} \bibinfo{person}{Ruggero~G. Pensa}.} \bibinfo{year}{2019}\natexlab{}.
\newblock \showarticletitle{A Social Network Simulation Game to Raise Awareness of Privacy Among School Children}.
\newblock \bibinfo{journal}{\emph{{IEEE} Transactions on Learning Technologies}} \bibinfo{volume}{12}, \bibinfo{number}{4} (\bibinfo{year}{2019}), \bibinfo{pages}{456--469}.
\newblock
\href{https://doi.org/10.1109/TLT.2018.2881193}{doi:\nolinkurl{10.1109/TLT.2018.2881193}}


\bibitem[Blinder et~al\mbox{.}(2024)]%
        {blinder_evaluating_2024}
\bibfield{author}{\bibinfo{person}{Elana~B. Blinder}, \bibinfo{person}{Marshini Chetty}, \bibinfo{person}{Jessica Vitak}, \bibinfo{person}{Zoe Torok}, \bibinfo{person}{Salina Fessehazion}, \bibinfo{person}{Jason Yip}, \bibinfo{person}{Jerry~Alan Fails}, \bibinfo{person}{Elizabeth Bonsignore}, {and} \bibinfo{person}{Tamara Clegg}.} \bibinfo{year}{2024}\natexlab{}.
\newblock \showarticletitle{Evaluating the Use of Hypothetical `Would You Rather' Scenarios to Discuss Privacy and Security Concepts with Children}. In \bibinfo{booktitle}{\emph{Proceedings of the ACM on Human--Computer Interaction}}, Vol.~\bibinfo{volume}{8}. \bibinfo{publisher}{ACM}, Article \bibinfo{articleno}{165}, \bibinfo{numpages}{32}~pages.
\newblock
\href{https://doi.org/10.1145/3641004}{doi:\nolinkurl{10.1145/3641004}}


\bibitem[Brink and Wellman(2020)]%
        {brink2020robot}
\bibfield{author}{\bibinfo{person}{Kimberly~A. Brink} {and} \bibinfo{person}{Henry~M. Wellman}.} \bibinfo{year}{2020}\natexlab{}.
\newblock \showarticletitle{Robot teachers for children? Young children trust robots depending on their perceived accuracy and agency.}
\newblock \bibinfo{journal}{\emph{Developmental Psychology}} \bibinfo{volume}{56}, \bibinfo{number}{7} (\bibinfo{year}{2020}), \bibinfo{pages}{1268--1277}.
\newblock
\href{https://doi.org/10.1037/dev0000884}{doi:\nolinkurl{10.1037/dev0000884}}


\bibitem[Cardoso et~al\mbox{.}(2022)]%
        {cardoso_playing_2022}
\bibfield{author}{\bibinfo{person}{Felipe Cardoso}, \bibinfo{person}{Davide Andreoletti}, \bibinfo{person}{Alessandro Ferrari}, \bibinfo{person}{Luca Botturi}, \bibinfo{person}{Tiffany Fioroni}, \bibinfo{person}{Chiara Beretta}, \bibinfo{person}{Anna Picco-Schwendener}, \bibinfo{person}{Suzanna Marazza}, {and} \bibinfo{person}{Silvia Giordano}.} \bibinfo{year}{2022}\natexlab{}.
\newblock \showarticletitle{Playing for Privacy Awareness: Learning from a ``Wow-Moment'' with {iBuddy}}.
\newblock \bibinfo{journal}{\emph{Proceedings of the 16th European Conference on Games-based Learning}} (\bibinfo{year}{2022}), \bibinfo{pages}{128--138}.
\newblock
\href{https://doi.org/10.34190/ecgbl.16.1.490}{doi:\nolinkurl{10.34190/ecgbl.16.1.490}}


\bibitem[Casey et~al\mbox{.}(2023)]%
        {casey_motivating_2023}
\bibfield{author}{\bibinfo{person}{Eoghan Casey}, \bibinfo{person}{Jennifer Jocz}, \bibinfo{person}{Karen~A. Peterson}, \bibinfo{person}{Daryl Pfeif}, {and} \bibinfo{person}{Cassy Soden}.} \bibinfo{year}{2023}\natexlab{}.
\newblock \showarticletitle{Motivating youth to learn {STEM} through a gender inclusive digital forensic science program}.
\newblock \bibinfo{journal}{\emph{Smart Learning Environments}} \bibinfo{volume}{10}, \bibinfo{number}{1} (\bibinfo{year}{2023}).
\newblock
\href{https://doi.org/10.1186/s40561-022-00213-x}{doi:\nolinkurl{10.1186/s40561-022-00213-x}}


\bibitem[Chattopadhyay et~al\mbox{.}(2024)]%
        {chattopadhyay_vpet_2024}
\bibfield{author}{\bibinfo{person}{Ankur Chattopadhyay}, \bibinfo{person}{Saumya Sharma}, \bibinfo{person}{Elisee Mbaya}, {and} \bibinfo{person}{James Rice}.} \bibinfo{year}{2024}\natexlab{}.
\newblock \showarticletitle{{VPET}: A Novel Visual Privacy Themed Cybersecurity Educational Game}. In \bibinfo{booktitle}{\emph{Proceedings of the 2024 {IEEE} Frontiers in Education Conference (FIE)}}. \bibinfo{publisher}{IEEE}.
\newblock
\href{https://doi.org/10.1109/FIE61694.2024.10892871}{doi:\nolinkurl{10.1109/FIE61694.2024.10892871}}


\bibitem[Chiou et~al\mbox{.}(2020)]%
        {chiou2020social}
\bibfield{author}{\bibinfo{person}{Yan-Ming Chiou}, \bibinfo{person}{Tia Barnes}, \bibinfo{person}{Chrystalla Mouza}, {and} \bibinfo{person}{Chien-Chung Shen}.} \bibinfo{year}{2020}\natexlab{}.
\newblock \showarticletitle{Social Robot Teaches Cybersecurity}. In \bibinfo{booktitle}{\emph{Proceedings of the 2020 ACM Interaction Design and Children Conference: Extended Abstracts}}. \bibinfo{publisher}{ACM}, \bibinfo{pages}{199--204}.
\newblock
\href{https://doi.org/10.1145/3397617.3397824}{doi:\nolinkurl{10.1145/3397617.3397824}}


\bibitem[Chiou et~al\mbox{.}(2021)]%
        {chiou_augmented_2021}
\bibfield{author}{\bibinfo{person}{Yan-Ming Chiou}, \bibinfo{person}{Chien-Chung Shen}, \bibinfo{person}{Chrystalla Mouza}, {and} \bibinfo{person}{Teomara Rutherford}.} \bibinfo{year}{2021}\natexlab{}.
\newblock \showarticletitle{Augmented Reality-Based Cybersecurity Education on Phishing}. In \bibinfo{booktitle}{\emph{Proceedings of the 2021 {IEEE} International Conference on Artificial Intelligence and Virtual Reality}}. \bibinfo{publisher}{IEEE}, \bibinfo{pages}{228--231}.
\newblock
\href{https://doi.org/10.1109/AIVR52153.2021.00052}{doi:\nolinkurl{10.1109/AIVR52153.2021.00052}}


\bibitem[Christensen et~al\mbox{.}(2023)]%
        {christensen2023privacy}
\bibfield{author}{\bibinfo{person}{Michael Christensen}, \bibinfo{person}{Daniel Britze}, \bibinfo{person}{Jacob Vejlin}, \bibinfo{person}{Lene~Tolstrup S{\o}rensen}, {and} \bibinfo{person}{Jens~Myrup Pedersen}.} \bibinfo{year}{2023}\natexlab{}.
\newblock \showarticletitle{The {Privacy Universe} -- a game-based learning platform for data protection, privacy and ethics}. In \bibinfo{booktitle}{\emph{Proceedings of the 2023 {IEEE} Global Engineering Education Conference}}. \bibinfo{publisher}{IEEE}, \bibinfo{numpages}{8}~pages.
\newblock
\href{https://doi.org/10.1109/EDUCON54358.2023.10125160}{doi:\nolinkurl{10.1109/EDUCON54358.2023.10125160}}


\bibitem[Clark et~al\mbox{.}(2016)]%
        {clark_digital_2016}
\bibfield{author}{\bibinfo{person}{Douglas~B. Clark}, \bibinfo{person}{Emily~E. Tanner-Smith}, {and} \bibinfo{person}{Stephen~S. Killingsworth}.} \bibinfo{year}{2016}\natexlab{}.
\newblock \showarticletitle{Digital Games, Design, and Learning: A Systematic Review and Meta-Analysis}.
\newblock \bibinfo{journal}{\emph{Review of Educational Research}} \bibinfo{volume}{86}, \bibinfo{number}{1} (\bibinfo{year}{2016}), \bibinfo{pages}{79--122}.
\newblock
\href{https://doi.org/10.3102/0034654315582065}{doi:\nolinkurl{10.3102/0034654315582065}}


\bibitem[Coenraad et~al\mbox{.}(0 10)]%
        {coenraad_experiencing_2020}
\bibfield{author}{\bibinfo{person}{Merijke Coenraad}, \bibinfo{person}{Anthony Pellicone}, \bibinfo{person}{Diane~Jass Ketelhut}, \bibinfo{person}{Michel Cukier}, \bibinfo{person}{Jan Plane}, {and} \bibinfo{person}{David Weintrop}.} \bibinfo{year}{2020-10}\natexlab{}.
\newblock \showarticletitle{Experiencing Cybersecurity One Game at a Time: A Systematic Review of Cybersecurity Digital Games}.
\newblock \bibinfo{journal}{\emph{Simulation \& Gaming}} \bibinfo{volume}{51}, \bibinfo{number}{5} (\bibinfo{year}{2020-10}), \bibinfo{pages}{586--611}.
\newblock
\href{https://doi.org/10.1177/1046878120933312}{doi:\nolinkurl{10.1177/1046878120933312}}


\bibitem[Costa et~al\mbox{.}(2020)]%
        {costa_nerd_2020}
\bibfield{author}{\bibinfo{person}{Gabriele Costa}, \bibinfo{person}{Martina Lualdi}, \bibinfo{person}{Marina Ribaudo}, {and} \bibinfo{person}{Andrea Valenza}.} \bibinfo{year}{2020}\natexlab{}.
\newblock \showarticletitle{{A NERD DOGMA}: Introducing {CTF} to Non-expert Audience}.
\newblock \bibinfo{journal}{\emph{Proceedings of the 21st Annual Conference on Information Technology Education}} (\bibinfo{year}{2020}), \bibinfo{pages}{413--418}.
\newblock
\href{https://doi.org/10.1145/3368308.3415405}{doi:\nolinkurl{10.1145/3368308.3415405}}


\bibitem[{DeCusatis} et~al\mbox{.}(2022a)]%
        {decusatis_gamification_2022}
\bibfield{author}{\bibinfo{person}{Casimer {DeCusatis}}, \bibinfo{person}{Erin Alvarico}, {and} \bibinfo{person}{Omar Dirahoui}.} \bibinfo{year}{2022}\natexlab{a}.
\newblock \showarticletitle{Gamification of cybersecurity training}. In \bibinfo{booktitle}{\emph{Proceedings of the 1st International Workshop on Gamification of Software Development, Verification, and Validation}}. \bibinfo{publisher}{ACM}, \bibinfo{pages}{10--13}.
\newblock
\href{https://doi.org/10.1145/3548771.3561409}{doi:\nolinkurl{10.1145/3548771.3561409}}


\bibitem[{DeCusatis} et~al\mbox{.}(2022b)]%
        {decusatis_cybersecurity_2022}
\bibfield{author}{\bibinfo{person}{C. {DeCusatis}}, \bibinfo{person}{B. Gormanly}, \bibinfo{person}{E. Alvarico}, \bibinfo{person}{O. Dirahoui}, \bibinfo{person}{J. {McDonough}}, \bibinfo{person}{B. Sprague}, \bibinfo{person}{M. Maloney}, \bibinfo{person}{D. Avitable}, {and} \bibinfo{person}{B. Mah}.} \bibinfo{year}{2022}\natexlab{b}.
\newblock \showarticletitle{A Cybersecurity Awareness Escape Room using Gamification Design Principles}. In \bibinfo{booktitle}{\emph{Proceedings of the 2022 {IEEE} 12th Annual Computing and Communication Workshop and Conference}}. \bibinfo{publisher}{IEEE}, \bibinfo{pages}{765--770}.
\newblock
\href{https://doi.org/10.1109/CCWC54503.2022.9720748}{doi:\nolinkurl{10.1109/CCWC54503.2022.9720748}}


\bibitem[Del~Giudice et~al\mbox{.}(2023)]%
        {del2023sec}
\bibfield{author}{\bibinfo{person}{Nicola Del~Giudice}, \bibinfo{person}{Fausto Marcantoni}, \bibinfo{person}{Alessandro Marcelletti}, {and} \bibinfo{person}{Francesco Moschella}.} \bibinfo{year}{2023}\natexlab{}.
\newblock \showarticletitle{{SEC-GAME}: A Minigame Collection for Cyber Security Awareness}. In \bibinfo{booktitle}{\emph{Entertainment Computing -- ICEC 2023: 22nd IFIP TC 14 International Conference, ICEC 2023, Bologna, Italy, November 15--17, 2023, Proceedings}}. \bibinfo{publisher}{Springer}, \bibinfo{pages}{365--370}.
\newblock
\href{https://doi.org/10.1007/978-981-99-8248-6_33}{doi:\nolinkurl{10.1007/978-981-99-8248-6_33}}


\bibitem[Denning et~al\mbox{.}(2013)]%
        {denning_control-alt-hack_2013}
\bibfield{author}{\bibinfo{person}{Tamara Denning}, \bibinfo{person}{Adam Lerner}, \bibinfo{person}{Adam Shostack}, {and} \bibinfo{person}{Tadayoshi Kohno}.} \bibinfo{year}{2013}\natexlab{}.
\newblock \showarticletitle{{Control-Alt-Hack}: the design and evaluation of a card game for computer security awareness and education}. In \bibinfo{booktitle}{\emph{Proceedings of the 2013 {ACM} {SIGSAC} Conference on Computer and Communications Security}}. \bibinfo{publisher}{ACM}, \bibinfo{pages}{915--928}.
\newblock
\href{https://doi.org/10.1145/2508859.2516753}{doi:\nolinkurl{10.1145/2508859.2516753}}


\bibitem[Di~Gioia et~al\mbox{.}(2019)]%
        {gioia_cyber_2019}
\bibfield{author}{\bibinfo{person}{Rosanna Di~Gioia}, \bibinfo{person}{Stéphane Chaudron}, \bibinfo{person}{Monica Gemo}, {and} \bibinfo{person}{Ignacio Sanchez}.} \bibinfo{year}{2019}\natexlab{}.
\newblock \showarticletitle{{Cyber Chronix}, Participatory Research Approach to Develop and Evaluate a Storytelling Game on Personal Data Protection Rights and Privacy Risks}.
\newblock In \bibinfo{booktitle}{\emph{Games and Learning Alliance: 8th International Conference, GALA 2019, Athens, Greece, November 27--29, 2019, Proceedings}}. \bibinfo{series}{Lecture Notes in Computer Science}, Vol.~\bibinfo{volume}{11899}. \bibinfo{publisher}{Springer}, \bibinfo{pages}{221--230}.
\newblock
\href{https://doi.org/10.1007/978-3-030-34350-7_22}{doi:\nolinkurl{10.1007/978-3-030-34350-7_22}}


\bibitem[Doria et~al\mbox{.}(2024)]%
        {doria_designing_2024}
\bibfield{author}{\bibinfo{person}{Jan Doria}, \bibinfo{person}{Petra Grimm}, \bibinfo{person}{Michel Hohendanner}, {and} \bibinfo{person}{Susanne Kuhnert}.} \bibinfo{year}{2024}\natexlab{}.
\newblock \showarticletitle{Designing and Evaluating an Interactive Learning Technology to Foster Privacy Literacy}.
\newblock \bibinfo{journal}{\emph{{IEEE} Transactions on Learning Technologies}}  \bibinfo{volume}{17} (\bibinfo{year}{2024}), \bibinfo{pages}{827--840}.
\newblock
\href{https://doi.org/10.1109/TLT.2023.3333930}{doi:\nolinkurl{10.1109/TLT.2023.3333930}}


\bibitem[Drevin et~al\mbox{.}(2023)]%
        {drevin_story_2023}
\bibfield{author}{\bibinfo{person}{Günther Drevin}, \bibinfo{person}{Dirk Snyman}, \bibinfo{person}{Lynette Drevin}, \bibinfo{person}{Hennie Kruger}, {and} \bibinfo{person}{Johann Allers}.} \bibinfo{year}{2023}\natexlab{}.
\newblock \showarticletitle{The Story of Safety Snail and Her e-Mail: A Digital Wellness and Cybersecurity Serious Game for Pre-School Children}. In \bibinfo{booktitle}{\emph{Proceedings of the 9th International Conference on Information Systems Security and Privacy}}. \bibinfo{publisher}{SciTePress}, \bibinfo{pages}{519--527}.
\newblock
\href{https://doi.org/10.5220/0011682200003405}{doi:\nolinkurl{10.5220/0011682200003405}}


\bibitem[Faith et~al\mbox{.}(2022)]%
        {faith_intelligent_2022}
\bibfield{author}{\bibinfo{person}{B.~Fatokun Faith}, \bibinfo{person}{Zalizah~Awang Long}, \bibinfo{person}{Suraya Hamid}, \bibinfo{person}{O.~Fatokun Johnson}, \bibinfo{person}{Christopher~Ifeanyi Eke}, {and} \bibinfo{person}{Azah Norman}.} \bibinfo{year}{2022}\natexlab{}.
\newblock \showarticletitle{An Intelligent Gamification Tool to Boost Young Kids Cybersecurity Knowledge on {FB} Messenger}. In \bibinfo{booktitle}{\emph{Proceedings of the 2022 16th International Conference on Ubiquitous Information Management and Communication}}. \bibinfo{publisher}{IEEE}, \bibinfo{numpages}{8}~pages.
\newblock
\href{https://doi.org/10.1109/IMCOM53663.2022.9721733}{doi:\nolinkurl{10.1109/IMCOM53663.2022.9721733}}


\bibitem[Foley and Maddison(2010)]%
        {foley2010use}
\bibfield{author}{\bibinfo{person}{Louise Foley} {and} \bibinfo{person}{Ralph Maddison}.} \bibinfo{year}{2010}\natexlab{}.
\newblock \showarticletitle{Use of active video games to increase physical activity in children: a (virtual) reality?}
\newblock \bibinfo{journal}{\emph{Pediatric exercise science}} \bibinfo{volume}{22}, \bibinfo{number}{1} (\bibinfo{year}{2010}), \bibinfo{pages}{7--20}.
\newblock
\href{https://doi.org/10.1123/pes.22.1.7}{doi:\nolinkurl{10.1123/pes.22.1.7}}


\bibitem[Fountana et~al\mbox{.}(2011)]%
        {fountana_story_2011}
\bibfield{author}{\bibinfo{person}{Maria Fountana}, \bibinfo{person}{Dimitris Kalaitzis}, \bibinfo{person}{Eftychios Valeontis}, {and} \bibinfo{person}{Vasilis Delis}.} \bibinfo{year}{2011}\natexlab{}.
\newblock \showarticletitle{A Story on Internet Safety: Experiences from Developing a {VR} Gaming Environment}. In \bibinfo{booktitle}{\emph{New Horizons in Web‑Based Learning -- {ICWL} 2010 Workshops}} \emph{(\bibinfo{series}{Lecture Notes in Computer Science}, Vol.~\bibinfo{volume}{6537})}. \bibinfo{publisher}{Springer}, \bibinfo{pages}{21--27}.
\newblock
\href{https://doi.org/10.1007/978-3-642-20539-2_3}{doi:\nolinkurl{10.1007/978-3-642-20539-2_3}}


\bibitem[Fujikawa et~al\mbox{.}(2020)]%
        {fujikawa_sns_2020}
\bibfield{author}{\bibinfo{person}{Masaki Fujikawa}, \bibinfo{person}{Hajime Ikehara}, {and} \bibinfo{person}{Yoshie Abe}.} \bibinfo{year}{2020}\natexlab{}.
\newblock \showarticletitle{{SNS} Education Game for Upper-Grade Elementary School Students}. In \bibinfo{booktitle}{\emph{Proceedings of the 2020 8th International Conference on Information and Education Technology}}. \bibinfo{publisher}{ACM}, \bibinfo{pages}{137--141}.
\newblock
\href{https://doi.org/10.1145/3395245.3395248}{doi:\nolinkurl{10.1145/3395245.3395248}}


\bibitem[Gallud et~al\mbox{.}(2023)]%
        {gallud2023technology}
\bibfield{author}{\bibinfo{person}{Jose~A. Gallud}, \bibinfo{person}{Monica Carre{\~n}o}, \bibinfo{person}{Ricardo Tesoriero}, \bibinfo{person}{Andr{\'e}s Sandoval}, \bibinfo{person}{Mar{\'\i}a~D Lozano}, \bibinfo{person}{Israel Dur{\'a}n}, \bibinfo{person}{Victor M.~R. Penichet}, {and} \bibinfo{person}{Rafael Cosio}.} \bibinfo{year}{2023}\natexlab{}.
\newblock \showarticletitle{Technology-enhanced and game based learning for children with special needs: a systematic mapping study}.
\newblock \bibinfo{journal}{\emph{Universal Access in the Information Society}} \bibinfo{volume}{22}, \bibinfo{number}{1} (\bibinfo{year}{2023}), \bibinfo{pages}{227--240}.
\newblock
\href{https://doi.org/10.1007/s10209-021-00824-0}{doi:\nolinkurl{10.1007/s10209-021-00824-0}}


\bibitem[Ghazinour et~al\mbox{.}(2019)]%
        {ghazinour_digital-pass_2019}
\bibfield{author}{\bibinfo{person}{Kambiz Ghazinour}, \bibinfo{person}{Ken Messner}, \bibinfo{person}{Sean Scarnecchia}, {and} \bibinfo{person}{David Selinger}.} \bibinfo{year}{2019}\natexlab{}.
\newblock \showarticletitle{{Digital-PASS}: A Simulation-based Approach to Privacy Education}. In \bibinfo{booktitle}{\emph{Proceedings of the 18th {ACM} Workshop on Privacy in the Electronic Society}}. \bibinfo{publisher}{ACM}, \bibinfo{pages}{162--174}.
\newblock
\href{https://doi.org/10.1145/3338498.3358647}{doi:\nolinkurl{10.1145/3338498.3358647}}


\bibitem[Giannakas et~al\mbox{.}(2019)]%
        {giannakas_comprehensive_2019}
\bibfield{author}{\bibinfo{person}{Filippos Giannakas}, \bibinfo{person}{Andreas Papasalouros}, \bibinfo{person}{Georgios Kambourakis}, {and} \bibinfo{person}{Stefanos Gritzalis}.} \bibinfo{year}{2019}\natexlab{}.
\newblock \showarticletitle{A comprehensive cybersecurity learning platform for elementary education}.
\newblock \bibinfo{journal}{\emph{Information Security Journal: A Global Perspective}} \bibinfo{volume}{28}, \bibinfo{number}{3} (\bibinfo{year}{2019}), \bibinfo{pages}{81--106}.
\newblock
\href{https://doi.org/10.1080/19393555.2019.1657527}{doi:\nolinkurl{10.1080/19393555.2019.1657527}}


\bibitem[Gibson and O'Rawe(2018)]%
        {gibson2018virtual}
\bibfield{author}{\bibinfo{person}{Adam Gibson} {and} \bibinfo{person}{Martha O'Rawe}.} \bibinfo{year}{2018}\natexlab{}.
\newblock \showarticletitle{Virtual Reality as a Travel Promotional Tool: Insights from a Consumer Travel Fair}.
\newblock In \bibinfo{booktitle}{\emph{Virtual Reality and Tourism}}. \bibinfo{publisher}{Springer}, \bibinfo{pages}{93--107}.
\newblock
\href{https://doi.org/10.1007/978-3-319-64027-3_7}{doi:\nolinkurl{10.1007/978-3-319-64027-3_7}}


\bibitem[Gkioulos et~al\mbox{.}(2017)]%
        {gkioulos2017security}
\bibfield{author}{\bibinfo{person}{Vasileios Gkioulos}, \bibinfo{person}{Gaute Wangen}, \bibinfo{person}{Sokratis~K. Katsikas}, \bibinfo{person}{Georgios Kavallieratos}, {and} \bibinfo{person}{Panayiotis Kotzanikolaou}.} \bibinfo{year}{2017}\natexlab{}.
\newblock \showarticletitle{Security Awareness of the Digital Natives}.
\newblock \bibinfo{journal}{\emph{Information}} \bibinfo{volume}{8}, \bibinfo{number}{2}, Article \bibinfo{articleno}{42} (\bibinfo{year}{2017}), \bibinfo{numpages}{13}~pages.
\newblock
\href{https://doi.org/10.3390/info8020042}{doi:\nolinkurl{10.3390/info8020042}}


\bibitem[González-Tablas et~al\mbox{.}(2020)]%
        {gonzalez-tablas_shuffle_2020}
\bibfield{author}{\bibinfo{person}{Ana~I. González-Tablas}, \bibinfo{person}{María~I. González~Vasco}, \bibinfo{person}{Ignacio Cascos}, {and} \bibinfo{person}{Álvaro~Planet Palomino}.} \bibinfo{year}{2020}\natexlab{}.
\newblock \showarticletitle{Shuffle, Cut, and Learn: {Crypto Go}, a Card Game for Teaching Cryptography}.
\newblock \bibinfo{journal}{\emph{Mathematics}} \bibinfo{volume}{8}, \bibinfo{number}{11}, Article \bibinfo{articleno}{1993} (\bibinfo{year}{2020}), \bibinfo{numpages}{13}~pages.
\newblock
\href{https://doi.org/10.3390/math8111993}{doi:\nolinkurl{10.3390/math8111993}}


\bibitem[Guha et~al\mbox{.}(2004)]%
        {Guha2004mixing}
\bibfield{author}{\bibinfo{person}{Mona~Leigh Guha}, \bibinfo{person}{Allison Druin}, \bibinfo{person}{Gene Chipman}, \bibinfo{person}{Jerry~Alan Fails}, \bibinfo{person}{Sante Simms}, {and} \bibinfo{person}{Allison Farber}.} \bibinfo{year}{2004}\natexlab{}.
\newblock \showarticletitle{Mixing ideas: a new technique for working with young children as design partners}. In \bibinfo{booktitle}{\emph{Proceedings of the 2004 Conference on Interaction Design and Children: Building a Community}}. \bibinfo{publisher}{ACM}, \bibinfo{pages}{35--42}.
\newblock
\href{https://doi.org/10.1145/1017833.1017838}{doi:\nolinkurl{10.1145/1017833.1017838}}


\bibitem[Hardin and Dalsen(2020)]%
        {hardin_digital_2020}
\bibfield{author}{\bibinfo{person}{Caroline~D. Hardin} {and} \bibinfo{person}{Jen Dalsen}.} \bibinfo{year}{2020}\natexlab{}.
\newblock \showarticletitle{{Digital Privacy Detectives}: An Interactive Game for Classrooms}. In \bibinfo{booktitle}{\emph{Proceedings of the 2020 {IEEE} 44th Annual Computers, Software, and Applications Conference}}. \bibinfo{publisher}{IEEE}, \bibinfo{pages}{184--189}.
\newblock
\href{https://doi.org/10.1109/COMPSAC48688.2020.00033}{doi:\nolinkurl{10.1109/COMPSAC48688.2020.00033}}


\bibitem[Hartikainen et~al\mbox{.}(2019)]%
        {hartikainen_childrens_2019}
\bibfield{author}{\bibinfo{person}{Heidi Hartikainen}, \bibinfo{person}{Netta Iivari}, {and} \bibinfo{person}{Marianne Kinnula}.} \bibinfo{year}{2019}\natexlab{}.
\newblock \showarticletitle{Children's design recommendations for online safety education}.
\newblock \bibinfo{journal}{\emph{International Journal of Child-Computer Interaction}}  \bibinfo{volume}{22}, Article \bibinfo{articleno}{100146} (\bibinfo{year}{2019}), \bibinfo{numpages}{19}~pages.
\newblock
\href{https://doi.org/10.1016/j.ijcci.2019.100146}{doi:\nolinkurl{10.1016/j.ijcci.2019.100146}}


\bibitem[Heintz and Law(2015)]%
        {heintz2015game}
\bibfield{author}{\bibinfo{person}{Stephanie Heintz} {and} \bibinfo{person}{Effie Lai-Chong Law}.} \bibinfo{year}{2015}\natexlab{}.
\newblock \showarticletitle{The Game Genre Map: A Revised Game Classification}. In \bibinfo{booktitle}{\emph{Proceedings of the 2015 Annual Symposium on Computer-Human Interaction in Play}}. \bibinfo{publisher}{ACM}, \bibinfo{pages}{175--184}.
\newblock
\href{https://doi.org/10.1145/2793107.27931}{doi:\nolinkurl{10.1145/2793107.27931}}


\bibitem[Hodhod et~al\mbox{.}(2023)]%
        {hodhod_cyberhero_2023}
\bibfield{author}{\bibinfo{person}{Rania Hodhod}, \bibinfo{person}{Harlie Hardage}, \bibinfo{person}{Safia Abbas}, {and} \bibinfo{person}{Eman~Abdullah Aldakheel}.} \bibinfo{year}{2023}\natexlab{}.
\newblock \showarticletitle{{CyberHero}: An Adaptive Serious Game to Promote Cybersecurity Awareness}.
\newblock \bibinfo{journal}{\emph{Electronics}} \bibinfo{volume}{12}, \bibinfo{number}{17}, Article \bibinfo{articleno}{3544} (\bibinfo{year}{2023}), \bibinfo{numpages}{18}~pages.
\newblock
\href{https://doi.org/10.3390/electronics12173544}{doi:\nolinkurl{10.3390/electronics12173544}}


\bibitem[Huang and Schmidt(2022)]%
        {huang2022systematic}
\bibfield{author}{\bibinfo{person}{Rui Huang} {and} \bibinfo{person}{Matthew Schmidt}.} \bibinfo{year}{2022}\natexlab{}.
\newblock \showarticletitle{A systematic review of theory-informed design and implementation of digital game-based language learning}.
\newblock In \bibinfo{booktitle}{\emph{Digital Games in Language Learning}}. \bibinfo{publisher}{Taylor \& Francis}, \bibinfo{pages}{14--34}.
\newblock
\href{https://doi.org/10.4324/9781003240075-2}{doi:\nolinkurl{10.4324/9781003240075-2}}


\bibitem[Jaafar et~al\mbox{.}(2024)]%
        {jaafar_empowering_2024}
\bibfield{author}{\bibinfo{person}{Nur~Huda Jaafar}, \bibinfo{person}{Zuriati Ismail}, \bibinfo{person}{Nur~Diana Zamani}, \bibinfo{person}{Mazlyda Abd~Rahman}, {and} \bibinfo{person}{Rashidah Mokhtar}.} \bibinfo{year}{2024}\natexlab{}.
\newblock \showarticletitle{Empowering Indigenous Pupils: Enhancing Cyber Security Awareness Through Interactive Gaming Experiences}. In \bibinfo{booktitle}{\emph{Proceedings of the 2024 10th International Conference on Frontiers of Educational Technologies}}. \bibinfo{publisher}{ACM}, \bibinfo{pages}{96--103}.
\newblock
\href{https://doi.org/10.1145/3678392.3678397}{doi:\nolinkurl{10.1145/3678392.3678397}}


\bibitem[Jin et~al\mbox{.}(2018)]%
        {jin_game_2018}
\bibfield{author}{\bibinfo{person}{Ge Jin}, \bibinfo{person}{Manghui Tu}, \bibinfo{person}{Tae-Hoon Kim}, \bibinfo{person}{Justin Heffron}, {and} \bibinfo{person}{Jonathan White}.} \bibinfo{year}{2018}\natexlab{}.
\newblock \showarticletitle{Game based Cybersecurity Training for High School Students}. In \bibinfo{booktitle}{\emph{Proceedings of the 49th {ACM} Technical Symposium on Computer Science Education}}. \bibinfo{publisher}{ACM}, \bibinfo{pages}{68--73}.
\newblock
\href{https://doi.org/10.1145/3159450.3159591}{doi:\nolinkurl{10.1145/3159450.3159591}}


\bibitem[Kim et~al\mbox{.}(2021)]%
        {kim_security_2021}
\bibfield{author}{\bibinfo{person}{Hyung-Jong Kim}, \bibinfo{person}{Soyeon Park}, {and} \bibinfo{person}{Jin~B. Hong}.} \bibinfo{year}{2021}\natexlab{}.
\newblock \showarticletitle{Security Education and Training for Non--technical School Students Using Games}. In \bibinfo{booktitle}{\emph{Proceedings of the 2021 International Conference on Software Security and Assurance ({ICSSA})}}. \bibinfo{publisher}{IEEE}, \bibinfo{pages}{1--6}.
\newblock
\href{https://doi.org/10.1109/ICSSA53632.2021.00009}{doi:\nolinkurl{10.1109/ICSSA53632.2021.00009}}


\bibitem[Kumar et~al\mbox{.}(2018)]%
        {kumar_co-designing_2018}
\bibfield{author}{\bibinfo{person}{Priya Kumar}, \bibinfo{person}{Jessica Vitak}, \bibinfo{person}{Marshini Chetty}, \bibinfo{person}{Tamara~L. Clegg}, \bibinfo{person}{Jonathan Yang}, \bibinfo{person}{Brenna {McNally}}, {and} \bibinfo{person}{Elizabeth Bonsignore}.} \bibinfo{year}{2018}\natexlab{}.
\newblock \showarticletitle{Co-Designing Online Privacy-Related Games and Stories with Children}.
\newblock \bibinfo{journal}{\emph{Proceedings of the 2018 {ACM} Conference on Interaction Design and Children}} (\bibinfo{year}{2018}), \bibinfo{pages}{67--79}.
\newblock
\href{https://doi.org/10.1145/3202185.3202735}{doi:\nolinkurl{10.1145/3202185.3202735}}


\bibitem[Lazarinis et~al\mbox{.}(2015)]%
        {lazarinis_raising_2015}
\bibfield{author}{\bibinfo{person}{Fotis Lazarinis}, \bibinfo{person}{Kyriaki Alexandri}, \bibinfo{person}{Vasileios~S. Verykios}, {and} \bibinfo{person}{Chris Panagiotakopoulos}.} \bibinfo{year}{2015}\natexlab{}.
\newblock \showarticletitle{Raising safer internet awareness through a mobile application based on contrasting visual stories}. In \bibinfo{booktitle}{\emph{Proceedings of the 2015 International Conference on Interactive Mobile Communication Technologies and Learning}}. \bibinfo{publisher}{IEEE}, \bibinfo{pages}{88--90}.
\newblock
\href{https://doi.org/10.1109/IMCTL.2015.7359561}{doi:\nolinkurl{10.1109/IMCTL.2015.7359561}}


\bibitem[Maqsood and Chiasson(2021)]%
        {maqsood_design_2021}
\bibfield{author}{\bibinfo{person}{Sana Maqsood} {and} \bibinfo{person}{Sonia Chiasson}.} \bibinfo{year}{2021}\natexlab{}.
\newblock \showarticletitle{Design, Development, and Evaluation of a Cybersecurity, Privacy, and Digital Literacy Game for Tweens}.
\newblock \bibinfo{journal}{\emph{ACM Transactions on Privacy and Security}} \bibinfo{volume}{24}, \bibinfo{number}{4}, Article \bibinfo{articleno}{28} (\bibinfo{year}{2021}), \bibinfo{numpages}{37}~pages.
\newblock
\href{https://doi.org/10.1145/3469821}{doi:\nolinkurl{10.1145/3469821}}


\bibitem[Masadeh(2012)]%
        {masadeh2012training}
\bibfield{author}{\bibinfo{person}{Mousa Masadeh}.} \bibinfo{year}{2012}\natexlab{}.
\newblock \showarticletitle{Training, Education, Development and Learning: What is the difference?}
\newblock \bibinfo{journal}{\emph{European Scientific Journal}} \bibinfo{volume}{8}, \bibinfo{number}{10} (\bibinfo{year}{2012}).
\newblock
\urldef\tempurl%
\url{https://eujournal.org/index.php/esj/article/view/163}
\showURL{%
\tempurl}


\bibitem[{Meta}(2024)]%
        {metaquest_age_limit}
\bibfield{author}{\bibinfo{person}{{Meta}}.} \bibinfo{year}{2024}\natexlab{}.
\newblock \bibinfo{title}{{Meta} {Quest} Parent Information}.
\newblock
\urldef\tempurl%
\url{https://www.meta.com/gb/quest/parent-info/}
\showURL{%
\tempurl}
\newblock
\shownote{Accessed: 2024-07-11}.


\bibitem[{Microsoft}(2024)]%
        {hololens2_age_limit}
\bibfield{author}{\bibinfo{person}{{Microsoft}}.} \bibinfo{year}{2024}\natexlab{}.
\newblock \bibinfo{title}{Where {HoloLens} 2 is supported}.
\newblock
\urldef\tempurl%
\url{https://www.microsoft.com/en-gb/d/hololens-2/91pnzzznzwcp}
\showURL{%
\tempurl}
\newblock
\shownote{Accessed: 2024-07-11}.


\bibitem[Mikka-Muntuumo and Peters(2021)]%
        {mikka-muntuumo_designing_2021}
\bibfield{author}{\bibinfo{person}{Josephina Mikka-Muntuumo} {and} \bibinfo{person}{Anicia~Nicola Peters}.} \bibinfo{year}{2021}\natexlab{}.
\newblock \showarticletitle{Designing an Interactive Game for Preventing Online Abuse in Namibia}.
\newblock \bibinfo{journal}{\emph{Proceedings of the 2021 3rd International Multidisciplinary Information Technology and Engineering Conference}} (\bibinfo{year}{2021}), \bibinfo{numpages}{12}~pages.
\newblock
\href{https://doi.org/10.1109/IMITEC52926.2021.9714592}{doi:\nolinkurl{10.1109/IMITEC52926.2021.9714592}}


\bibitem[Mnisi et~al\mbox{.}(2024)]%
        {mnisi_digital_2024}
\bibfield{author}{\bibinfo{person}{Given Mnisi}, \bibinfo{person}{Gunther Drevin}, \bibinfo{person}{Lynette Drevin}, \bibinfo{person}{Joshua Esterhuizen}, {and} \bibinfo{person}{Christo Croucamp}.} \bibinfo{year}{2024}\natexlab{}.
\newblock \showarticletitle{Digital Wellness of Preschool Children: The Story of {Cyber--cat} and the Consequences of Hacking}. In \bibinfo{booktitle}{\emph{Proceedings of Ninth International Congress on Information and Communication Technology}} \emph{(\bibinfo{series}{Lecture Notes in Networks and Systems}, Vol.~\bibinfo{volume}{1013})}. \bibinfo{publisher}{Springer}, \bibinfo{pages}{635–646}.
\newblock
\urldef\tempurl%
\url{https://link.springer.com/chapter/10.1007/978-981-97-3559-4_51}
\showURL{%
\tempurl}


\bibitem[Mohammed et~al\mbox{.}(2024)]%
        {mohammed_designing_2024}
\bibfield{author}{\bibinfo{person}{Shafi~Parvez Mohammed}, \bibinfo{person}{Gahangir Hossain}, \bibinfo{person}{Syed~Yaseen Quadri}, \bibinfo{person}{Steven Keosouvanh}, {and} \bibinfo{person}{Mikyung Shin}.} \bibinfo{year}{2024}\natexlab{}.
\newblock \showarticletitle{Designing Ethical Hacking Training Through Game‑Based Design for High School Students (9th--12th Grade)}. In \bibinfo{booktitle}{\emph{Proceedings of the 2024 {IEEE} Integrated STEM Education Conference ({ISEC})}}. \bibinfo{publisher}{IEEE}.
\newblock
\href{https://doi.org/10.1109/ISEC61299.2024.10664724}{doi:\nolinkurl{10.1109/ISEC61299.2024.10664724}}


\bibitem[Murray(2020)]%
        {murray2020more}
\bibfield{author}{\bibinfo{person}{Jack Murray}.} \bibinfo{year}{2020}\natexlab{}.
\newblock \showarticletitle{More Than Just the Table: Analog Games as Computational Platforms}. In \bibinfo{booktitle}{\emph{Proceedings of the 15th International Conference on the Foundations of Digital Games}}. \bibinfo{publisher}{ACM}, Article \bibinfo{articleno}{42}, \bibinfo{numpages}{4}~pages.
\newblock
\href{https://doi.org/10.1145/3402942.3402974}{doi:\nolinkurl{10.1145/3402942.3402974}}


\bibitem[Natalia et~al\mbox{.}(2023)]%
        {natalia2023gamification}
\bibfield{author}{\bibinfo{person}{Monica~Christy Natalia}, \bibinfo{person}{Setiyo Cayhono}, \bibinfo{person}{Rahmat Purwoko}, {and} \bibinfo{person}{I~Gede~Maha Putra}.} \bibinfo{year}{2023}\natexlab{}.
\newblock \showarticletitle{Gamification Design as Learning Media to Motivate Students to Increase Cyber Security Awareness towards Phishing}. In \bibinfo{booktitle}{\emph{Proceedings of the 2023 International Conference on Informatics, Multimedia, Cyber and Informations System}}. \bibinfo{publisher}{IEEE}, \bibinfo{pages}{252--256}.
\newblock
\href{https://doi.org/10.1109/ICIMCIS60089.2023.10349069}{doi:\nolinkurl{10.1109/ICIMCIS60089.2023.10349069}}


\bibitem[{National Cyber Security Centre, UK}(2023)]%
        {NCSC2023what}
\bibfield{author}{\bibinfo{person}{{National Cyber Security Centre, UK}}.} \bibinfo{year}{2023}\natexlab{}.
\newblock \bibinfo{title}{What is Cyber Security?}
\newblock
\urldef\tempurl%
\url{https://www.ncsc.gov.uk/section/about-ncsc/what-is-cyber-security}
\showURL{%
\tempurl}
\newblock
\shownote{Accessed: 2024-07-11}.


\bibitem[{National Institute of Standards and Technology (NIST)}(2003)]%
        {nist2003special800-50}
\bibfield{author}{\bibinfo{person}{{National Institute of Standards and Technology (NIST)}}.} \bibinfo{year}{2003}\natexlab{}.
\newblock \bibinfo{booktitle}{\emph{Building an Information Technology Security Awareness and Training Program}}.
\newblock \bibinfo{type}{NIST Special Publication} 800-50. \bibinfo{institution}{National Institute of Standards and Technology ({NIST})}.
\newblock
\urldef\tempurl%
\url{https://nvlpubs.nist.gov/nistpubs/Legacy/SP/nistspecialpublication800-50.pdf}
\showURL{%
\tempurl}


\bibitem[Neo et~al\mbox{.}(2021)]%
        {neo_safe_2021}
\bibfield{author}{\bibinfo{person}{Han-Foon Neo}, \bibinfo{person}{Chuan-Chin Teo}, {and} \bibinfo{person}{Chee~Lim Peng}.} \bibinfo{year}{2021}\natexlab{}.
\newblock \showarticletitle{Safe Internet: An Edutainment Tool for Teenagers}.
\newblock In \bibinfo{booktitle}{\emph{Information Science and Applications: Proceedings of ICISA 2020}}. \bibinfo{series}{Lecture Notes in Electrical Engineering}, Vol.~\bibinfo{volume}{739}. \bibinfo{publisher}{Springer}, \bibinfo{pages}{53--70}.
\newblock
\href{https://doi.org/10.1007/978-981-33-6385-4_6}{doi:\nolinkurl{10.1007/978-981-33-6385-4_6}}


\bibitem[{NHS}(2024)]%
        {age_nhs}
\bibfield{author}{\bibinfo{person}{{NHS}}.} \bibinfo{year}{2024}\natexlab{}.
\newblock \bibinfo{title}{Age}.
\newblock
\urldef\tempurl%
\url{https://service-manual.nhs.uk/content/inclusive-content/age}
\showURL{%
\tempurl}
\newblock
\shownote{Accessed: 2024-07-11}.


\bibitem[{Ofcom, UK}(2024)]%
        {Ofcom2024}
\bibfield{author}{\bibinfo{person}{{Ofcom, UK}}.} \bibinfo{year}{2024}\natexlab{}.
\newblock \bibinfo{title}{Children's Media Use and Attitudes Report 2024}.
\newblock
\urldef\tempurl%
\url{https://www.ofcom.org.uk/siteassets/resources/documents/research-and-data/media-literacy-research/children/children-media-use-and-attitudes-2024/childrens-media-literacy-report-2024.pdf}
\showURL{%
\tempurl}
\newblock
\shownote{Accessed: 2024-06-30}.


\bibitem[Olano et~al\mbox{.}(2014)]%
        {olano_securityempire_2014}
\bibfield{author}{\bibinfo{person}{Marc Olano}, \bibinfo{person}{Alan Sherman}, \bibinfo{person}{Linda Oliva}, \bibinfo{person}{Ryan Cox}, \bibinfo{person}{Deborah Firestone}, \bibinfo{person}{Oliver Kubik}, \bibinfo{person}{Milind Patil}, \bibinfo{person}{John Seymour}, \bibinfo{person}{Isaac Sohn}, {and} \bibinfo{person}{Donna Thomas}.} \bibinfo{year}{2014}\natexlab{}.
\newblock \showarticletitle{{SecurityEmpire}: Development and Evaluation of a Digital Game to Promote Cybersecurity Education}. In \bibinfo{booktitle}{\emph{Proceedings of the 2014 USENIX Summit on Gaming, Games, and Gamification in Security Education}}. \bibinfo{publisher}{USENIX}, \bibinfo{numpages}{10}~pages.
\newblock
\urldef\tempurl%
\url{https://www.usenix.org/conference/3gse14/summit-program/presentation/olano}
\showURL{%
\tempurl}


\bibitem[Page et~al\mbox{.}(2021)]%
        {prisma2020}
\bibfield{author}{\bibinfo{person}{Matthew~J. Page}, \bibinfo{person}{Joanne~E. McKenzie}, \bibinfo{person}{Patrick~M. Bossuyt}, \bibinfo{person}{Isabelle Boutron}, \bibinfo{person}{Tammy~C. Hoffmann}, \bibinfo{person}{Cynthia~D. Mulrow}, \bibinfo{person}{Larissa Shamseer}, \bibinfo{person}{Jennifer~M. Tetzlaff}, \bibinfo{person}{Elie~A. Akl}, \bibinfo{person}{Sue~E. Brennan}, \bibinfo{person}{Roger Chou}, \bibinfo{person}{Julie Glanville}, \bibinfo{person}{Jeremy~M. Grimshaw}, \bibinfo{person}{Asbjørn Hróbjartsson}, \bibinfo{person}{Manoj~M. Lalu}, \bibinfo{person}{Tianjing Li}, \bibinfo{person}{Elizabeth~W Loder}, \bibinfo{person}{Evan Mayo-Wilson}, \bibinfo{person}{Steve McDonald}, \bibinfo{person}{Luke~A. McGuinness}, \bibinfo{person}{Lesley~A. Stewart}, \bibinfo{person}{James Thomas}, \bibinfo{person}{Andrea~C. Tricco}, \bibinfo{person}{Vivian~A. Welch}, \bibinfo{person}{Penny Whiting}, {and} \bibinfo{person}{David Moher}.} \bibinfo{year}{2021}\natexlab{}.
\newblock \showarticletitle{The {PRISMA} 2020 statement: an updated guideline for reporting systematic reviews}.
\newblock \bibinfo{journal}{\emph{BMJ}}  \bibinfo{volume}{372}, Article \bibinfo{articleno}{n71} (\bibinfo{year}{2021}), \bibinfo{numpages}{9}~pages.
\newblock
\href{https://doi.org/10.1136/bmj.n71}{doi:\nolinkurl{10.1136/bmj.n71}}


\bibitem[Panga et~al\mbox{.}(2022)]%
        {panga2022game}
\bibfield{author}{\bibinfo{person}{Rosemary Cosmas~Tlatlaa Panga}, \bibinfo{person}{Janeth Marwa}, {and} \bibinfo{person}{Jema~David Ndibwile}.} \bibinfo{year}{2022}\natexlab{}.
\newblock \showarticletitle{A Game or Notes? The Use of a Customized Mobile Game to Improve Teenagers' Phishing Knowledge, Case of Tanzania}.
\newblock \bibinfo{journal}{\emph{Journal of Cybersecurity and Privacy}} \bibinfo{volume}{2}, \bibinfo{number}{3} (\bibinfo{year}{2022}), \bibinfo{pages}{466--489}.
\newblock
\href{https://doi.org/10.3390/jcp2030024}{doi:\nolinkurl{10.3390/jcp2030024}}


\bibitem[Pellicone et~al\mbox{.}(2022)]%
        {pellicone_designing_2022}
\bibfield{author}{\bibinfo{person}{Anthony Pellicone}, \bibinfo{person}{Diane~Jass Ketelhut}, \bibinfo{person}{Ekta Shokeen}, \bibinfo{person}{David Weintrop}, \bibinfo{person}{Mchel Cukier}, {and} \bibinfo{person}{Jandelyn~Dawn Plane}.} \bibinfo{year}{2022}\natexlab{}.
\newblock \showarticletitle{Designing a Game to Promote Equity in Cybersecurity}.
\newblock \bibinfo{journal}{\emph{Proceedings of the 16th European Conference on Games-based Learning}} (\bibinfo{year}{2022}), \bibinfo{pages}{448--454}.
\newblock
\href{https://doi.org/10.34190/ecgbl.16.1.825}{doi:\nolinkurl{10.34190/ecgbl.16.1.825}}


\bibitem[Piccolo et~al\mbox{.}(2021)]%
        {piccolo2021chatbots}
\bibfield{author}{\bibinfo{person}{Lara Schibelsky~Godoy Piccolo}, \bibinfo{person}{Pinelopi Troullinou}, {and} \bibinfo{person}{Harith Alani}.} \bibinfo{year}{2021}\natexlab{}.
\newblock \showarticletitle{Chatbots to support children in coping with online threats: Socio-technical requirements}. In \bibinfo{booktitle}{\emph{Proceedings of the 2021 ACM Designing Interactive Systems Conference}}. \bibinfo{publisher}{ACM}, \bibinfo{pages}{1504--1517}.
\newblock
\href{https://doi.org/10.1145/3461778.3462114}{doi:\nolinkurl{10.1145/3461778.3462114}}


\bibitem[Quayyum(2023)]%
        {quayyum_collaboration_2023}
\bibfield{author}{\bibinfo{person}{Farzana Quayyum}.} \bibinfo{year}{2023}\natexlab{}.
\newblock \showarticletitle{Collaboration between parents and children to raise cybersecurity awareness}. In \bibinfo{booktitle}{\emph{Proceedings of the 2023 European Interdisciplinary Cybersecurity Conference}}. \bibinfo{publisher}{ACM}, \bibinfo{pages}{149--152}.
\newblock
\href{https://doi.org/10.1145/3590777.3590802}{doi:\nolinkurl{10.1145/3590777.3590802}}


\bibitem[Quayyum et~al\mbox{.}(2021)]%
        {quayyum2021cyber}
\bibfield{author}{\bibinfo{person}{Farihah Quayyum}, \bibinfo{person}{Daniela~S. Cruzes}, {and} \bibinfo{person}{Letizia Jaccheri}.} \bibinfo{year}{2021}\natexlab{}.
\newblock \showarticletitle{Cyber security awareness for children: A systematic literature review}.
\newblock \bibinfo{journal}{\emph{International Journal of Child-Computer Interaction}}  \bibinfo{volume}{30}, Article \bibinfo{articleno}{100343} (\bibinfo{year}{2021}), \bibinfo{numpages}{25}~pages.
\newblock
\href{https://doi.org/10.1016/j.ijcci.2021.100343}{doi:\nolinkurl{10.1016/j.ijcci.2021.100343}}


\bibitem[Qusa and Tarazi(2021)]%
        {qusa_cyber-hero_2021}
\bibfield{author}{\bibinfo{person}{Hani Qusa} {and} \bibinfo{person}{Jumana Tarazi}.} \bibinfo{year}{2021}\natexlab{}.
\newblock \showarticletitle{{Cyber-Hero}: A Gamification framework for Cyber Security Awareness for High Schools Students}.
\newblock \bibinfo{journal}{\emph{Proceedings of the 2021 {IEEE} 11th Annual Computing and Communication Workshop and Conference}} (\bibinfo{year}{2021}), \bibinfo{pages}{677--682}.
\newblock
\href{https://doi.org/10.1109/CCWC51732.2021.9375847}{doi:\nolinkurl{10.1109/CCWC51732.2021.9375847}}


\bibitem[Raihana et~al\mbox{.}(2024)]%
        {raihana_implementation_2024}
\bibfield{author}{\bibinfo{person}{Fikra~Amalia Raihana}, \bibinfo{person}{Hermawan Setiawan}, \bibinfo{person}{Herman Kabetta}, {and} \bibinfo{person}{Nurul Qomariasih}.} \bibinfo{year}{2024}\natexlab{}.
\newblock \showarticletitle{Implementation of Game‑Based Learning to Enhance Security Awareness Against Child Cyber Grooming Attacks}. In \bibinfo{booktitle}{\emph{Proceedings of the 2024 8th International Conference on Information Technology, Information Systems and Electrical Engineering ({ICITISEE})}}. \bibinfo{publisher}{IEEE}, \bibinfo{pages}{238--243}.
\newblock
\href{https://doi.org/10.1109/ICITISEE63424.2024.10729941}{doi:\nolinkurl{10.1109/ICITISEE63424.2024.10729941}}


\bibitem[Raynes-Goldie and Allen(2014)]%
        {raynes-goldie_gaming_2014}
\bibfield{author}{\bibinfo{person}{Kate Raynes-Goldie} {and} \bibinfo{person}{Matthew Allen}.} \bibinfo{year}{2014}\natexlab{}.
\newblock \showarticletitle{{Gaming Privacy}: a {Canadian} case study of a children's co-created privacy literacy game}.
\newblock \bibinfo{journal}{\emph{Surveillance \& Society}} \bibinfo{volume}{12}, \bibinfo{number}{3} (\bibinfo{year}{2014}), \bibinfo{pages}{414--426}.
\newblock
\href{https://doi.org/10.24908/ss.v12i3.4958}{doi:\nolinkurl{10.24908/ss.v12i3.4958}}


\bibitem[Roepke and Schroeder(2020)]%
        {roepke_teaching_2020}
\bibfield{author}{\bibinfo{person}{Rene Roepke} {and} \bibinfo{person}{Ulrik Schroeder}.} \bibinfo{year}{2020}\natexlab{}.
\newblock \showarticletitle{Teaching Defence Against the Dark Arts Using Game-Based Learning: A Review of Learning Games for Cybersecurity Education}. In \bibinfo{booktitle}{\emph{Computer Supported Education: 11th International Conference, CSEDU 2019, Heraklion, Crete, Greece, May 2-4, 2019, Revised Selected Papers}}. \bibinfo{publisher}{Springer}, \bibinfo{pages}{71--87}.
\newblock
\href{https://doi.org/10.1007/978-3-030-58459-7_4}{doi:\nolinkurl{10.1007/978-3-030-58459-7_4}}


\bibitem[Salazar et~al\mbox{.}(2013)]%
        {salazar_augmenting_2013}
\bibfield{author}{\bibinfo{person}{Mikel Salazar}, \bibinfo{person}{José Gaviria}, \bibinfo{person}{Carlos Laorden}, {and} \bibinfo{person}{Pablo García~Bringas}.} \bibinfo{year}{2013}\natexlab{}.
\newblock \showarticletitle{Enhancing cybersecurity learning through an augmented reality-based serious game}. In \bibinfo{booktitle}{\emph{Proceedings of the 2013 {IEEE} Global Engineering Education Conference ({EDUCON})}}. \bibinfo{publisher}{IEEE}, \bibinfo{pages}{602--607}.
\newblock
\href{https://doi.org/10.1109/EduCon.2013.6530167}{doi:\nolinkurl{10.1109/EduCon.2013.6530167}}


\bibitem[Saunders et~al\mbox{.}(2019)]%
        {saunders2019validating}
\bibfield{author}{\bibinfo{person}{Jonathan Saunders}, \bibinfo{person}{Steffi Davey}, \bibinfo{person}{Petra~Saskia Bayerl}, {and} \bibinfo{person}{Philipp Lohrmann}.} \bibinfo{year}{2019}\natexlab{}.
\newblock \showarticletitle{Validating Virtual Reality as an Effective Training Medium in the Security Domain}. In \bibinfo{booktitle}{\emph{Proceedings of the 26th {IEEE} Conference on Virtual Reality and 3D User Interfaces}}. \bibinfo{publisher}{IEEE}, \bibinfo{pages}{1908--1911}.
\newblock
\href{https://doi.org/10.1109/VR.2019.8798371}{doi:\nolinkurl{10.1109/VR.2019.8798371}}


\bibitem[Sağlam et~al\mbox{.}(3 06)]%
        {saglam_systematic_2023}
\bibfield{author}{\bibinfo{person}{Rahime~Belen Sağlam}, \bibinfo{person}{Vincent Miller}, {and} \bibinfo{person}{Virginia N.~L. Franqueira}.} \bibinfo{year}{2023-06}\natexlab{}.
\newblock \showarticletitle{A Systematic Literature Review on Cyber Security Education for Children}.
\newblock \bibinfo{journal}{\emph{IEEE Transactions on Education}} \bibinfo{volume}{66}, \bibinfo{number}{3} (\bibinfo{year}{2023-06}), \bibinfo{pages}{274--286}.
\newblock
\href{https://doi.org/10.1109/TE.2022.3231019}{doi:\nolinkurl{10.1109/TE.2022.3231019}}


\bibitem[Scholl and Schuktomow(2020)]%
        {scholl_information_2020}
\bibfield{author}{\bibinfo{person}{Margit Scholl} {and} \bibinfo{person}{Regina Schuktomow}.} \bibinfo{year}{2020}\natexlab{}.
\newblock \showarticletitle{Information Security at Schools: A Practical Game-Based Application with Sustained Impact}.
\newblock \bibinfo{journal}{\emph{Journal of Systemics, Cybernetics and Informatics}} \bibinfo{volume}{18}, \bibinfo{number}{5} (\bibinfo{year}{2020}), \bibinfo{pages}{74--85}.
\newblock
\urldef\tempurl%
\url{https://www.iiisci.org/journal/sci/Abstract.asp?var=&id=SA989TJ20}
\showURL{%
\tempurl}


\bibitem[Schuktomow et~al\mbox{.}(2020)]%
        {schuktomow_play_2020}
\bibfield{author}{\bibinfo{person}{Regina Schuktomow}, \bibinfo{person}{Margit Scholl}, \bibinfo{person}{Peter Koppatz}, {and} \bibinfo{person}{Denis Edich}.} \bibinfo{year}{2020}\natexlab{}.
\newblock \showarticletitle{Play the Game and be Aware: Information Security Project with Schools}.
\newblock \bibinfo{journal}{\emph{Proceedings of the 14th IADIS International Conference Interfaces and Human Computer Interaction and the 13th IADIS International Conference Game and Entertainment Technologies}} (\bibinfo{year}{2020}), \bibinfo{pages}{59--66}.
\newblock
\urldef\tempurl%
\url{https://www.iadisportal.org/digital-library/play-the-game-and-be-aware-information-security-project-with-schools}
\showURL{%
\tempurl}


\bibitem[Serrano et~al\mbox{.}(2012)]%
        {serrano2012framework}
\bibfield{author}{\bibinfo{person}{{\'A}ngel Serrano}, \bibinfo{person}{Eugenio~J Marchiori}, \bibinfo{person}{{\'A}ngel del Blanco}, \bibinfo{person}{Javier Torrente}, {and} \bibinfo{person}{Baltasar Fern{\'a}ndez-Manj{\'o}n}.} \bibinfo{year}{2012}\natexlab{}.
\newblock \showarticletitle{A framework to improve evaluation in educational games}. In \bibinfo{booktitle}{\emph{Proceedings of the 2012 {IEEE} Global Engineering Education Conference}}. \bibinfo{publisher}{IEEE}, \bibinfo{numpages}{8}~pages.
\newblock
\href{https://doi.org/10.1109/EDUCON.2012.6201154}{doi:\nolinkurl{10.1109/EDUCON.2012.6201154}}


\bibitem[Shen et~al\mbox{.}(2021a)]%
        {shen_work_progress_design_2021}
\bibfield{author}{\bibinfo{person}{Chien~Chung Shen}, \bibinfo{person}{Yan-Ming Chiou}, \bibinfo{person}{Chrystalla Mouza}, {and} \bibinfo{person}{Teomara Rutherford}.} \bibinfo{year}{2021}\natexlab{a}.
\newblock \showarticletitle{Work-in-Progress---Design and Evaluation of Mixed Reality Programs for Cybersecurity Education}. In \bibinfo{booktitle}{\emph{Proceedings of the 2021 7th International Conference of the Immersive Learning Research Network}}. \bibinfo{publisher}{IEEE}.
\newblock
\href{https://doi.org/10.23919/iLRN52045.2021.9459309}{doi:\nolinkurl{10.23919/iLRN52045.2021.9459309}}


\bibitem[Shen et~al\mbox{.}(2021b)]%
        {shen_cyber_2021}
\bibfield{author}{\bibinfo{person}{Low~Wan Shen}, \bibinfo{person}{Hazinah~Kutty Mammi}, {and} \bibinfo{person}{Mazura~Mat Din}.} \bibinfo{year}{2021}\natexlab{b}.
\newblock \showarticletitle{Cyber Security Awareness Game ({CSAG}) for Secondary School Students}. In \bibinfo{booktitle}{\emph{Proceedings of the 2021 International Conference on Data Science and Its Applications}}. \bibinfo{publisher}{IEEE}, \bibinfo{pages}{48--53}.
\newblock
\href{https://doi.org/10.1109/ICoDSA53588.2021.9617548}{doi:\nolinkurl{10.1109/ICoDSA53588.2021.9617548}}


\bibitem[Silva(2019)]%
        {silva2019practical}
\bibfield{author}{\bibinfo{person}{Frutuoso G.~M. Silva}.} \bibinfo{year}{2019}\natexlab{}.
\newblock \showarticletitle{Practical Methodology for the Design of Educational Serious Games}.
\newblock \bibinfo{journal}{\emph{Information}} \bibinfo{volume}{11}, \bibinfo{number}{1} (\bibinfo{year}{2019}), \bibinfo{pages}{1--13}.
\newblock
\href{https://doi.org/10.3390/info11010014}{doi:\nolinkurl{10.3390/info11010014}}


\bibitem[Snyman et~al\mbox{.}(2021)]%
        {snyman_wolf_2021}
\bibfield{author}{\bibinfo{person}{Dirk~P. Snyman}, \bibinfo{person}{Gunther~R. Drevin}, \bibinfo{person}{Hennie~A. Kruger}, \bibinfo{person}{Lynette Drevin}, {and} \bibinfo{person}{Johann Allers}.} \bibinfo{year}{2021}\natexlab{}.
\newblock \showarticletitle{A Wolf, Hyena, and Fox Game to Raise Cybersecurity Awareness Among Pre-school Children}.
\newblock In \bibinfo{booktitle}{\emph{Information Security Education for Cyber Resilience: 14th IFIP WG 11.8 World Conference, WISE 2021, Virtual Event, June 22–24, 2021, Proceedings}}. \bibinfo{series}{IFIP Advances in Information and Communication Technology}, Vol.~\bibinfo{volume}{615}. \bibinfo{publisher}{Springer}, \bibinfo{pages}{91--101}.
\newblock
\href{https://doi.org/10.1007/978-3-030-81111-2_8}{doi:\nolinkurl{10.1007/978-3-030-81111-2_8}}


\bibitem[Sudha et~al\mbox{.}(2023)]%
        {sudha_impact_2023}
\bibfield{author}{\bibinfo{person}{Sai~Sushmitha Sudha}, \bibinfo{person}{Jyothi~Priyanka Bandreddi}, \bibinfo{person}{Laxmi~Mounika Podila}, \bibinfo{person}{Ramesh Govindula}, \bibinfo{person}{Austin Richardson}, \bibinfo{person}{Quamar Niyaz}, \bibinfo{person}{Xiaoli Yang}, {and} \bibinfo{person}{Ahmad~Y. Javaid}.} \bibinfo{year}{2023}\natexlab{}.
\newblock \showarticletitle{Impact of Smartphone-Based Interactive Learning Modules on Cybersecurity Learning at the High-School Level}. In \bibinfo{booktitle}{\emph{Proceedings of the 2023 {IEEE} Global Engineering Education Conference}}. \bibinfo{publisher}{IEEE}, \bibinfo{numpages}{10}~pages.
\newblock
\href{https://doi.org/10.1109/EDUCON54358.2023.10125124}{doi:\nolinkurl{10.1109/EDUCON54358.2023.10125124}}


\bibitem[Thomas et~al\mbox{.}(2019)]%
        {thomas2019educational}
\bibfield{author}{\bibinfo{person}{Michael~K. Thomas}, \bibinfo{person}{Andria Shyjka}, \bibinfo{person}{Skip Kumm}, {and} \bibinfo{person}{Rigel Gjomemo}.} \bibinfo{year}{2019}\natexlab{}.
\newblock \showarticletitle{Educational Design Research for the Development of a Collectible Card Game for Cybersecurity Learning}.
\newblock \bibinfo{journal}{\emph{Journal of Formative Design in Learning}}  \bibinfo{volume}{3} (\bibinfo{year}{2019}), \bibinfo{pages}{27--38}.
\newblock
\href{https://doi.org/10.1007/s41686-019-00027-0}{doi:\nolinkurl{10.1007/s41686-019-00027-0}}


\bibitem[Tjostheim et~al\mbox{.}(2022)]%
        {tjostheim_dark_2022}
\bibfield{author}{\bibinfo{person}{Ingvar Tjostheim}, \bibinfo{person}{Vanessa Ayres-Pereira}, \bibinfo{person}{Chris Wales}, \bibinfo{person}{Angela Manna}, {and} \bibinfo{person}{Simon Egenfeldt-Nielsen}.} \bibinfo{year}{2022}\natexlab{}.
\newblock \showarticletitle{Dark Pattern: A Serious Game for Learning About the Dangers of Sharing Data}.
\newblock \bibinfo{journal}{\emph{Proceedings of the 16th European Conference on Games-based Learning}} (\bibinfo{year}{2022}), \bibinfo{pages}{774--783}.
\newblock
\href{https://doi.org/10.34190/ecgbl.16.1.872}{doi:\nolinkurl{10.34190/ecgbl.16.1.872}}


\bibitem[Toledo et~al\mbox{.}(2022)]%
        {toledo_netdefense_2022}
\bibfield{author}{\bibinfo{person}{William Toledo}, \bibinfo{person}{Sushil~J. Louis}, {and} \bibinfo{person}{Shamik Sengupta}.} \bibinfo{year}{2022}\natexlab{}.
\newblock \showarticletitle{{NetDefense}: A Tower Defense Cybersecurity Game for Middle and High School Students}. In \bibinfo{booktitle}{\emph{Proceedings of the 2022 {IEEE} Frontiers in Education Conference}}. \bibinfo{publisher}{IEEE}, \bibinfo{numpages}{6}~pages.
\newblock
\href{https://doi.org/10.1109/FIE56618.2022.9962410}{doi:\nolinkurl{10.1109/FIE56618.2022.9962410}}


\bibitem[Tschakert and Ngamsuriyaroj(2019)]%
        {tschakert2019effectiveness}
\bibfield{author}{\bibinfo{person}{K.~F. Tschakert} {and} \bibinfo{person}{S. Ngamsuriyaroj}.} \bibinfo{year}{2019}\natexlab{}.
\newblock \showarticletitle{{Effectiveness of and user preferences for security awareness training methodologies}}.
\newblock \bibinfo{journal}{\emph{Heliyon}} \bibinfo{volume}{5}, \bibinfo{number}{6}, Article \bibinfo{articleno}{e02010} (\bibinfo{year}{2019}), \bibinfo{numpages}{10}~pages.
\newblock
\href{https://doi.org/10.1016/j.heliyon.2019.e02010}{doi:\nolinkurl{10.1016/j.heliyon.2019.e02010}}


\bibitem[Tseng et~al\mbox{.}(2024)]%
        {tseng_building_2024}
\bibfield{author}{\bibinfo{person}{Shian‑Shyong Tseng}, \bibinfo{person}{Tsung-Yu Yang}, \bibinfo{person}{Wen-Chung Shih}, {and} \bibinfo{person}{Bo-Yang Shan}.} \bibinfo{year}{2024}\natexlab{}.
\newblock \showarticletitle{Building a Self--Evolving {iMonsters} Board Game for Cyber--Security Education}.
\newblock \bibinfo{journal}{\emph{Interactive Learning Environments}} \bibinfo{volume}{32}, \bibinfo{number}{4} (\bibinfo{year}{2024}), \bibinfo{pages}{1300--1318}.
\newblock
\href{https://doi.org/10.1080/10494820.2022.2120015}{doi:\nolinkurl{10.1080/10494820.2022.2120015}}


\bibitem[Tuparova et~al\mbox{.}(2021)]%
        {tuparova2021learning}
\bibfield{author}{\bibinfo{person}{Daniela Tuparova}, \bibinfo{person}{Georgi Tuparov}, {and} \bibinfo{person}{Krista Mehandzhiyska}.} \bibinfo{year}{2021}\natexlab{}.
\newblock \showarticletitle{Learning Topic ``{Safe Internet}'' in Low Secondary School through Games}. In \bibinfo{booktitle}{\emph{Proceedings of the 2021 44th International Convention on Information, Communication and Electronic Technology}}. \bibinfo{publisher}{IEEE}, \bibinfo{pages}{716--723}.
\newblock
\href{https://doi.org/10.23919/MIPRO52101.2021.9597032}{doi:\nolinkurl{10.23919/MIPRO52101.2021.9597032}}


\bibitem[Tyner and Rajabion(2022)]%
        {tyner_framework_2022}
\bibfield{author}{\bibinfo{person}{Georgia Tyner} {and} \bibinfo{person}{Lila Rajabion}.} \bibinfo{year}{2022}\natexlab{}.
\newblock \showarticletitle{Framework for Developing Cybersecurity Activities for Children in Grades {K}-5}.
\newblock \bibinfo{journal}{\emph{Proceedings of the 2022 International Conference on Computational Science and Computational Intelligence}} (\bibinfo{year}{2022}), \bibinfo{pages}{2034--2040}.
\newblock
\href{https://doi.org/10.1109/CSCI58124.2022.00365}{doi:\nolinkurl{10.1109/CSCI58124.2022.00365}}


\bibitem[Ulsamer et~al\mbox{.}(2021)]%
        {ulsamer2021immersive}
\bibfield{author}{\bibinfo{person}{Philipp Ulsamer}, \bibinfo{person}{Andreas Schütz}, \bibinfo{person}{Tobias Fertig}, {and} \bibinfo{person}{Lisa Keller}.} \bibinfo{year}{2021}\natexlab{}.
\newblock \showarticletitle{Immersive Storytelling for Information Security Awareness Training in Virtual Reality}. In \bibinfo{booktitle}{\emph{Proceedings of the 54th Annual Hawaii International Conference on System Sciences}}. \bibinfo{publisher}{University of Hawaii at Mānoa}, \bibinfo{pages}{7153--7162}.
\newblock
\href{https://doi.org/10.24251/hicss.2021.861}{doi:\nolinkurl{10.24251/hicss.2021.861}}


\bibitem[({UNICEF})(2016)]%
        {unicef_uncrc_2016}
\bibfield{author}{\bibinfo{person}{{United Nations International Children's Emergency Fund} ({UNICEF})}.} \bibinfo{year}{2016}\natexlab{}.
\newblock \bibinfo{title}{{UN Convention on the Rights of the Child}}.
\newblock
\urldef\tempurl%
\url{https://www.unicef.org.uk/wp-content/uploads/2016/08/unicef-convention-rights-child-uncrc.pdf}
\showURL{%
\tempurl}
\newblock
\shownote{Accessed: 2024-07-02}.


\bibitem[van~der Stappen et~al\mbox{.}(2019)]%
        {stappen2019mathbuilder}
\bibfield{author}{\bibinfo{person}{Almar van~der Stappen}, \bibinfo{person}{Yunjie Liu}, \bibinfo{person}{Jiangxue Xu}, \bibinfo{person}{Xiaoyu Yu}, \bibinfo{person}{Jingya Li}, {and} \bibinfo{person}{Erik~D. van~der Spek}.} \bibinfo{year}{2019}\natexlab{}.
\newblock \showarticletitle{{MathBuilder}: A Collaborative {AR} Math Game for Elementary School Students}. In \bibinfo{booktitle}{\emph{Extended Abstracts: Extended Abstracts of the Annual Symposium on Computer-Human Interaction in Play Companion Extended Abstracts}}. \bibinfo{publisher}{ACM}, \bibinfo{pages}{731--738}.
\newblock
\href{https://doi.org/10.1145/3341215.3356295}{doi:\nolinkurl{10.1145/3341215.3356295}}


\bibitem[van Schaik et~al\mbox{.}(2018)]%
        {vanSchaik2018security}
\bibfield{author}{\bibinfo{person}{Paul van Schaik}, \bibinfo{person}{Jurjen Jansen}, \bibinfo{person}{Joseph Onibokun}, \bibinfo{person}{Jean Camp}, {and} \bibinfo{person}{Petko Kusev}.} \bibinfo{year}{2018}\natexlab{}.
\newblock \showarticletitle{Security and privacy in online social networking: Risk perceptions and precautionary behaviour}.
\newblock \bibinfo{journal}{\emph{Computers in Human Behavior}}  \bibinfo{volume}{78} (\bibinfo{year}{2018}), \bibinfo{pages}{283--297}.
\newblock
\href{https://doi.org/10.1016/j.chb.2017.10.007}{doi:\nolinkurl{10.1016/j.chb.2017.10.007}}


\bibitem[Videnovik et~al\mbox{.}(2024)]%
        {videnovik_novel_2024}
\bibfield{author}{\bibinfo{person}{Maja Videnovik}, \bibinfo{person}{Sonja Filiposka}, {and} \bibinfo{person}{Vladimir Trajkovik}.} \bibinfo{year}{2024}\natexlab{}.
\newblock \showarticletitle{A Novel Methodological Approach for Learning Cybersecurity Topics in Primary Schools}.
\newblock \bibinfo{journal}{\emph{Multimedia Tools and Applications}} (\bibinfo{year}{2024}).
\newblock
\href{https://doi.org/10.1007/s11042-024-20077-2}{doi:\nolinkurl{10.1007/s11042-024-20077-2}}


\bibitem[Visoottiviseth et~al\mbox{.}(2018)]%
        {visoottiviseth_pomega_2018}
\bibfield{author}{\bibinfo{person}{Vasaka Visoottiviseth}, \bibinfo{person}{Rossarin Sainont}, \bibinfo{person}{Thanatorn Boonnak}, {and} \bibinfo{person}{Vorapas Thammakulkrajang}.} \bibinfo{year}{2018}\natexlab{}.
\newblock \showarticletitle{{POMEGA}: Security Game for Building Security Awareness}. In \bibinfo{booktitle}{\emph{Proceedings of the 2018 Seventh {ICT} International Student Project Conference ({ICT--ISPC 2018})}}. \bibinfo{publisher}{IEEE}, \bibinfo{pages}{71--76}.
\newblock
\href{https://doi.org/10.1109/ICT-ISPC.2018.8523965}{doi:\nolinkurl{10.1109/ICT-ISPC.2018.8523965}}


\bibitem[Vogel et~al\mbox{.}(2006)]%
        {vogel2006computer}
\bibfield{author}{\bibinfo{person}{Jennifer~J. Vogel}, \bibinfo{person}{David~S. Vogel}, \bibinfo{person}{Jan Cannon-Bowers}, \bibinfo{person}{Clint~A. Bowers}, \bibinfo{person}{Kathryn Muse}, {and} \bibinfo{person}{Michelle Wright}.} \bibinfo{year}{2006}\natexlab{}.
\newblock \showarticletitle{Computer Gaming and Interactive Simulations for Learning: A Meta-Analysis}.
\newblock \bibinfo{journal}{\emph{Journal of Educational Computing Research}} \bibinfo{volume}{34}, \bibinfo{number}{3} (\bibinfo{year}{2006}), \bibinfo{pages}{229--243}.
\newblock
\href{https://doi.org/10.2190/FLHV-K4WA-WPVQ-H0YM}{doi:\nolinkurl{10.2190/FLHV-K4WA-WPVQ-H0YM}}


\bibitem[Von~Solms and Van~Niekerk(2013)]%
        {vonsolms2013information}
\bibfield{author}{\bibinfo{person}{Rossouw Von~Solms} {and} \bibinfo{person}{Johan Van~Niekerk}.} \bibinfo{year}{2013}\natexlab{}.
\newblock \showarticletitle{From information security to cyber security}.
\newblock \bibinfo{journal}{\emph{Computers \& Security}}  \bibinfo{volume}{38} (\bibinfo{year}{2013}), \bibinfo{pages}{97--102}.
\newblock
\href{https://doi.org/10.1016/j.cose.2013.04.004}{doi:\nolinkurl{10.1016/j.cose.2013.04.004}}


\bibitem[Waldock et~al\mbox{.}(2022)]%
        {waldock2022pre-univeristy}
\bibfield{author}{\bibinfo{person}{Krysia~Emily Waldock}, \bibinfo{person}{Vince Miller}, \bibinfo{person}{Shujun Li}, {and} \bibinfo{person}{Virginia N.~L. Franqueira}.} \bibinfo{year}{2022}\natexlab{}.
\newblock \bibinfo{booktitle}{\emph{Pre-University Cyber Security Education: A report on developing cyber skills amongst children and young people}}.
\newblock \bibinfo{type}{{T}echnical {R}eport}. \bibinfo{institution}{Global Forum on Cyber Expertise (GFCE) and Institute of Cyber Security for Society (iCSS), University of Kent}.
\newblock
\urldef\tempurl%
\url{https://cybilportal.org/publications/pre-university-cyber-security-education-a-report-on-developing-cyber-skills-amongst-children-and-young-people/}
\showURL{%
\tempurl}


\bibitem[Yamin et~al\mbox{.}(2021)]%
        {yamin_serious_2021}
\bibfield{author}{\bibinfo{person}{Muhammad~Mudassar Yamin}, \bibinfo{person}{Basel Katt}, {and} \bibinfo{person}{Mariusz Nowostawski}.} \bibinfo{year}{2021}\natexlab{}.
\newblock \showarticletitle{Serious games as a tool to model attack and defense scenarios for cyber-security exercises}.
\newblock \bibinfo{journal}{\emph{Computers \& Security}}  \bibinfo{volume}{110}, Article \bibinfo{articleno}{102450} (\bibinfo{year}{2021}), \bibinfo{numpages}{22}~pages.
\newblock
\href{https://doi.org/10.1016/j.cose.2021.102450}{doi:\nolinkurl{10.1016/j.cose.2021.102450}}


\bibitem[Yeoh et~al\mbox{.}(2025)]%
        {yeoh2025immersive}
\bibfield{author}{\bibinfo{person}{William Yeoh}, \bibinfo{person}{Sophie McKenzie}, \bibinfo{person}{Robin Doss}, \bibinfo{person}{Graeme Pye}, \bibinfo{person}{Radhika Gorur}, {and} \bibinfo{person}{Jeb Webb}.} \bibinfo{year}{2025}\natexlab{}.
\newblock \showarticletitle{Immersive Metaverse Learning for Children’s Cyber Harm Awareness}.
\newblock \bibinfo{journal}{\emph{Journal of Computer Information Systems}} \bibinfo{volume}{0}, \bibinfo{number}{0} (\bibinfo{year}{2025}), \bibinfo{pages}{1--11}.
\newblock
\href{https://doi.org/10.1080/08874417.2025.2537110}{doi:\nolinkurl{10.1080/08874417.2025.2537110}}


\bibitem[Yuan et~al\mbox{.}(2024)]%
        {yuan_redcapes_2024}
\bibfield{author}{\bibinfo{person}{Xiaowen Yuan}, \bibinfo{person}{Hongni Ye}, \bibinfo{person}{Ziheng Tang}, \bibinfo{person}{Xiangrong Zhu}, \bibinfo{person}{Yaxing Yao}, {and} \bibinfo{person}{Xin Tong}.} \bibinfo{year}{2024}\natexlab{}.
\newblock \showarticletitle{{RedCapes}: the Design and Evaluation of a Game Towards Improving Autistic Children's Privacy Awareness}. In \bibinfo{booktitle}{\emph{Proceedings of the Eleventh International Symposium of {Chinese CHI}}}. \bibinfo{publisher}{ACM}, \bibinfo{pages}{110--126}.
\newblock
\href{https://doi.org/10.1145/3629606.3629618}{doi:\nolinkurl{10.1145/3629606.3629618}}


\bibitem[Zahed et~al\mbox{.}(2019)]%
        {zahed_play_2019}
\bibfield{author}{\bibinfo{person}{Bushra~Tasnim Zahed}, \bibinfo{person}{Gregory White}, {and} \bibinfo{person}{John Quarles}.} \bibinfo{year}{2019}\natexlab{}.
\newblock \showarticletitle{Play It Safe: An Educational Cyber Safety Game for Children in Elementary School}. In \bibinfo{booktitle}{\emph{Proceedings of the 2019 11th International Conference on Virtual Worlds and Games for Serious Applications}}. \bibinfo{publisher}{IEEE}, \bibinfo{numpages}{4}~pages.
\newblock
\href{https://doi.org/10.1109/VS-Games.2019.8864594}{doi:\nolinkurl{10.1109/VS-Games.2019.8864594}}


\end{thebibliography}

\newpage

\appendix

\section{Appendix: List of Papers Included in the SLR (Supplemental Content)}
\label{appendix_A}

\footnotesize
\begin{longtable}{p{1cm}p{2.5cm}p{8.5cm}p{1cm}p{1cm}}
\midrule
\textbf{ID} & \textbf{Authors} & \textbf{Title} & \textbf{Year} & \textbf{Paper}\\
\midrule
\endfirsthead

\midrule
\textbf{ID} & \textbf{Authors} & \textbf{Title} & \textbf{Year} & \textbf{Paper}\\
\midrule
\endhead

P1 & \citeauthor{azzahra_socenggo_2024} & SocengGo: Social Engineering Educational Application Based on Attack-Defense Multiplayer Card Game & 2024 & \cite{azzahra_socenggo_2024}\\
\midrule
P2 & \citeauthor{blinder_evaluating_2024} & Evaluating the Use of Hypothetical `Would You Rather' Scenarios to Discuss Privacy and Security Concepts with Children & 2024 & \cite{blinder_evaluating_2024}\\
\midrule
P3 & \citeauthor{chattopadhyay_vpet_2024} & VPET: A Novel Visual Privacy Themed Cybersecurity Educational Game & 2024 & \cite{chattopadhyay_vpet_2024}\\
\midrule
P4 & \citeauthor{doria_designing_2024} & Designing and Evaluating an Interactive Learning Technology to Foster Privacy Literacy & 2024 & \cite{doria_designing_2024}\\
\midrule
P5 & \citeauthor{giuseppe_blue_2024} & Blue and Red team quiz game to train high school students & 2024 & \cite{giuseppe_blue_2024}\\
\midrule
P6 & \citeauthor{jaafar_empowering_2024} & Empowering Indigenous Pupils: Enhancing Cyber Security Awareness Through Interactive Gaming Experiences & 2024 & \cite{jaafar_empowering_2024}\\
\midrule
P7 & \citeauthor{mnisi_digital_2024} & Digital Wellness of Preschool Children: The Story of {Cyber-cat} and the Consequences of Hacking & 2024 & \cite{mnisi_digital_2024}\\
\midrule
P8 & \citeauthor{mohammed_designing_2024} & Designing Ethical Hacking Training Through Game-Based Design for High School Students (9th--12th Grade) & 2024 & \cite{mohammed_designing_2024}\\
\midrule
P9 & \citeauthor{raihana_implementation_2024} & Implementation of Game-Based Learning to Enhance Security Awareness Against Child Cyber Grooming Attacks & 2024 & \cite{raihana_implementation_2024}\\
\midrule
P10 & \citeauthor{tseng_building_2024} & Building a self-evolving {iMonsters} board game for cyber-security education & 2024 & \cite{tseng_building_2024}\\
\midrule
P11 & \citeauthor{videnovik_novel_2024} & A novel methodological approach for learning cybersecurity topics in primary schools & 2024 & \cite{videnovik_novel_2024}\\
\midrule
P12 & \citeauthor{bassi2023serious} & A Serious Video Game on Cybersecurity & 2023 & \cite{bassi2023serious}\\
\midrule
P13  & \citeauthor{casey_motivating_2023} & Motivating youth to learn {STEM} through a gender inclusive digital forensic science program & 2023 & \cite{casey_motivating_2023}\\
\midrule
P14 & \citeauthor{christensen2023privacy} & {The Privacy Universe} -- A game-Based learning platform for data protection, privacy and ethics & 2023 & \cite{christensen2023privacy}\\
\midrule
P15 & \citeauthor{del2023sec} & {SEC--GAME}: A Minigame Collection for Cyber Security Awareness & 2023 & \cite{del2023sec}\\
\midrule
P16  & \citeauthor{drevin_story_2023} & The Story of Safety Snail and Her e-Mail: A Digital Wellness and Cybersecurity Serious Game for Pre-School Children & 2023 & \cite{drevin_story_2023}\\
\midrule
P17  & \citeauthor{hodhod_cyberhero_2023} & {CyberHero}: An Adaptive Serious Game to Promote Cybersecurity Awareness & 2023 & \cite{hodhod_cyberhero_2023}\\
\midrule
P18 & \citeauthor{natalia2023gamification} & Gamification Design as Learning Media to Motivate Students to Increase Cyber Security Awareness towards Phishing & 2023 & \cite{natalia2023gamification}\\
\midrule
P19  & \citeauthor{quayyum_collaboration_2023} & Collaboration between parents and children to raise cybersecurity awareness & 2023 & \cite{quayyum_collaboration_2023}\\
\midrule
P20  & \citeauthor{sudha_impact_2023} & Impact of Smartphone-Based Interactive Learning Modules on Cybersecurity Learning at the High-School Level & 2023 & \cite{sudha_impact_2023}\\
\midrule
P21 & \citeauthor{alma2022soceng} & Soceng Warriors: Game-Based Learning to Increase Security Awareness Against Social Engineering Attacks & 2022 & \cite{alma2022soceng}\\
\midrule
P22 & \citeauthor{cardoso_playing_2022} & Playing for Privacy Awareness: Learning from a ``Wow-Moment'' with {iBuddy} & 2022 & \cite{cardoso_playing_2022}\\
\midrule
P23 & \citeauthor{decusatis_cybersecurity_2022} & A Cybersecurity Awareness Escape Room using Gamification Design Principles & 2022 & \cite{decusatis_cybersecurity_2022}\\
\midrule
P24 & \citeauthor{decusatis_gamification_2022} & Gamification of cybersecurity training & 2022 & \cite{decusatis_gamification_2022}\\
\midrule
P25 & \citeauthor{faith_intelligent_2022} & An Intelligent Gamification Tool to Boost Young Kids Cybersecurity Knowledge on {FB} Messenger & 2022 & \cite{faith_intelligent_2022}\\
\midrule
P26 & \citeauthor{panga2022game} & A Game or Notes? The Use of a Customized Mobile Game to Improve Teenagers' Phishing Knowledge & 2022 & \cite{panga2022game}\\
\midrule
P27 & \citeauthor{pellicone_designing_2022} & Designing a Game to Promote Equity in Cybersecurity & 2022 & \cite{pellicone_designing_2022}\\
\midrule
P28 & \citeauthor{tjostheim_dark_2022} & {Dark Pattern}: A Serious Game for Learning About the Dangers of Sharing Data & 2022 & \cite{tjostheim_dark_2022}\\
\midrule
P29 & \citeauthor{toledo_netdefense_2022} & {NetDefense}: A Tower Defense Cybersecurity Game for Middle and High School Students & 2022 & \cite{toledo_netdefense_2022}\\
\midrule
P30 & \citeauthor{tyner_framework_2022} & Framework for Developing Cybersecurity Activities for Children in Grades K--5 & 2022 & \cite{tyner_framework_2022}\\
\midrule
P31 & \citeauthor{allers_childrens_2021} & Children’s Awareness of Digital Wellness: A Serious Games Approach & 2021 & \cite{allers_childrens_2021}\\
\midrule
P32 & \citeauthor{balakrishna_design_2021} & Design considerations for developing a game-based learning resource for cyber security education & 2021 & \cite{balakrishna_design_2021}\\
\midrule
P33 & \citeauthor{chiou_augmented_2021} & Augmented Reality-Based Cybersecurity Education on Phishing & 2021 & \cite{chiou_augmented_2021}\\
\midrule
P34 & \citeauthor{kim_security_2021} & Security Education and Training for Non-technical School Students using Games & 2021 & \cite{kim_security_2021}\\
\midrule
P35 & \citeauthor{maqsood_design_2021} & Design, Development, and Evaluation of a Cybersecurity, Privacy, and Digital Literacy Game for Tweens & 2021 & \cite{maqsood_design_2021}\\
\midrule
P36 & \citeauthor{mikka-muntuumo_designing_2021} & Designing an Interactive Game for Preventing Online Abuse in Namibia & 2021 & \cite{mikka-muntuumo_designing_2021}\\
\midrule
P37 & \citeauthor{qusa_cyber-hero_2021} & {Cyber-Hero}: A Gamification framework for Cyber Security Awareness for High Schools Students & 2021 & \cite{qusa_cyber-hero_2021}\\
\midrule
P38 & \citeauthor{shen_cyber_2021} & Cyber Security Awareness Game ({CSAG}) for Secondary School Students & 2021 & \cite{shen_cyber_2021}\\
\midrule
P39 & \citeauthor{shen_work_progress_design_2021} & Work-in-progress---design and evaluation of mixed reality programs for cybersecurity education & 2021 & \cite{shen_work_progress_design_2021}\\
\midrule
P40 & \citeauthor{snyman_wolf_2021} & A Wolf, Hyena, and Fox Game to Raise Cybersecurity Awareness Among Pre-school Children & 2021 & \cite{snyman_wolf_2021}\\
\midrule
P41 & \citeauthor{tuparova2021learning} & Learning Topic `Safe Internet' in Low Secondary School through Games & 2021 & \cite{tuparova2021learning}\\
\midrule
P42 & \citeauthor{yamin_serious_2021} & Serious games as a tool to model attack and defense scenarios for cyber-security exercises & 2021 & \cite{yamin_serious_2021}\\
\midrule
P43 & \citeauthor{neo_safe_2021} & Safe Internet: An Edutainment Tool for Teenagers & 2021 & \cite{neo_safe_2021}\\
\midrule
P44 & \citeauthor{alemany_assessing_2020} & Assessing the Effectiveness of a Gamified Social Network for Applying Privacy Concepts: An Empirical Study With Teens & 2020 & \cite{alemany_assessing_2020}\\
\midrule
P45 & \citeauthor{alsadhan_manar_2020} & {Manar}: An Arabic Game-based Application Aimed for Teaching Cybersecurity using Image Processing & 2020 & \cite{alsadhan_manar_2020}\\
\midrule
P46 & \citeauthor{costa_nerd_2020} & A {NERD DOGMA}: Introducing {CTF} to Non-expert Audience & 2020 & \cite{costa_nerd_2020}\\
\midrule
P47 & \citeauthor{fujikawa_sns_2020} & {SNS} Education Game for Upper-Grade Elementary School Students & 2020 & \cite{fujikawa_sns_2020}\\
\midrule
P48 & \citeauthor{gonzalez-tablas_shuffle_2020} & Shuffle, cut, and learn: Crypto go, a card game for teaching cryptography & 2020 & \cite{gonzalez-tablas_shuffle_2020}\\
\midrule
P49 & \citeauthor{hardin_digital_2020} & Digital Privacy Detectives: An Interactive Game for Classrooms & 2020 & \cite{hardin_digital_2020}\\
\midrule
P50 & \citeauthor{scholl_information_2020} & Information security at schools: A practical game-Based application with sustained impact & 2020 & \cite{scholl_information_2020}\\
\midrule
P51 & \citeauthor{schuktomow_play_2020} & Play the game and be aware: Information security project with schools & 2020 & \cite{schuktomow_play_2020}\\
\midrule
P52 & \citeauthor{berger_privacity_2019} & {PrivaCity} A Chatbot Game to Raise Privacy Awareness Among Teenagers & 2019 & \cite{berger_privacity_2019}\\
\midrule
P53 & \citeauthor{bioglio_social_2019} & A Social Network Simulation Game to Raise Awareness of Privacy Among School Children & 2019 & \cite{bioglio_social_2019}\\
\midrule
P54 & \citeauthor{ghazinour_digital-pass_2019} & {Digital-PASS}: A simulation-based approach to privacy education & 2019 & \cite{ghazinour_digital-pass_2019}\\
\midrule
P55 & \citeauthor{giannakas_comprehensive_2019} & A comprehensive cybersecurity learning platform for elementary education & 2019 & \cite{giannakas_comprehensive_2019}\\
\midrule
P56 & \citeauthor{gioia_cyber_2019} & {Cyber Chronix}, Participatory Research Approach to Develop and Evaluate a Storytelling Game on Personal Data Protection Rights and Privacy Risks & 2019 & \cite{gioia_cyber_2019}\\
\midrule
P57 & \citeauthor{thomas2019educational} & Educational Design Research for the Development of a Collectible Card Game for Cybersecurity Learning & 2019 & \cite{thomas2019educational}\\
\midrule
P58 & \citeauthor{zahed_play_2019} & Play It Safe: An Educational Cyber Safety Game for Children in Elementary School & 2019 & \cite{zahed_play_2019}\\
\midrule
P59 & \citeauthor{jin_game_2018} & Game based Cybersecurity Training for High School Students & 2018 & \cite{jin_game_2018}\\
\midrule
P60 & \citeauthor{kumar_co-designing_2018} & Co-designing online privacy-related games and stories with children & 2018 & \cite{kumar_co-designing_2018}\\
\midrule
P61 & \citeauthor{visoottiviseth_pomega_2018} & {POMEGA}: Security Game for Building Security Awareness & 2018 & \cite{visoottiviseth_pomega_2018}\\
\midrule
P62 & \citeauthor{yuan_redcapes_2024} & {RedCapes}: the Design and Evaluation of a Game Towards Improving Autistic Children's Privacy Awareness & 2018 & \cite{yuan_redcapes_2024}\\
\midrule
P63 & \citeauthor{lazarinis_raising_2015} & Raising safer internet awareness through a mobile application based on contrasting visual stories & 2015 & \cite{lazarinis_raising_2015}\\
\midrule
P64 & \citeauthor{olano_securityempire_2014} & {SecurityEmpire}: Development and evaluation of a digital game to promote cybersecurity education & 2014 & \cite{olano_securityempire_2014}\\
\midrule
P65 & \citeauthor{raynes-goldie_gaming_2014} & Gaming Privacy: a Canadian case study of a children’s co-created privacy literacy game & 2014 & \cite{raynes-goldie_gaming_2014}\\
\midrule
P66 & \citeauthor{denning_control-alt-hack_2013} & {Control--Alt--Hack}: The design and evaluation of a card game for computer security awareness and education & 2013 & \cite{denning_control-alt-hack_2013}\\
\midrule
P67 & \citeauthor{salazar_augmenting_2013} & Enhancing cybersecurity learning through an augmented reality-based serious game & 2013 & \cite{salazar_augmenting_2013}\\
\midrule
P68 & \citeauthor{fountana_story_2011} & A story on Internet safety: Experiences from developing a {VR} gaming environment & 2011 & \cite{fountana_story_2011}\\
\bottomrule
\end{longtable}

\end{document}